%% file: main.tex
\documentclass[11pt]{article}
\pdfoutput=1
\usepackage{amsmath}
\usepackage{amssymb}
\usepackage{graphicx,bbm,mathrsfs}
\usepackage{nicefrac}
\usepackage{slashed}
\usepackage{bbm}
\usepackage{geometry}
\usepackage{empheq}
\usepackage{stackrel}
\usepackage{ulem}
\usepackage{jheppub}
\usepackage{setspace}
\usepackage{relsize}
\usepackage{empheq}
\usepackage{pifont}
\usepackage{mathtools}
\usepackage{enumitem}
\usepackage{dcolumn}   
\usepackage{bm}        
\usepackage{graphicx,mathrsfs}
\usepackage{nicefrac}
\usepackage{multirow}
\usepackage{color}
\usepackage{mathtools}
\usepackage{makecell}
\usepackage{environ} 
\usepackage{lipsum} 
\usepackage{wasysym}
\usepackage{hhline,colortbl}  
\usepackage{tikz}
\usepackage{graphicx}
\usepackage{tensor}
\usepackage{dsfont}
\usepackage{simpler-wick}

\usepackage{booktabs}
\usepackage{mdframed}
\usepackage{xcolor}

\usepackage{physics}
\usepackage{breqn} 

\usepackage{bbold}

 \NewEnviron{Smaller11}{
           \scalebox{1.1}{$\BODY$} 
 } 
 \NewEnviron{Smaller08}{
           \scalebox{0.8}{$\BODY$} 
 }

\newcommand{\be}{\begin{equation}}
\newcommand{\ee}{\end{equation}}
\newcommand{\bea}{\begin{eqnarray}}
\newcommand{\eea}{\end{eqnarray}}

\newcommand{\vv}{\kappa}
\newcommand{\vsd}{\alpha}
\newcommand{\vhat}{\hat \alpha}

\newcommand{\x}{x}

\newcommand{\y}{y}

\renewcommand\labelenumi{{(\roman{enumi}})}
\renewcommand\theenumi\labelenumi

\renewcommand{\arraystretch}{1.1} 

\newcommand{\nn}{\nonumber}

\newcommand{\h}{h}

\newcommand{\diagram}[2][0.23]{%
    \hspace{-1em}\vcenter{\hbox{\includegraphics[scale=#1]{#2}}}\hspace{-1em}%
}

\usepackage{titlesec}

\titleformat*{\section}{\Large\bfseries}
\titleformat*{\subsection}{\large\bfseries}
\titleformat*{\subsubsection}{\large\bfseries}
\titleformat*{\paragraph}{\large\bfseries}
\titleformat*{\subparagraph}{\large\bfseries}

\makeatletter
\newcommand*{\prodsym}{%
  \DOTSB
  \mathop{
    \mathchoice
      {\rlap{\kern.3em\rotatebox[origin=c]{-90}{}}{\prod}}
      {\vcenter{\rlap{\kern.2em\rotatebox[origin=c]{-90}{}}}{\prod}}
      {\sum}{\sum}
  }\slimits@
}
\makeatother

\makeatletter
\DeclareFontFamily{OMX}{MnSymbolE}{}
\DeclareSymbolFont{MnLargeSymbols}{OMX}{MnSymbolE}{m}{n}
\SetSymbolFont{MnLargeSymbols}{bold}{OMX}{MnSymbolE}{b}{n}
\DeclareFontShape{OMX}{MnSymbolE}{m}{n}{
    <-6>  MnSymbolE5
   <6-7>  MnSymbolE6
   <7-8>  MnSymbolE7
   <8-9>  MnSymbolE8
   <9-10> MnSymbolE9
  <10-12> MnSymbolE10
  <12->   MnSymbolE12
}{}
\DeclareFontShape{OMX}{MnSymbolE}{b}{n}{
    <-6>  MnSymbolE-Bold5
   <6-7>  MnSymbolE-Bold6
   <7-8>  MnSymbolE-Bold7
   <8-9>  MnSymbolE-Bold8
   <9-10> MnSymbolE-Bold9
  <10-12> MnSymbolE-Bold10
  <12->   MnSymbolE-Bold12
}{}

\let\llangle\@undefined
\let\rrangle\@undefined
\DeclareMathDelimiter{\llangle}{\mathopen}%
                     {MnLargeSymbols}{'164}{MnLargeSymbols}{'164}
\DeclareMathDelimiter{\rrangle}{\mathclose}%
                     {MnLargeSymbols}{'171}{MnLargeSymbols}{'171}
\makeatother

\begin{document}

\vspace*{4mm}

\thispagestyle{empty}

\begin{center}

\begin{minipage}{20cm}
\begin{center}
\hspace{-5cm }
\Large
\sc
$N$-Photon Amplitudes in  EFT 
\\  \hspace{-5cm }    
from Recursion Relations and Effective Vertices
\end{center}
\end{minipage}
\\[30mm]

\renewcommand{\thefootnote}{\fnsymbol{footnote}}

{\large  
Gustavo~Bispo$^{\, a}$ \footnote{gbispo@ymail.com},\,
Sylvain~Fichet$^{\, a,b}$ \footnote{sylvain.fichet@unesp.br}\,, 
}\\[12mm]
\end{center} 
\noindent

\noindent
\quad\quad\quad \textit{$^a$ CCNH - Universidade Federal do ABC}

\noindent \quad\quad\quad \textit{Avenida dos Estados, 5001, Santo Andre, 09210-580 SP, Brazil} \\

\noindent
\quad\quad\quad \textit{$^b$  IFT - UNESP \&  ICTP South American Institute for Fundamental Research  }

\noindent \quad\quad\quad \textit{R. Dr. Bento Teobaldo Ferraz 271, S\~ao Paulo,  2019-032 SP, Brazil}

\addtocounter{footnote}{-2}

\vspace*{12mm}
 
\begin{center}
{  \bf  Abstract }
\end{center}
\begin{minipage}{15cm}
\setstretch{0.95}

We present a recursive framework for computing arbitrary tree-level helicity amplitudes in the general effective field theory (EFT) of electromagnetism. We show that  photon amplitudes can be constructed using a CSW-like recursion, whose building blocks consist of purely local subamplitudes. 
For a fixed number of external legs, only a finite number of these contact amplitudes  need to be computed, and these admit compact expressions in terms of Hafnians.  
We further show that the contact amplitudes are encoded in an effective contact Lagrangian, that can be  computed systematically from the general EFT Lagrangian using a suitable generating functional. We provide this contact Lagrangian up to $N=14$ photons. 
We argue that the contact Lagrangian reveals the hidden simplicity of helicity-conserving amplitudes,  providing a simple proof of the  equivalence between off-shell electromagnetic duality and helicity conservation. We further show how to resum the contact Lagrangian of a helicity-conserving theory to all $N$, and illustrate the procedure for Born-Infeld (BI) electromagnetism. 
Building on these methods and results,  
we compute the complete 6-point and 8-point tree-level helicity amplitudes generated by generic EFT operators, and the tree amplitudes up to 10-point in  BI electromagnetism.

    \vspace{0.5cm}
\end{minipage}

\newpage
\setcounter{tocdepth}{3}

\tableofcontents  

\vspace{1cm}
\hrule
\vspace{1cm}

\section{Introduction \label{se:Intro}}

\input{Sections/Intro.tex}
    \input{Sections/operators.tex}

    \input{Sections/self_dual_decomp.tex}

    \input{Sections/helicity_amplitudes}

    \input{Sections/csw_expansion}

    \input{Sections/contact_eff_lag}

    \input{Sections/BI_bootstrap}
    \input{Sections/applications}
\input{Sections/Summary.tex}

\appendix 

\input{Appendices/shf_review}

\input{Appendices/comparing_amps}

\input{Appendices/traditional_amps}

\input{Appendices/interactions}

\bibliographystyle{JHEP}
\normalem
\bibliography{biblio}

\end{document}

%% file: Sections/Intro.tex
On-shell scattering amplitudes exhibit remarkable factorization properties. An on-shell amplitude generally contains a factorizable part that satisfies a recursion relation involving amplitudes with fewer external states \cite{Cachazo:2004kj, Britto:2004ap, Britto:2005fq}.
Beyond their formal interest, recursion relations are especially valuable because, depending on the underlying theory and  the polarizations of the external states, they can be turned into efficient computational tools.  The factorizable part of an amplitude can then be reconstructed directly from lower-point data, bypassing the need for Feynman diagrams and cumbersome combinatorial factors. 
 See \cite{Berends:1987me, Bern:1994zx, Bern:1994cg}
for earlier seminal works and \cite{Ellis:2011cr,Elvang:2013cua, Dixon:2013uaa, Cheung:2017pzi, Feng:2011np} for  reviews and textbooks on on-shell methods and factorization. 

Some spectacular examples of such computations are known in Yang-Mills theory and QCD  (see e.g. \cite{Georgiou:2004by,Georgiou:2004wu, Bern:2005ji, Bern:2005cq, Berger:2006vq, Bern:2007dw}),   supersymmetric theories (see e.g. \cite{Britto:2005ha, Drummond:2008cr}), general relativity \cite{Bjerrum-Bohr:2005xoa,Bjerrum-Bohr:2013bxa} and the Standard Model \cite{Berger:2006sh, Badger:2009hw}.

While recursion methods provide an efficient framework for computing factorizable contributions, they cannot, by definition, determine the non-factorizable parts of scattering amplitudes (see e.g. \cite{Benincasa:2007xk,Schuster:2008nh,Cohen:2010mi, Franken:2019wqr, Cheung:2015cba, Arkani-Hamed:2017jhn} for related studies on amplitude constructibility).
In fact, the examples quoted above are particularly successful because the non-factorizable terms vanish. In a generic effective field theory (EFT), however, such terms are generically present, hindering the application of recursion methods.\,\footnote{In the case of general relativity, a nonfactorizable  term is known to arise for $n\geq 12$ gravitons \cite{Bianchi:2008pu, Elvang:2013cua}.}

Since EFTs govern  the physics of the real world, it is certainly important to develop recursion techniques capable of efficiently computing EFT scattering amplitudes. One possible route is to incorporate additional constraints on the infrared behavior of amplitudes. In particular, imposing enhanced soft behavior can severely restrict — and in some cases completely determine — the allowed non-factorizable contributions. Such approaches have been  developed for scalar EFTs (see \cite{Cheung:2015ota} and  e.g. \cite{Cachazo:2016njl,Bianchi:2016viy,Cheung:2016drk,Padilla:2016mno,Rodina:2016jyz,Rodina:2018pcb,CarrilloGonzalez:2019fzc,Kampf:2019mcd,Brauner:2022ymm,Low:2025agh} for developments), and, to some extent, for  electrodynamics \cite{Cheung:2018oki,Kampf:2021bet,Kampf:2021tbk}.\,\footnote{Other developments involving on-shell amplitudes in EFT include
\cite{Fu:2017uzt,   Rodina:2018pcb, Low:2019ynd, Aoude:2019tzn,  Durieux:2019eor, Gu:2020thj, EliasMiro:2020tdv, Baratella:2020lzz,  Balkin:2021dko,  Baratella:2021guc, Aoude:2020onz,  AccettulliHuber:2021uoa, Liu:2023jbq,  DeAngelis:2022qco,   Goldberg:2024eot,  Low:2025agh, Zhou:2026isc}.
}

In this work, we pursue a different strategy. Working within the general EFT of electromagnetism, we show that all non-factorizable contributions can be encoded into contact amplitudes, which then serve as the fundamental building blocks for the construction of general amplitudes. 
A key ingredient of our analysis is to work in the self-dual basis, in which the photon field strength propagator reveals its simplicity. 

Our recursive approach is somewhat reminiscent of the CSW construction for computing maximally helicity-violating (MHV) amplitudes in Yang--Mills theory \cite{Cachazo:2004kj}. In the CSW framework, the fundamental building blocks are the MHV amplitudes, given by the Parke--Taylor formula \cite{Parke:1986gb}. The sewing of our contact amplitudes proceeds analogously to the sewing of the MHV vertices in the CSW expansion.

Moreover, just like CSW admits a MHV Lagrangian viewpoint \cite{Mansfield:2005yd, Gorsky:2005sf, Ettle:2006bw}, our approach also benefits from a functional formalism. We will show that the contact amplitudes can be encapsulated into an effective Lagrangian, that we refer to as the  self-dual contact Lagrangian. This contact Lagrangian can be computed systematically from the original EFT through an appropriate functional method.

The paper is organized as follows. Section \ref{se:EFT}  reviews the general EFT of electromagnetism and important examples. 
Section \ref{se:SD_basis} introduces the self-dual basis and analyses the field strength propagator. 
Section \ref{se:Diagrammatics} introduces the spinor helicity formalism, shows that the EFT vertices have a pairing structure, and shows that the associated amplitudes 
can be written in terms of weighted Hafnians. 
Section \ref{se:CSW} provides a diagrammatic analysis of the amplitudes, showing that a substructure made of contact amplitudes arises in the self-dual basis. A  a CSW-like recursion method to construct generic amplitudes from the contact amplitudes is then outlined and exemplified.  
Section \ref{se:Contact_lagrangian} presents a functional approach that systematically provide the contact amplitudes, through the reformulation of the theory in terms of unconstrained fields. 
Section \ref{se:hel_conserving_lagrangian} focusses on the contact Lagrangian of helicity-conserving theories. It outlines a method to resum the contact Lagrangian, which is then exemplified  in Born-Infeld electrodynamics. 
 Section \ref{se:further_results} presents some higher-point results. The appendices contain details on the spinor helicity formalism  (\ref{app:SHF}), details on symmetries and bases on amplitudes (\ref{app:symmetries}), and an independent computation of a 6-pt amplitude that serves as as sanity check  (\ref{app:traditional_amps}). 

%% file: Sections/operators.tex
\section{The EFT of Electromagnetism (Review)}
\label{se:EFT}

Let us briefly remind the basics of EFT. An effective field theory is a local QFT that provides predictions within a finite domain of energies or distances. The EFT is defined by a field content, symmetries, and by an effective Lagrangian taking the form of a series of local operators. These effective operators are combinations of fields and derivatives. They can be organized by scaling dimension, and there is a finite number of operators of each dimension. Finally, the truncation of the series controls the precision of the predictions of the EFT.

In the EFT of electromagnetism, the building blocks for the operators are the derivatives $\partial_\mu$ and the photon field strength $F_{\mu\nu}$. 
Two independent scales $\Lambda'$, $\Lambda$ can be associated with $\partial$ and $F$, such that $\partial/\Lambda'$ and $F/\Lambda^2$ can be viewed as two independent expansion parameters organizing the effective Lagrangian. The former corresponds to the long distance expansion, while the latter corresponds to the weak field expansion.

In the present work, we consider only the lowest order in the derivative expansion.
 In this limit, the most general effective Lagrangian of electromagnetism, ${\cal L}_{\rm EFT}$, can be written as 
\be
{\cal L}_{\rm EFT}[F, \partial] = {\cal L}_{F}[F] \times \left(1+O\left(\frac{\partial^2}{{\Lambda'}^2}\right)\right) \,. 
\ee

In the weak field expansion, ${\cal L}_{F}$ is expanded in Lorentz invariant monomials of $F_{\mu\nu}$.  In four-dimensional spacetime, the group theoretical properties of the photon field strength imply that all Lorentz-invariant  can be reduced to monomials of only two Lorentz scalars ${\cal F}=F^{\mu\nu} F_{\mu\nu}$ and ${\cal G}=F^{\mu\nu} \tilde F_{\mu\nu}$ with $\tilde F_{\mu\nu} =\frac{1}{2}\epsilon_{\mu\nu\rho\sigma} F^{\rho\sigma} $.\,\footnote{
The tensor  $F_{\mu\nu}$ transforms as the representation  $(1,0)\oplus (0,1)$ of the Lorentz algebra.  One scalar can be built from powers of each of the irreps $(1,0)$ and $(0,1)$. Therefore, all Lorentz invariant powers of $F_{\mu\nu}$ can be expressed in terms of only two scalars, that can be chosen to be ${\cal F}$, ${\cal G}$. The fact that all scalars built from  $F_{\mu\nu}$ can be expressed as polynomials of ${\cal F}$, ${\cal G}$ can also be  shown algebraically \cite{Escobar:2013rsa}. This  is  a consequence of the Cayley-Hamilton theorem applied to $4\times 4$ matrices \cite{Cheung:2018oki}. }

Requiring invariance under parity, the EFT Lagrangian can only depend on $\cal G$ through ${\cal G}^2$.
The general EFT Lagrangian is thus expressed as a sum of monomials of the form
${\cal F}^{n-2i}{\cal G}^{2i}$, with $n,i$ integers and $i\leq \lfloor n \rfloor$,
\begin{equation}
    \mathcal{L}_F = \sum^\infty_{n=0}\sum^{\lfloor n \rfloor}_{i=0 }\vv_i^{(2n)} \mathcal{F}^{n-2i} \mathcal{G}^{2i}\,.
\label{eq:LF_standardbasis}
\end{equation}
The lowest dimensional effective operators are 
\begin{equation}\label{eq:EFTlagrangian}
    \begin{aligned}
        \mathcal{L}_{F} = &-\tfrac{1}{4}\mathcal{F} 
        + \vv_0^{(4)}\mathcal{F}^2 + \vv_1^{(4)}\mathcal{G}^2 
        + \vv_0^{(6)}\mathcal{F}^3 + \vv_1^{(6)}\mathcal{F}\mathcal{G}^2 
        + \vv_0^{(8)}\mathcal{F}^4 + \vv_1^{(8)}\mathcal{F}^2\mathcal{G}^2 + \vv_2^{(8)}\mathcal{G}^4 
        + \mathcal{O}(F^{10}) \,.
    \end{aligned}
\end{equation}
For operators with $2n$ field strengths, there are $\lfloor\frac{n}{2}\rfloor$ independent operators, with $\lfloor x\rfloor$  the floor function. 
It is also possible to define a complete basis written solely with cyclic contractions of $F_{\mu\nu}$, such as $F_{\mu\nu}F^{\nu\rho}F_{\rho\sigma}F^{\sigma \mu}$. 
The translation between the two  bases is done through the identity \be \left( F\cdot F \right)^n =  \frac{1}{2}{\cal F}\,
\left(F\cdot F \right)^{n-1} 
+ \frac{1}{16}{\cal G}^2\,
\left(F\cdot F \right)^{n-2}
\label{eq:cyclic_to_dual}\,, \ee 
with $F\cdot F =F_\mu^{~\nu} F_\nu^{~\rho}$. 

\subsection{Examples}

\begin{table}[ht]
\centering
\renewcommand{\arraystretch}{2.2}

\begin{tabular}{
>{\raggedright\arraybackslash}m{1cm}
>{\raggedleft\arraybackslash}m{3cm}
>{\raggedleft\arraybackslash}m{3.5cm}  
>{\raggedleft\arraybackslash}m{3cm} 
}

\toprule

Coeff. & Born-Infeld & Scalar QED & Spinor QED \\
\midrule

$\vv^{(4)}_0$ & $\dfrac{1}{32\,\Lambda^2}$ & $\dfrac{7}{23040\,m^4\pi^2}$ & $\dfrac{1}{1440\,m^4\pi^2}$ \\

$\vv^{(4)}_1$ & $\dfrac{1}{32\,\Lambda^2}$ & $\dfrac{1}{23040\,m^4\pi^2}$ & $\dfrac{7}{5760\,m^4\pi^2}$ \\[0.5em]

\midrule[0.03em]

$\vv^{(6)}_0$ & $-\dfrac{1}{128\,\Lambda^4}$ & $-\dfrac{31}{322560\,m^8\pi^2}$ & $-\dfrac{1}{5040\,m^8\pi^2}$ \\

$\vv^{(6)}_1$ & $-\dfrac{1}{128\,\Lambda^4}$ & $-\dfrac{11}{322560\,m^8\pi^2}$ & $-\dfrac{13}{40320\,m^8\pi^2}$ \\[0.5em]

\midrule[0.03em]

$\vv^{(8)}_0$ & $\dfrac{5}{2048\,\Lambda^6}$ & $\dfrac{127}{1290240\,m^{12}\pi^2}$ & $\dfrac{1}{5040\,m^{12}\pi^2}$ \\

$\vv^{(8)}_1$ & $\dfrac{3}{1024\,\Lambda^6}$ & $\dfrac{113}{1935360\,m^{12}\pi^2}$ & $\dfrac{11}{30240\,m^{12}\pi^2}$ \\

$\vv^{(8)}_2$ & $\dfrac{1}{2048\,\Lambda^6}$ & $\dfrac{13}{3870720\,m^{12}\pi^2}$ & $\dfrac{19}{241920\,m^{12}\pi^2}$ \\[0.5em]

\midrule[0.03em]

$\vv^{(10)}_0$ & $-\dfrac{7}{8192\,\Lambda^8}$ & $-\dfrac{511}{2433024\,m^{16}\pi^2}$ & $-\dfrac{1}{2376\,m^{16}\pi^2}$ \\

$\vv^{(10)}_1$ & $-\dfrac{5}{4096\,\Lambda^8}$ & $-\dfrac{1073}{6082560\,m^{16}\pi^2}$ & $-\dfrac{83}{95040\,m^{16}\pi^2}$ \\

$\vv^{(10)}_2$ & $-\dfrac{3}{8192\,\Lambda^8}$ & $-\dfrac{31}{12165120\,m^{16}\pi^2}$ & $-\dfrac{127}{380160\,m^{16}\pi^2}$ \\[0.5em]

\bottomrule

\end{tabular}
\caption{Coefficients of the low-energy EFT Lagrangian \eqref{eq:EFTlagrangian} for Born--Infeld theory, scalar QED, and spinor QED, up to $\mathcal{O}(F^{10})$.} 
\label{tab:EFT_coefs}
\end{table}

An important example for ${\cal L}_F$ is the Euler-Heisenberg effective Lagrangian, that encodes loops of massive charged matter particles with mass $m$ and charge $q$ at energy scales much below $m$ \cite{Dunne:2012vv}. In that case the two EFT scales are $\Lambda'=m$, $\Lambda = \frac{m}{\sqrt{q}} $. The values of the first coefficients for a scalar and a Dirac fermion at one loop are given in Tab.\,\ref{tab:EFT_coefs}. 

Another important example is Born-Infeld electromagnetism, given in 4d by the effective Lagrangian 
\begin{equation}
    \mathcal{L}_{F,{\rm BI}} = -{\Lambda'}^2 \sqrt{1+\frac{\mathcal{F}}{2{\Lambda'}^2} - \frac{\mathcal{G}^2}{16{\Lambda'}^2}} + {\Lambda'}^2\,.
    \label{eq:BI_def}
 \end{equation}
While first written heuristically \cite{Born:1934gh}, 
this effective Lagrangian emerged in a variety of string-related context. It appears in models where the photon arises from a D-brane, with the charged particles corresponding to open strings attached to the brane \cite{Fradkin:1985qd}. 
It also arises in the partial spontaneous breaking of $N=2$ supersymmetry to $N=1$ \cite{Bagger:1996wp}. 
The value of the $\Lambda'$ scale is tied to the specific UV completion.

%% file: Sections/self_dual_decomp.tex
\section{Self-Dual Basis and Propagators}
\label{se:SD_basis}

The field strength tensor can be decomposed into irreducible representations (irreps) of the Lorentz group:  the self-dual $(1,0)$ and anti-self-dual $(0,1)$. To this end, we introduce the projectors 
\be
P_{\pm,\mu\nu}{}^{\alpha\beta}= \frac{1}{4}\left(\delta_\mu^\alpha \delta_\nu^\beta- \delta_\mu^\beta \delta_\nu^\alpha \pm i \varepsilon_{\mu\nu}{}^{\alpha\beta}\right)\,, \label{eq:Pdef}
\ee
with $\varepsilon$ the totally antisymmetric tensor. 
The projector properties $P_\pm \cdot P_\pm= P_\pm$, $P_+ \cdot P_- =0$ are verified due to the identity $\varepsilon_{\mu\nu}^{~~~ \alpha\beta} \varepsilon_{\alpha\beta}^{~~~ \rho\sigma}= 2(\delta_\mu^\rho \delta_\nu^\sigma-\delta_\mu^\sigma \delta_\nu^\rho) $. 
The sum $P_+ + P_-$ is the identity tensor on the antisymmetric rank-two tensors, while the difference $P_+ - P_- = \frac{i}{2}\varepsilon$ is tied  to the duality operator (i.e. the Hodge star).

Using these projectors,  we decompose the field strength as
\be
P_{\pm,\mu\nu}{}^{\alpha\beta} F_{\alpha\beta} = F_{\pm, \mu\nu}\,,
\ee
where the  $F_+$ and $F_-$ components satisfy
\be
F_\pm = \frac{1}{2}(F\pm i \tilde F)\,. \label{eq:Fpm_def}
\ee
These tensors are eigenstates of the duality operator, with  eigenvalues $\mp i$. Therefore $F_-$ and $F_+$ are conventionally referred to as the \textit{self-dual} and \textit{anti-self-dual} tensors, respectively. 

\subsubsection*{EFT Lagrangian}

The $\mathcal{F}$ and $\mathcal{G}$ invariants introduced  in section \ref{se:EFT} 
take the form
\begin{equation}
    \begin{gathered}
        \mathcal{F} = F_+^2 + F_-^2 \\[0.5em]
        \mathcal{G} = -i(F_+^2 - F_-^2)\,
    \end{gathered}
\end{equation}
when using the decomposition \eqref{eq:Fpm_def}.
Introducing the notation 
\be
F^{2p}_\pm \equiv (F_{\pm,\mu\nu} F^{\mu\nu}_\pm )^p \,, \label{eq:def_F2p}
\ee
the monomial of $\cal F$, $\cal G$ are expressed in terms of the \textit{Krawtchouk matrices} as follows, 
\be
{\cal F}^{n-2i}{\cal G}^{2i} = \sum_{k=0}^n (-1)^i K^{(n)}_{k, 2i}~ F_-^{2n-2k} F_+^{2k} \,,\quad\quad~
K^{(n)}_{k,\, 2i} = \sum_{r=0}^{n} (-1)^r   {2i \choose r}{n-2i \choose k-r} \,.
\ee
As a result, the Lagrangian \eqref{eq:LF_standardbasis} expressed with self-dual fields 
takes the form 
\be
{\cal L}_F = \sum_{n=0}^\infty\sum_{k=0}^n \alpha^{(2n)}_{2k} F_-^{2n-2k}F^{2k}_+ \,, \quad \vsd_{2k}^{(2n)} = \sum_{i=0}^{\lfloor n/2 \rfloor} (-1)^i~ \vv_i^{(2n)} \, K^{(n)}_{k,\,2i} \,,
\label{eq:L_F_def}
\ee
with the lowest dimension operators given by 
\begin{equation}
    \begin{aligned}
        \mathcal{L}_{F}[F_\pm] = &-\tfrac{1}{4}(F_+^2+F_-^2)\\[0.5em]
        & + (\vv^{(4)}_0-\vv^{(4)}_1)(F_+^4+F_-^4) + 2(\vv^{(4)}_0+\vv^{(4)}_1)F_+^2F_-^2 \\[0.5em]
        & + (\vv^{(6)}_0-\vv^{(6)}_1)(F_+^6+F_-^6) + (3\vv^{(6)}_0+\vv^{(6)}_1)(F_+^4F_-^2+F_+^2F_-^4) \\[0.5em]
        & + (\vv^{(8)}_0-\vv^{(8)}_1+\vv^{(8)}_2)(F_+^8+F_-^8) + 4(\vv^{(8)}_0-\vv^{(8)}_2)(F_+^6F_-^2 + F_+^2F_-^6) \\
        & \hspace{8cm} 
        + 2(3\vv^{(8)}_0+\vv^{(8)}_1+3\vv^{(8)}_2)F_+^4 F_-^4 \\[0.5em]
        & +\mathcal{O}(F^{10}) \,. \label{eq:Lag_SD}
    \end{aligned}
\end{equation}

\subsection{(Non-)Propagation in the Self-Dual Basis }
\label{se:Prop_SD_Lorentz}

To perform perturbative computations in the self-dual basis, we need the Feynman propagators for the self-dual fields $F_\pm$, denoted $\ev{F^+_{\mu\nu} F^+_{\rho\sigma}}$, $\ev{F^-_{\mu\nu} F^-_{\rho\sigma}}$, and $\ev{F^+_{\mu\nu} F^-_{\rho\sigma}}= \ev{F^-_{\mu\nu} F^+_{\rho\sigma}}$.  

A direct derivation in terms of the field strength  is tricky, because $F_{\mu\nu}$ is a \textit{constrained} field: it must identically  satisfy the Bianchi identity 
\be
\partial^\mu \widetilde F_{\mu\nu}=0\,. \label{eq:Bianchi}
\ee
The consequences of the Bianchi constraint are discussed later in section \ref{se:bianchi_constraint}.

The self-dual propagators are most easily computed
starting from the propagator of the vector potential in some gauge, e.g. $\ev{A_\mu A_\rho}=\frac{-i \eta_{\mu\rho}}{q^2}$. 
    To proceed we first express the self-dual propagators in terms of $F$, $\tilde F$ using \eqref{eq:Fpm_def}, which gives 
\begin{equation}
    \begin{gathered}\label{eq:LorentzianExpansion}
        \ev{F^+_{\mu\nu} F^+_{\rho\sigma}} = \frac{1}{4}\qty[\qty(\langle FF \rangle - \langle\tilde F \tilde F\rangle) +i\qty(\langle F \widetilde F \rangle + \langle \widetilde F F \rangle )]\,, \\[0.5em]
        \ev{F^-_{\mu\nu} F^-_{\rho\sigma}} = \frac{1}{4}\qty[\qty(\langle FF \rangle - \langle \widetilde F \widetilde F\rangle) -i\qty(\langle F \widetilde F\rangle + \langle \widetilde F F\rangle )] \,, \\[0.5em]
        \ev{F^-_{\mu\nu} F^+_{\rho\sigma}} = \frac{1}{4}\qty[\qty(\langle FF \rangle + \langle\widetilde F \widetilde F\rangle ) +i\qty(\langle F \widetilde F \rangle - \langle \widetilde F F\rangle )]\,. 
    \end{gathered}
\end{equation}
The $\ev{FF}$  propagators are then obtained from the $\ev{AA}$ propagator using $F_{\mu\nu}=i(q_\mu A_\nu-q_\nu A_\mu)$\,\footnote{When computing $\ev{FF}$, the gauge-dependent piece of the $\ev{AA}$ propagator  vanishes due to antisymmetrization.}. The result is\,\footnote{We can verify that taking the divergence of $\langle F F\rangle$ yields a contact term, which is expected since the classical equation of motion is $\partial^\mu F_{\mu\nu}=0$. In contrast, taking the divergence of the dual field in any of the other propagators gives zero. This is expected because the Bianchi identity should hold irrespective of the field's dynamics. } 
\begin{equation}
    \begin{aligned}
        &\langle F_{\mu\nu} F_{\rho\sigma} \rangle  = -\tfrac{i}{q^2}\qty(q_\mu q_\rho \eta_{\nu\sigma} + q_\nu q_\sigma \eta_{\mu\rho} - q_\mu q_\sigma \eta_{\nu\rho} - q_\nu q_\rho \eta_{\mu\sigma}) \\[0.5em]
        &\langle \widetilde F_{\mu\nu} F_{\rho\sigma} \rangle = -\tfrac{i}{q^2}\qty(q_\sigma \varepsilon_{\mu\nu\rho\lambda} - q_\rho \varepsilon_{\mu\nu\sigma\lambda})q^\lambda\\[0.5em]
        &\langle F_{\mu\nu} \widetilde F_{\rho\sigma} \rangle = -\tfrac{i}{q^2}\qty(q_\nu \varepsilon_{\rho\sigma\mu\lambda} - q_\mu \varepsilon_{\rho\sigma\nu\lambda})q^\lambda\\[0.5em]
        &\langle \widetilde F_{\mu\nu} \widetilde F_{\rho\sigma} \rangle = -\tfrac{i}{q^2}\qty(q_\mu q_\rho \eta_{\nu\sigma} + q_\nu q_\sigma \eta_{\mu\rho} - q_\mu q_\sigma \eta_{\nu\rho} - q_\nu q_\rho \eta_{\mu\sigma}) +i(\eta_{\mu\rho}\eta_{\nu\sigma} - \eta_{\mu\sigma}\eta_{\nu\rho})\,.
    \end{aligned} \label{eq:FF_Lorentz}
\end{equation}

While the combinations of terms may seem to be complicated at first view, simplifications occur in the self-dual basis. 
First, we observe that $\ev{FF}$ and $\langle \widetilde F \widetilde F \rangle$ are related by a contact term,
\begin{equation}
        \ev{FF} - \langle \widetilde F \widetilde F \rangle = -i(\eta_{\mu\rho}\eta_{\nu\sigma} - \eta_{\mu\sigma}\eta_{\nu\rho})\,. \label{eq:combination1}
\end{equation}

Second, and subtler, a combination of $\langle \widetilde F  F \rangle $ and $\langle  F \widetilde F \rangle $
simplifies due to the fact that in 4d space, any fully antisymmetric tensor with rank $>4$ must vanish. This implies that the rank-5 tensor $ T_{\mu\nu\rho\sigma\lambda} = q_\mu \varepsilon_{\nu\rho\sigma\lambda} $ satisfies the cyclic identity 
\be T_{[\mu\nu\rho\sigma\lambda]} =
q_\mu\varepsilon_{\nu\rho\sigma\lambda}+ q_\nu\varepsilon_{\rho\sigma\lambda\mu}+
q_\rho\varepsilon_{\sigma\lambda\mu\nu}+
q_\sigma\varepsilon_{\lambda\mu\nu\rho}+ 
q_\lambda\varepsilon_{\mu\nu\rho\sigma}
= 0\,. \ee
This identity implies that the combination $\langle{F \widetilde F}\rangle + \langle{\widetilde F F}\rangle$ is also local, with 
\begin{equation}
        \langle{F_{\mu\nu}\widetilde F_{\rho\sigma}}\rangle + \langle{\widetilde F_{\mu\nu} F_{\rho\sigma}}\rangle = -i\varepsilon_{\mu\nu\rho\sigma}\,. \label{eq:combination2}
\end{equation}
The relations \eqref{eq:combination1} and \eqref{eq:combination2} imply that the propagators  $\langle{F_+  F_+}\rangle$ and $\langle{F_-  F_-}\rangle$ are \textit{purely local}, with 
\begin{equation}
    \begin{aligned}
        \ev{F^+_{\mu\nu} F^+_{\rho\sigma}} =& \frac{-i}{4}\qty[(\eta_{\mu\rho}\eta_{\nu\sigma}-\eta_{\mu\sigma}\eta_{\nu\rho}) +i\varepsilon_{\mu\nu\rho\sigma}] \,, \\[0.5em]
        \ev{F^-_{\mu\nu} F^-_{\rho\sigma}} =& \frac{-i}{4}\qty[(\eta_{\mu\rho}\eta_{\nu\sigma}-\eta_{\mu\sigma}\eta_{\nu\rho}) -i\varepsilon_{\mu\nu\rho\sigma}]  \,. 
    \end{aligned}
\end{equation}
Notice also that the self-dual structure implies that the contact terms are proportional to the projectors, $\ev{F_\pm F_\pm}=-i P_\pm$.

The $\ev{F_\pm F_\pm}$ propagators being local, the only nontrivial propagation occurs from  the $\langle{F_-  F_+}\rangle$, which does have a $\frac{1}{q^2}$ pole. We will see further below that this remarkable fact   is directly tied to the Bianchi identity. 

The expression for  $\langle{F_-  F_+}\rangle$ is rather cumbersome and will  not be displayed  here. As we shall see,  the simplicity of the photon propagator in the self-dual basis becomes fully manifest  in the spinor  formalism.

\subsection{Propagators in Spinor   Representation } 

\label{se:Spinor_Rep}

The representations of the Lorentz group can be mapped onto those of its double cover, the $SL(2,\mathbb{ C})$ group. In doing so, each vectorial index of a Lorentz tensor is traded for a pair of spinor indexes --- corresponding to the $(\frac{1}{2},\frac{1}{2})$ of the Lorentz group. 

This is one of the key ingredients of the spinor helicity formalism, used in the next sections to compute on-shell amplitudes. 
However we will see here that, even at the off-shell level, the spinor representation greatly simplifies the calculations and brings new insights.

The map between  Lorentz  and spinors indexes can be done using the Pauli matrices as
\be
X_{\mu \ldots } \to X_{a \dot a \ldots } =  (\sigma_{a\dot a})^{\mu}   \,\,X_{\mu \ldots }\,,
\label{eq:SL2_map}
\ee \be
X_{a \dot a \ldots } \to  X_{\mu \ldots }  = \frac{1}{2} (\bar \sigma_{\mu})^{\dot a a } X_{a \dot a \ldots }
\label{eq:SL2_map_inv}
\ee
where $\sigma^\mu = (1, \sigma^i ) $, $\bar \sigma^\mu = (1, -\sigma^i ) $, and $a=1,2$, $\dot a =\dot{1},\dot{2}$ transform respectively in the $(\frac{1}{2},0)$ and $(0,\frac{1}{2})$ representations. 
The inverse map \eqref{eq:SL2_map_inv} is ensured by the identity Tr$(\sigma^\mu \bar \sigma^\nu) = 2 \eta^{\mu\nu}$.  Further conventions and identities for spinors are collected in App.\,\ref{app:SHF}. 

 Since it identifies $(\mu,\nu,\ldots)\to (a\dot a, b\dot b, \ldots)$, the map \eqref{eq:SL2_map} produces tensors with  spinor indexes. Depending on the representation, such tensors may contain the invariant symbols  $\varepsilon_{ab}$, $\varepsilon_{\dot a \dot b}$, $\delta_a^b$, $\delta_{\dot a}^{\dot b}$ of the  
$(0,\frac{1}{2})$ and $(\frac{1}{2},0)$ irreps.

\subsubsection{Propagators}
\label{se:propagators}

The photon field strength under the \eqref{eq:SL2_map} map becomes the spinor tensor
\be
F_{\mu\nu} \to F_{a\dot a b\dot b}= F^+_{a\dot a b\dot b }+F^-_{a\dot a b\dot b}\,\,. 
\ee
Here we have used  the decomposition into self-dual  components from (\ref{eq:Fpm_def}). 

Due to the (anti)symmetry properties of the spinor indices in the $(\frac{1}{2},0)$ and $(0,\frac{1}{2})$ irreps, 
we further know that the $F^\pm$ must decompose as 
$$F^+_{a\dot a b\dot b}= \varepsilon_{ab}F^+_{\dot{a}\dot{b}}\,, \quad F^-_{a\dot a b\dot b}=\varepsilon_{\dot{a}\dot{b}}F^-_{ab}\,,$$
where $F^-_{ab}$ and $F^+_{\dot a\dot b}$ are symmetric tensors. 
Hence we can write the field strength in spinor representation as 
\begin{equation}
    F_{ a\dot a b\dot b} = \varepsilon_{ab}F^+_{\dot{a}\dot{b}} + \varepsilon_{\dot{a}\dot{b}}F^-_{ab} \,.
    \label{eq:F_decomp}
\end{equation}
Notice that the 6 degrees of freedom of $F_{\mu\nu}$ are encoded into the 3 degrees of freedom of $F^-_{ab}$ and $F^+_{\dot a \dot b}$, that we refer to as the \textit{reduced} self-dual fields.

Let us compute the propagators of the reduced self-dual  fields $F^\pm_{ab}$. We first write the $\ev{FF}$ correlator given in \eqref{eq:FF_Lorentz} in spinor indexes, using the map \eqref{eq:SL2_map}. We obtain
\begin{equation}
    \begin{aligned}\label{eq:FF_spinor}
        &\langle F_{a\dot a b\dot b} F_{ c\dot c d\dot d } \rangle = \tfrac{2i}{q^2}\left( q_{a\dot{a}} q_{c\dot{c}} \varepsilon_{bd} \varepsilon_{\dot{b}\dot{d}} - q_{a\dot{a}} q_{d \dot{d}} \varepsilon_{bc} \varepsilon_{\dot{b}\dot{c}} - q_{b\dot{b}} q_{c \dot{c}} \varepsilon_{ad} \varepsilon_{\dot{a}\dot{d}} + q_{b\dot{b}} q_{d \dot{d}} \varepsilon_{ac} \varepsilon_{\dot{a}\dot{c}} \right)  \,.  
    \end{aligned}
\end{equation}
We then introduce the self-dual decomposition \eqref{eq:F_decomp} into \eqref{eq:FF_spinor}. The reduced fields are easily extracted by contracting with $\frac{1}{2}\varepsilon_{ab}$ and $\frac{1}{2}\varepsilon_{\dot{a}\dot{b}}$. 
Using the identity 
\be
q^{b}_{\dot{a}}\, q_{b\dot{b}} = \varepsilon_{\dot{a}\dot{b}}q^2 \,, \label{eq:Idq2}
\ee
we find 
the propagators 
\begin{align}
    \langle F^+_{\dot{a}\dot{b}} F^+_{\dot{c}\dot{d}} \rangle & = -i \left( \varepsilon_{\dot{a}\dot{c}}\varepsilon_{\dot{b}\dot{d}} + \varepsilon_{\dot{a}\dot{d}}\varepsilon_{\dot{b}\dot{c}} \right) \,, \label{eq:PP}\\[0.5em]
    \langle F^-_{ab} F^-_{cd} \rangle & = -i\left( \varepsilon_{ac}\varepsilon_{bd} + \varepsilon_{ad}\varepsilon_{bc}\right) \,, \label{eq:MM}
    \\[0.5em]
    \langle F^-_{a b} F^+_{\dot{c}\dot{d}} \rangle & = \tfrac{i}{q^2}\left( q_{a\dot{c}}q_{b\dot{d}}+q_{a\dot{d}}q_{b\dot{c}} \right) \,.\label{eq:MP}
\end{align}

The propagators $\langle{F_+  F_+}\rangle$ and $\langle{F_-  F_-}\rangle$ are purely local, similarly to the result in Sec.\,\ref{se:Prop_SD_Lorentz}. Moreover,   the $\langle{F_+  F_-}\rangle$ propagator also takes a very simple form in the spinor  representation\footnote{The same results are found in \cite{Boels:2008fc}, but with a flipped sign from conventions.}. 
More properties of the propagators are discussed below.

\subsubsection{Spinor structures}
\label{se:spinor_structures}

The structure of the propagators \eqref{eq:PP} -- \eqref{eq:MP} 
motivates the introduction of specific tensors: the projectors $P_\pm$, and a tensor  $Q$ that links the $+$ and $-$ subspaces. These spinor structures are useful to understand the properties of the propagator, and appear also in the functional formalism of  Sec.\,\ref{se:Contact_lagrangian}. 

The spinor structures are defined as
\begin{equation}
(P_+)_{ab\dot{a}\dot{b}}^{\quad ~\,\dot{c}\dot{d}cd} = \frac{1}{4} \varepsilon_{ab}\,\varepsilon^{cd}\qty(\delta_{\dot{a}}^{\dot{c}} \, \delta_{\dot{b}}^{\dot{d}} + \delta_{\dot{a}}^{\dot{d}} \, \delta_{\dot{b}}^{\dot{c}})\,, \quad \ P_- = P_+\big|_{\dot{o}\leftrightarrow o} 
\label{eq:P_def}
\end{equation}
\begin{equation}
         (Q_{-+})_{ab\dot{a}\dot{b}}^{\quad ~\, \dot{c}\dot{d}cd} = \frac{1}{4}\varepsilon_{\dot{a}\dot{b}}\,\varepsilon^{cd}\frac{\qty(q_{a}^{\dot{c}} \, q_{b}^{\dot{d}} + q_{a}^{\dot{d}} \, q_{b}^{\dot{c}})}{q^2}\,, \quad  Q_{+-} = Q_{-+}\big|_{\dot{o}\leftrightarrow o}
         \label{eq:Q_def}
\end{equation}
Multiplying these objects between themselves by contracting upper block of indexes with lower block yields the set of relations
\begin{equation}\label{eq:OpsAlgebra}
    \begin{gathered}
         P_+  P_- = 0 \,, \quad  P_\pm^2 =  P_\pm\,,\\[0.5em] 
         Q_{\pm\mp} Q_{\mp\pm} =  P_\pm\,,\quad  Q_{\pm\mp}^2 = 0 \,,\\[0.5em] \quad  P_\pm  Q_{\pm\mp} =  Q_{\pm\mp}\,,\quad  P_\mp  Q_{\pm\mp} = 0 \,,\quad   Q_{\pm\mp} P_\pm = 0 \,,\quad   Q_{\pm\mp} P_\mp =  Q_{\pm\mp}\,.
    \end{gathered}
\end{equation}
Using the  $P$, $Q$ tensors, we  can express the propagators of the non-reduced self-dual fields as 
\begin{equation}
    \mqty( \langle F_+F_+\rangle & \langle F_+F_-\rangle \\ \langle F_-F_+\rangle & \langle F_-F_-\rangle )=4i\mqty( - P_+ &  Q_{+-} \\  Q_{-+} & - P_- )\,. \label{eq:prop_matrix}
\end{equation}

\subsection{Consequences of the Bianchi Constraint}

\label{se:bianchi_constraint}

In this subsection we clarify the interplay between the Bianchi identity, the propagators, and the kinetic Lagrangian. 

\subsubsection{Bianchi  in the spinor representation}

We first have to study the Bianchi identity in the spinor formalism. 
The dual field strength decomposes as 
\be
\widetilde F_{\mu\nu} \to \widetilde F_{a\dot a b \dot b } = -i(\varepsilon_{ab}F^+_{\dot{a}\dot{b}}- \varepsilon_{\dot{a}\dot{b}}F^-_{ab})\,.
\ee
This can be obtained using the inverse of \eqref{eq:Fpm_def} and decomposition \eqref{eq:F_decomp}. 
In turn, the Bianchi identity $  q_\mu \widetilde F^{\mu\nu} = 0$ translates on the reduced self-dual fields as 
 \begin{equation}\label{eq:BianchiIdentity}
    q^{~a}_{~\dot{b}} \, F_+^{\dot{b}\dot{a}} = q^{\dot a}_{b}\, F_-^{ba}\,.
\end{equation}

The  Bianchi identity takes an interesting form when using the $Q$ structure defined in \eqref{eq:Q_def}. 
 Let us apply the $Q_{+-}$ tensor to the self-dual tensor $F^-_{(a\dot a) (b \dot b)}$. In terms of the reduced fields, we have
\begin{equation}
     (Q_{+-})_{ab\dot{a}\dot{b}}{}^{AB\dot{A}\dot{B}} \varepsilon_{\dot{A}\dot{B}}F_{AB} = 
\varepsilon_{ab}\,\frac{q^{~A}_{~\dot{a}} \,q^{~B}_{~\dot{b}} }{q^2} \, F_{AB}      
     =\varepsilon_{ab}\,\frac{q^{~A}_{~\dot{a}}}{q^2} \, q_{~A}^{~ \dot{c}}F_{\dot{b} \dot{c} } =- \varepsilon_{ab} F_{\dot{a}\dot{b}}\,,
\end{equation}
where in the second step, we  use  \eqref{eq:BianchiIdentity}, and in the last step we use identity \eqref{eq:Idq2} to simplify the momenta. 
Expressed with the non-reduced fields, this form of the Bianchi identity becomes simply
\begin{equation}
     Q_{+-} F_- = -F_+ \label{eq:Bianchi2}\,.
\end{equation}
Finally,  \eqref{eq:Bianchi2} can be used to relate the propagators between themselves. In particular,  we have 
\be
Q_{-+} \langle F_+F_+\rangle = -\langle F_-F_+\rangle\,. \label{eq:Bianchi3}
\ee

The result \eqref{eq:Bianchi3} provides an interesting perspective.  It tells us that the nonlocality of the $\ev{F_\pm F_\mp}$ propagators is a direct consequence of the Bianchi constraint. Thus, in a sense, the Bianchi identity is responsible for the propagation of photons.

\subsubsection{Inversion of the Propagator}\label{se:L_kin_discussion}

We then study how the propagator relates to the kinetic Lagrangian. The Maxwell kinetic term is 
 \be 
 {\cal L}_{F, {\rm kin}} =  -\frac{1}{16}(F_+^2+F_-^2) \ee 
 when written in terms of non-reduced self-dual fields.
Let us start with two puzzles.  

When working with unconstrained fields, a  general way to get the propagator is to  take the second variation of the kinetic term with respect to the fields, which yields the kinetic operator $K$, and then invert $K$. In our case, 
 taking the second variation of ${\cal L}_{F, {\rm kin}}$  yields a kinetic operator that is diagonal and local, since no derivative acts on $F_\pm$ in the kinetic term. How, then, could the inversion of this matrix produce a result that is both non-diagonal and non-local?

Another puzzle appears from the propagator side. It is convenient to define the propagator matrix, 
\be
G= 4 i \begin{pmatrix}
    -P_+ & Q_{+-} \\ Q_{-+} & -P_-
\end{pmatrix} \,. 
\ee
Using the set of relations in \eqref{eq:OpsAlgebra}, we notice that $G$ is proportional to an idempotent matrix, i.e. $G\cdot G \propto G$. However, an idempotent matrix cannot be inverted, unless it is proportional to the identity. Given this, how can  $G$ be related to $K$? 

The resolution of both puzzles has to do with the Bianchi constraint. Let us first observe that the relations 
\eqref{eq:OpsAlgebra}  can be used to construct the following matrices,
\begin{equation}
    I = \mqty( P_+ & 0 \\[0.5em] 0 & P_- )\,, \quad J = \mqty( 0 & Q_{+-} \\[0.5em] Q_{-+} & 0 ) \,.
\end{equation}
Since $I^2=I$ and $J^2=I$, they form a representation of $\mathbb Z_2$.  It follows that the  projectors $(I\pm J)/2$  select the even and odd irreps of $Z_2$. These projectors are denoted as 
\begin{equation}
    \Pi_G = \frac{I-J}{2}\,, \quad\quad  \Pi_\o = \frac{I+J}{2} \,.
\end{equation}
In this language, the propagator is  proportional to one of the projectors,
\be
G = -8i \,\Pi_G \,. 
\ee

Observe then that, in our problem,  the vector space on which all these matrices act should be constrained by the Bianchi identity. Expressing  \eqref{eq:Bianchi2} in terms of projectors,  we find that the Bianchi identity is 
\be
\Pi_\o\mqty( F_+\\[0.5em] F_- ) = 0\,. 
\ee
Therefore, the Bianchi identity projects out the even subspace, leaving  only the odd one. It follows that the inversion of $G$ must be defined with $\Pi_G$ as the identity, 
\be
 G \cdot K = i \Pi_G\,. \label{eq:G_inversion}
\ee
Notice that, since $G\propto \Pi_G$, the $G$ matrix is  invertible in the odd subspace, thereby  solving the second puzzle.

The general solution to \eqref{eq:G_inversion}  can be found by considering
\begin{equation}
    K= \frac{1}{4}\qty(a\,I + b\, J)\,.
\end{equation}
Substituting in \eqref{eq:G_inversion}, we find $a-b = -\frac{1}{2}$, such that the coefficients satisfy the family of solutions $a=-\frac{t}{2} $, $b=\frac{1-t}{2} $. 
Therefore, the most general kinetic operator is given by
\begin{equation}
    K = \frac{1}{8} \mqty( -tP_+ & (1-t)Q_{+-} \\(1-t)Q_{-+} & -tP_- ) \,. \label{eq:K_gen}
\end{equation} 
The $t$ dependence corresponds to the kernel of $G$. One can verify that $K$ corresponds to the  \textit{generalized inverse} of $G$. 
Substituting $K$ in the kinetic term reproduces the correct  Maxwell kinetic term for any $t$, 
\begin{equation}
    \frac{1}{2} \mqty(F_+ & F_-) K \mqty(F_+ \\ F_-) = -\frac{1}{16}(F_+^2+F_-^2)\,. \label{eq:K_in_L}
\end{equation}

 We can finally resolve the first puzzle in light of this calculation. 
 The  kinetic operator  obtained by taking two derivatives of $ {\cal L}_{F, {\rm kin}} $    
 corresponds to a specific choice, namely $t=1$, in \eqref{eq:K_gen}. Although $K|_{t=1}$ involves only local operators, its inversion is done with $\Pi_G$ playing  the role of identity. As a result, the inverse is  non-local and non-diagonal, elucidating the first puzzle.

\subsection{Brief Summary}

In this section we  have shown that, when   expressed in the self-dual basis, the photon field strength propagator exhibits a remarkable locality property:  only the mixed $\langle\pm \mp\rangle$ propagators actually propagate, while the  $\langle\pm \pm\rangle$ are purely contact terms. The expressions take a remarkably simple   form in  the spinor representation.

The locality of the $\langle\pm \pm\rangle$ propagators can be viewed as a consequence of the kinetic Lagrangian $-\frac{1}{16}(F_+^2+ F_-^2)$, even though the exact relation between Lagrangian and propaga\-tor is somewhat subtle. In turn, the nonlocal structure of the $\langle\pm \mp\rangle$  propagators is a consequence of the Bianchi identity,  and is, in a sense, fixed by it. 
From this perspective, the Bianchi constraint is the reason why the photon propagates.

%% file: Sections/helicity_amplitudes.tex
\section{Helicity Amplitudes and the Hafnian}
\label{se:Diagrammatics}

We turn to the scattering of photons. This  phenomenon is described by elements of the $S$-matrix, which, when taken between helicity states, are referred to as \textit{helicity amplitudes}. 
As we shall see, a great advantage of the self-dual basis is that self-dual fields  map directly onto photon helicity states, as shown in \cite{Martin:2003gb}.

The helicity amplitudes will  eventually  be calculated using recursive methods in the next sections. However, in the present section, to understand their  structure, it is convenient to  imagine that the amplitudes are computed from the EFT Lagrangian via  standard diagrammatic methods:  Feynman rules and the LSZ amputation procedure.

\subsection{Helicity Amplitudes}
\label{se:SHF}

We adopt the all-incoming convention for the momenta of photons in the scattering amplitudes.  The  amplitudes have symmetry properties that govern the interchange of states:  parity, time-reversal, boson-exchange   and crossing symmetries, see  App.~\ref{app:symmetries} for details. These symmetries imply that a general helicity amplitude describing the scattering photons is simply characterized by 
the \textit{number} of  positive and negative helicity states --- and by the photon momenta. 
Given this, in our conventions, an $N$-photon helicity amplitude with  $K$ positive helicity states and $N-K$ negative helicity states is written as
\be
\mathcal{A}\left(1^+,2^+,\cdots,K^+;
(K+1)^-,\cdots, N^-\right)\equiv {\cal A }\left[K^+, (N-K)^-\right] \,. \label{eq:Amp_def}
\ee
Due to symmetries, all permutations of states are related to the representative \eqref{eq:Amp_def}, hence this amplitude contains all physical information.  For the same reason, it is  sufficient to focus on $K\geq \frac{N}{2}$,  and there are thus  $\frac{N}{2}+1$ independent representatives. This follows closely the notation in \cite{Martin:2003gb}.

\subsubsection{Spinor helicity for massless vectors}

\label{se:spinor_formalism}

The spinor representation of the Lorentz $\cong  SL(2,\mathbb{C})$ algebra discussed in section \ref{se:Spinor_Rep} becomes extra useful when applied to momenta of massless particles. 
Applying the map \eqref{eq:SL2_map}, a $4$-momentum becomes
\be
p^\mu \to p^{\dot a a}\,,\quad\quad{\rm with}\quad  {\rm det}(p^{\dot a  a})=  p^\mu p_\mu \,. \label{eq:det_p}
\ee
For a null momentum i.e.  $p^2=0$, \eqref{eq:det_p} implies that the $p^{\dot a  a}$  matrix has a single nonzero eigenvalue. This  in turn  implies that a null momentum can be written as the tensor product of two spinors,  denoted as  $|p\rangle_{ a}$, $[p|_{\dot a} $,  
\begin{equation}\label{eq:p_spinor_factorization}
    p^{\dot a a}\Big|_{p^2=0} = |p]^{\dot a} \bra{p}^a\,,\quad\quad 
    p_{a \dot a}\Big|_{p^2=0} = \ket{p}_a [p|_{\dot a}\,.
\end{equation}
The spinor indexes are raised as     $|p\rangle^{ a} = \varepsilon^{ a  b} \langle p|_{ b} $, $[p|^{\dot a} = \varepsilon^{\dot a \dot b} |p]_{ \dot b} $, and one conventionaly defines the spinor brackets $\langle p\, q\rangle = - \langle q\, p \rangle \equiv  \langle p|^{ a} | q \rangle_{ a} $,  $[ p\, q ] = - [ q\, p ] \equiv  [ p|_{\dot a} | q ]^{\dot a} $. The elementary properties of these spinors are summarized in App.\ref{app:SHF}.

In the spinor helicity formalism, the photon polarization 4-vectors $\epsilon_\mu^\pm =(0,\bm \epsilon^\pm )$ become the spinor tensors $ \epsilon_{a \dot a}^\pm$ under the map \eqref{eq:SL2_map}. They  can be written as 
\begin{equation}\label{eq:epsilon_spinor}
    \epsilon_{a\dot{a}}^-(p,\xi)= \frac{\sqrt{2}\ket{p}_a[\xi|_{\dot a}}{[p \xi ]} \,, \quad \epsilon_{a \dot{a}}^+(p,\xi)= \frac{\sqrt{2}\ket{\xi}_a[p|_{\dot a}}{\langle \xi p\rangle}\,,
\end{equation}
where $\xi\neq p$ is an arbitrary null momentum. This reference momentum  $\xi$ is tied to the $U(1)$ gauge redundancy,  hence  the amplitudes must be independent of it.

The polarization tensors  are null i.e.  
$\epsilon_{a\dot a}^\pm \epsilon^{\pm, \dot a  a}=0$, and they satisfy  $(\epsilon_{a\dot a}^\pm)^*=\epsilon_{a\dot a}^\mp$. They are normalized as $\epsilon_{a\dot a}^\pm (\epsilon^{\pm,a\dot a})^*=-2$,
which is consistent with the unit normalization of the polarization vectors. 

\subsubsection{From field strengths to photons}

 When computing an amplitude  through the LSZ procedure
  in our EFT of electromagnetism,  the photon states are  connected to  vertices through   field strengths. This amounts to performing the substitution 
\be
F_{\mu\nu}(p)\Big|_{\rm on-shell} \equiv  F_{\mu\nu}(\epsilon,p) = i( p_\mu \epsilon_\nu- p_\nu \epsilon_\mu ) \,,
\label{eq:Fmunu_OS}
\ee
with arbitrary photon polarization $\epsilon$. 
The self-dual combinations $F_\pm^{\mu\nu}(\epsilon,p)$  can also be obtained via application of the projector \eqref{eq:Pdef}, see section \ref{se:SD_basis}. 

In the spinor representation, the  map \eqref{eq:SL2_map} yields $  F_{a\dot{a} b\dot{b} }(\epsilon,p) \to i\left( p_{a\dot{a}} \epsilon_{b\dot{b}} - p_{b\dot{b}} \epsilon_{a\dot{a}} \right)$. Specifying the polarization to be, for instance, the positive helicity $\epsilon^+$ given in \eqref{eq:epsilon_spinor}, we obtain \begin{equation}\label{eq:F_onshell_spinor}
    F_{a\dot{a} b\dot{b} }(\epsilon^+,p) = \frac{\sqrt{2} i\, }{\ev{\xi p}} \left( \ket{\xi}_b\, [p|_{\dot{b}} \ket{p}_a\, [p|_{\dot{a}}  - 
    \ket{\xi}_a\, [p|_{\dot{a}} \ket{p}_b\, [p|_{\dot{b}} 
    \right)  \,,
\end{equation}
and similarly for $\epsilon^-$. 

Furthermore, since the decomposition of the field strength into symmetric and antisymmetric factors shown in section \ref{se:propagators} (see \eqref{eq:F_decomp}) is general,  it applies to our on-shell field strength. We have thus
\be
    F_{ a\dot a b\dot b}(\epsilon,p) = \varepsilon_{ab}F^+_{\dot{a}\dot{b}}(\epsilon,p) + \varepsilon_{\dot{a}\dot{b}}F^-_{ab}(\epsilon,p) \,,
    \label{eq:F_decomp1}
\ee
where  $F^-_{\dot{a}\dot{b}}(\epsilon,p)$,  $F^+_{\dot{a}\dot{b}}(\epsilon,p)$   are the  on-shell reduced self-dual and anti-self-dual field strengths. 
The explicit expressions for these  on-shell reduced fields for a given helicity can be derived by applying  the  projectors from \eqref{eq:P_def} to \eqref{eq:F_onshell_spinor}. 
We obtain 
\begin{align}
    &F_{\dot a \dot b}^+(\epsilon^+) = -\sqrt{2} i[p|_{\dot{a}} [p|_{\dot{b}} \,, \quad &&
    F_{\dot a \dot b}^+(\epsilon^+) = 0 \,, \label{eq:Fp_hel}  \\[0.5em]
    &F_{ a b}^-(\epsilon^+) = 0 \,, \quad &&F_{ a b}^-(\epsilon^-) = -\sqrt{2} i\ket{p}_{a} \ket{p}_{b} \,.\,\label{eq:Fm_hel}
\end{align}

These remarkably simple results are manifestly independent of the reference spinor, as expected since the $U(1)$  field strength is a gauge invariant. 
These results  connect the duality eigenvalue to the sign of helicity: the self-dual $F^-$
correspond to negative helicities, and the anti self-dual $F^+$ to positive helicity (hence the labelling choice).

\subsection{Diagrammatic Features}
\label{se:diags}

Here we study the structure of the diagrams. 
We first notice that the interactions in the self-dual Lagrangian \eqref{eq:Lag_SD}, \eqref{eq:def_F2p}
exhibit a substructure. The $+$ and $-$ legs of a vertex  are organized  in pairs due to the powers of $F_{\pm,\mu\nu} F_\pm^{\mu\nu}$. 
It is convenient to represent this substructure diagrammatically. We write each $(+,+)$ pair using a \textit{square} slot, and each $(-,-)$ pair using an \textit{angle} slot, such that the vertices  are represented as follows,
\be
\begin{aligned} \nn
    \includegraphics[width=0.9\linewidth,trim={0cm 2cm 0cm 1.5cm},clip]{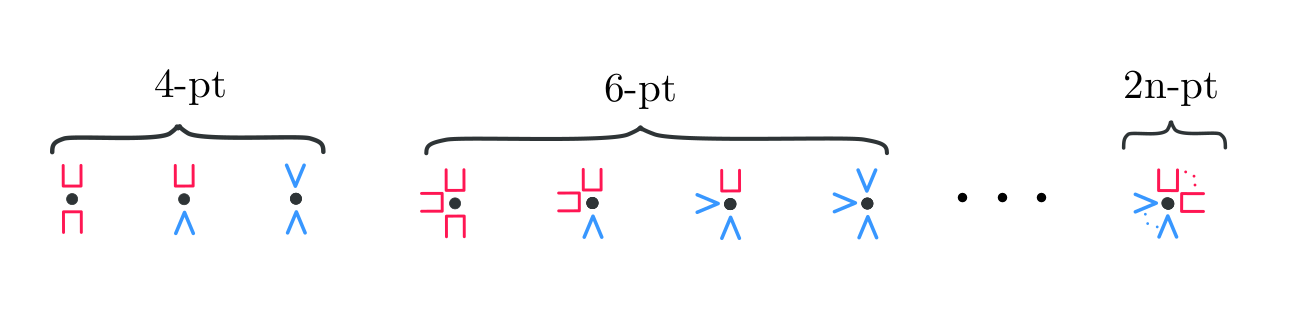}
    \label{eq:slots}
        \end{aligned}
    \ee
    \\
With this notation,  the reader can also  immediately identify/count the number of $+$ and $-$ legs of a given vertex. 

When a slot---either angle or square---is connected to photon states,  the results \eqref{eq:Fp_hel}, \eqref{eq:Fm_hel} apply. 
Put together with   property \eqref{eq:tensor_contraction_map}, the on-shell $F^2_\pm$ take the simple form 
\\[-1em]
\begin{equation}
    \begin{aligned}\label{eq:onshell_contractions}
       &\diagram{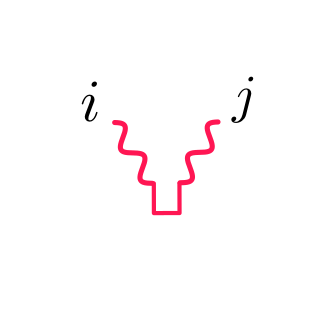} \hspace{-1em}\propto~~ F_+^{\mu\nu}(p_i) F_{+,\mu\nu}(p_j) \Big|_{\rm on-shell} = - [ij]^2 \,, \\[-3em]
       &\,\,\diagram{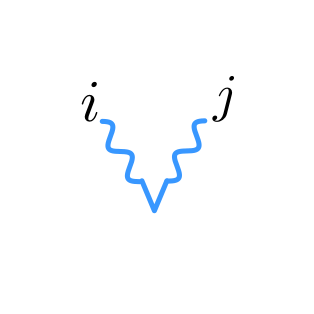} \hspace{-1em}\propto~~ F_-^{\mu\nu}(p_i) F_{-,\mu\nu}(p_j) \Big|_{\rm on-shell} = - \ev{ij}^2 \,.
    \end{aligned}
\end{equation}
These relations  play a crucial role in the representation of amplitudes further on.

We turn to internal lines. The propagator computed in section \ref{se:SD_basis}, see \eqref{eq:PP} -- \eqref{eq:MP}, also provide intuitive results. 
Consider  a  Wick contraction between two vertices
\begin{equation}\label{eq:propagator}
    F_{h_i}^{\mu \nu}\langle F_{h_i; \mu \nu} F_{h_j; \rho \sigma} \rangle F_{h_j}^{\rho \sigma} \,,
\end{equation}
with the external legs identified as on-shell photons and $h_{i,j}=\pm$. The  $++$, $--$, and $+-$  propagators contribute respectively as
\\[-2em]
\begin{gather}\label{eqs:PropDiagramsLocal}
    \diagram{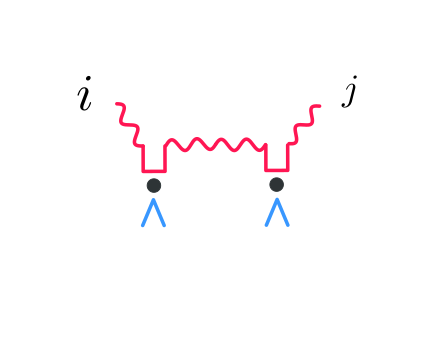} \hspace{-1em}\propto  \ [ij]^2 \,,
    \diagram[0.17]{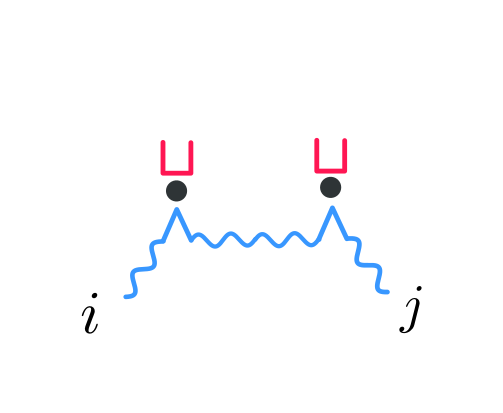} \hspace{-1em}\propto  \  \langle ij \rangle^2 \,, \\[-2em]
    \diagram{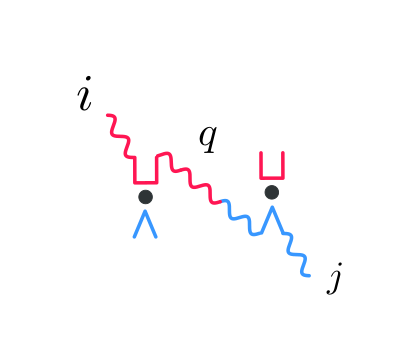} \hspace{-1em}\propto  \  \frac{[i|q\ket{j}^2}{q^2} \,.\label{eqs:PropDiagramsNonLocal}
\end{gather}
We can see that the $++$, $--$ contribution take  a form analogous to \eqref{eq:onshell_contractions} because the propagator is purely contact. The non-local $+-$ contribution is easily derived using the on-shell reduced self-dual fields \eqref{eq:Fm_hel}. Here we use the notation $[i|q\ket{j}=[i|_{\dot a}\,q^{\dot a a }\ket{j}_a$, consistent with section \ref{se:spinor_formalism}.

We then  consider diagrams in which two $+-$ lines are tied   by a vertex.
Two possibilities appear, depending whether the lines connect to two different slots or go through the same slot,
\\[-3em]
\begin{align}
    &\diagram{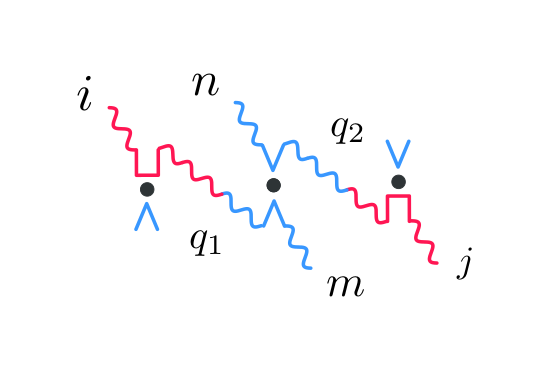} \hspace{-1em}\propto  \ \frac{[i|q_1\ket{m}^2}{q_1^2}\frac{[j|q_2\ket{n}^2}{q_2^2}\,, \label{eq:PM_PM_diagram}
    \\[-1em]
    &\diagram{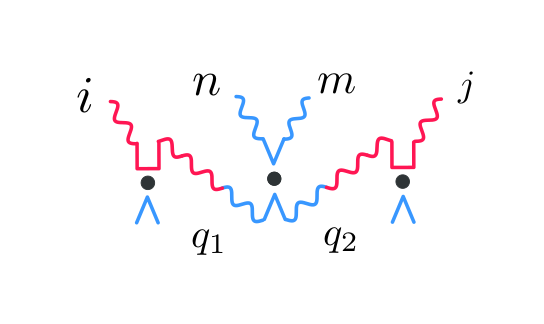}  \hspace{-1em}\propto \frac{[i|q_1\cdot q_2|j]}{q_1^2 q^2_2}^2\ev{mn}^2 = \frac{(q_1\cdot q_2)^2}{q_1^2 q_2^2} [ij]^2 \ev{mn}^2 \label{eq:PMMP_diagram} 
    \,.
\end{align}
This can be computed using \eqref{eq:Fp_hel}, \eqref{eq:Fm_hel} and \eqref{eq:MP}. 
We can see that, while diagram \eqref{eq:PM_PM_diagram} is just the squaring of \eqref{eqs:PropDiagramsNonLocal}, diagram  \eqref{eq:PMMP_diagram} connects the photons at the opposite ends of the diagram, giving a $[ij]^2$ bracket. The last expression is obtained from the identity \eqref{eq:scalar_product_identity}.   The fact that the internal momenta  get factored into a scalar product  is tied to the fact that $Q_{+-} Q_{-+}$ reduces to the projector $P_+$, see \eqref{eq:OpsAlgebra}.

An advantage of our improved diagrammatic representation is that the brackets of photon states appearing in a diagram can be intuitively obtained by following the photon lines.  For \eqref{eq:PM_PM_diagram}, we can see that the uninterrupted lines connect $i$ to $m$ and $j$ to $n$, while in  \eqref{eq:PMMP_diagram}, the line flows uninterrupted through the middle vertex, hence connecting $i$ to $j$. In both cases, the corresponding expressions naturally reflect these features. 

Calculations using standard Feynman diagram techniques would require to carefully take into account diagrams with different structures such as \eqref{eq:PM_PM_diagram} and \eqref{eq:PMMP_diagram}. The computation technique introduced throughout the rest of this paper conveniently bypasses this tricky task. Ultimately, with these techniques, specifying the square and angle slots will not be needed to compute the amplitudes.

\subsection{Irreducible Amplitudes from Hafnians}
\label{se:hafnians}
{
The vertices of the EFT of electromagnetism have a pairing structure, which leaves its imprint at the level of amplitudes (see section \ref{se:diags}).  
The natural combinatorial object associated with such pairings is the \textit{Hafnian}. }
It is related to the Pfaffian in analogy to how the permanent is related to the determinant, through removing the alternating sign structure\,\footnote{
The Hafnian also arises in the field of quantum optics, through the protocol of Gaussian boson sampling. See \cite{Hamilton:2017cbx, Bradler:2017jda} for details. }.

Here we introduce the Hafnian through the simplest example available: an irreducible tree diagram, i.e. a diagram with no internal line, produced  by a single $N$-photon operator,\,\footnote{We purposefully do not use the term ``contact'' here, because certain internal lines lead to contact interactions. The actual contact amplitudes are  discussed in section \ref{se:CSW}. 
There it will be shown that all contact  amplitudes are described by a single Hafnian, and furthermore, \textit{all}  photon helicity amplitudes take the form of  combinations of Hafnians. } 
\be \alpha^{(2n)}_{2k}=  (F^2_-)^{n-k} (F_+^2)^k\,. \label{eq:operator_example} \ee
Based on the definition of  helicity amplitudes given in \eqref{eq:Amp_def}, the numbers of photons with $+$ and $-$ polarizations is respectively $K$ and $N-K$, with 
\be N=2n\, \quad\quad K=2k \,. \ee

Due to the helicity pairing structure discussed in section \ref{se:diags}, we know that the amplitude should {reduce to products}  of $n-k$  angle brackets and $k$ square brackets . Crucially, each momentum appears exactly once {in each product}. These {products} have the form of two \textit{perfect matchings} ---  the pairing of all elements in a given set --- made separately in the sets of photons with positive and negative helicities.\,\footnote{The term ``matching'' originates from graph theory. } Moreover, the photons being identical, all permutations contribute. For example, for $N=4$, $K=4$, the  pairings are
\begin{equation}
\begin{aligned}
    \diagram{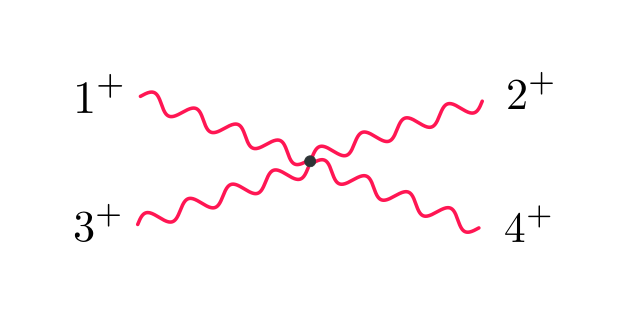} &= \diagram{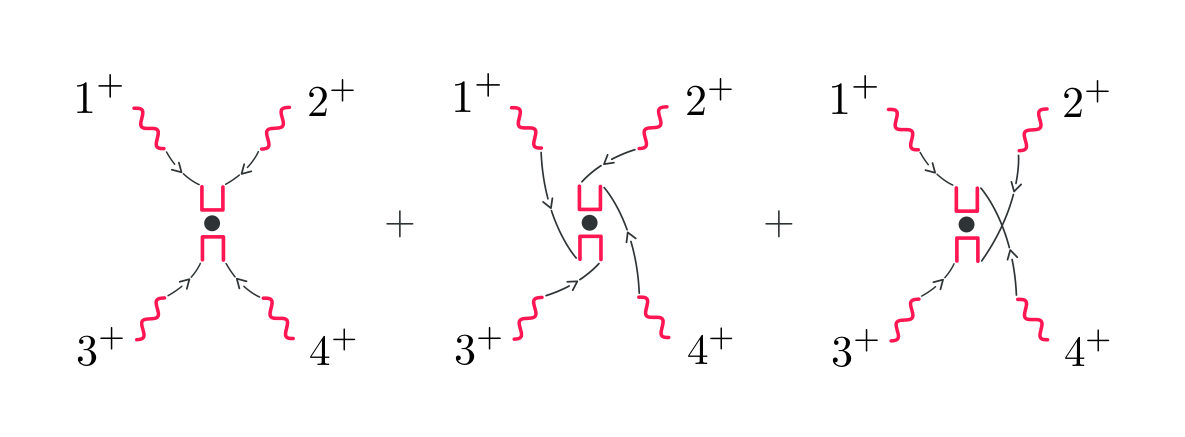}\\ 
    \mathcal{A}[4^+,0^-] \hspace{3.5em}& \propto ~ \hspace{1.4em}[12]^2[34]^2 \hspace{1.1em}+ \hspace{1.4em}[13]^2[24]^2\hspace{1.1em}+\hspace{1.4em}[14]^2[23]^2\,\,
\end{aligned}
\end{equation}
There are  three different perfect matchings  in this case.

To generate the perfect matchings in the general case with arbitrary $N$ and $K$, it is  convenient to define the matrix of all possible angle and square brackets, 
\begin{equation}
    \chi_K^{(N)} = \mqty(\chi^+ & 0 \\ 0 & \chi^-)\,: \ \begin{cases} \chi^+= [ij]^2\,, & i,j \in \{1,\cdots,K\} \\[0.5em] \chi^-= \langle mn \rangle^2\,, & m,n \in \{K+1,\cdots,N\}\end{cases} \,.
\end{equation}
The  Hafnian of this matrix  is  given by  ( see e.g. \cite{barvinok2017combinatorics})
\begin{equation}
    {\text{Hf}}(\chi_K^{(N)}) = \sum_{\rho \in P^2_{N}} \prod_{\{i,j\} \in \rho } (\chi_K^{(N)})_{ij}\,\,,
\end{equation}
where   $P^2_N$ is the set of all partitions of the set $\{ 1 , 2 ,\cdots , N \}$ into subsets of size 2, which produces all  {perfect matchings.}

Let us determine the overall factor of the amplitude. From standard Feynman rules for identical particles, we know that the vertex picks a $K!$ factor from the $F^K_+$ power and $(N-K)!$ factor from the $F^{N-K}_-$ power. On the other hand, the number of perfect matchings in a set of size $2k$ is $(2k-1){!}{!}$   Since the Hafnian captures every  inequivalent perfect matchings, the overall factor is 
\begin{equation}\label{eq:external_multiplicity}
C^{(N)}_K = \frac{K{!} (N-K){!}}{(K-1){!}{!} (N-K-1){!}{!}}=  K{!}{!} (N-K){!}{!}\,     
\end{equation}
In  the example of ${\cal A}[4^+,0^-]$, we have $C^{(4)}_0=8$, which correctly reproduces  the $4!=3\times 8$ permutations of the photons.

Using the above results, the tree amplitude produced by the operator \eqref{eq:operator_example} is simply written  as
\begin{equation}\label{eq:irred_amps}
    \diagram[0.25]{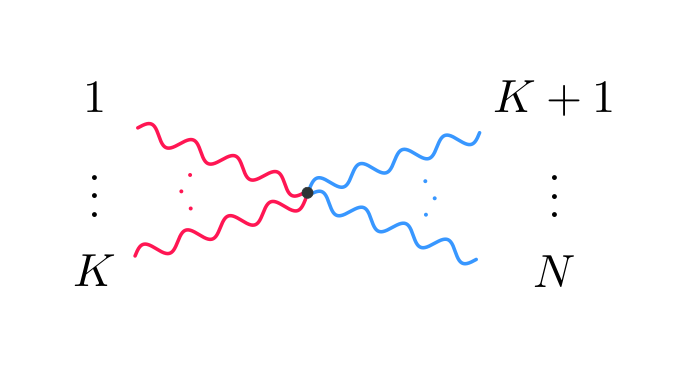} =~  (-1)^{N/2}\,i ~ \vsd^{(N)}_K  \, \widehat{\text{Hf}}\left(\chi_K^{(N)}\right)\,,
\end{equation}
where $\widehat{\text{Hf}}$ is the {\it weigthed Hafnian} defined as 
\begin{equation}
    \widehat{\text{Hf}}\left(\chi_K^{(N)}\right) =  C_K^{(N)}\,\text{Hf}\left(\chi_K^{(N)}\right) \,. 
\end{equation}

It is worth mentioning that the Hafnian could also be defined  using a sum over all the transformations of  the  symmetric groups $S_{K}$, $S_{N-K}$. 
With such choice, the  $C^{(N)}_K$ coefficient gets hidden within the sum over  $S_{K}$, $S_{N-K}$ --- which are much larger than the $P^2_K$, $P^2_{N-K}$ groups. It is also necessary
to introduce the $\chi^+$ and $\chi^-$ matrices separately. Summing over the symmetric group is thus  less convenient for practical purposes.

\paragraph{Examples.}Using the general formula \eqref{eq:irred_amps}, we obtain that the two independent 4-pt helicity amplitudes produced by the operator \eqref{eq:operator_example} are 
\be
{\cal A}[2^+, 2^-]= 2^2 i \,\alpha^{(4)}_2 \,\langle3 4\rangle^2\left[1 2\right]^2\,,
\ee
\be
{\cal A}[4^+, 0^-]=  2^3 i\, \alpha^{(4)}_4 \,\left([12]^2[34]^2+[13]^2[24]^2+[14]^2[23]^2\right)\,.
\ee 
The two independent 6-pt helicity amplitudes are
\be
{\cal A}[4^+, 2^-]= - 2^4 i \,\alpha^{(6)}_4 \,\left(\, \langle5 6\rangle^2\left[1 2\right]^2\left[3 4\right]^2 + \langle5 6\rangle^2\left[1 3\right]^2\left[2 4\right]^2 + \langle5 6\rangle^2\left[1 4\right]^2\left[2 3\right]^2\right)\,,
\ee 
\begin{align}
{\cal A}[6^+, 0^-]= -  3\cdot 2^4 i\, \alpha^{(6)}_6 \, \Big( & \quad\left[1 2\right]^2\left[3 4\right]^2\left[5 6\right]^2 + \left[1 2\right]^2\left[3 5\right]^2\left[4 6\right]^2 + \left[1 2\right]^2\left[3 6\right]^2\left[4 5\right]^2 \nn \\  \nn
& + \left[1 3\right]^2\left[2 4\right]^2\left[5 6\right]^2 + \left[1 3\right]^2\left[2 5\right]^2\left[4 6\right]^2 + \left[1 3\right]^2\left[2 6\right]^2\left[4 5\right]^2
\\ \nn 
& + 
\left[1 4\right]^2\left[2 3\right]^2\left[5 6\right]^2 + \left[1 4\right]^2\left[2 5\right]^2\left[3 6\right]^2 + \left[1 4\right]^2\left[2 6\right]^2\left[3 5\right]^2 
\\ \nn 
& + \left[1 5\right]^2\left[2 3\right]^2\left[4 6\right]^2 + \left[1 5\right]^2\left[2 4\right]^2\left[3 6\right]^2 + \left[1 5\right]^2\left[2 6\right]^2\left[3 4\right]^2 
\\ 
& + \left[1 6\right]^2\left[2 3\right]^2\left[4 5\right]^2 + \left[1 6\right]^2\left[2 4\right]^2\left[3 5\right]^2 + \left[1 6\right]^2\left[2 5\right]^2\left[3 4\right]^2 \Big)\,.
\end{align}
The three 8-pt amplitudes are 
\begin{gather}
    \mathcal{A}[4^+,4^-] =  3\cdot 2^7 i\, \alpha^{(8)}_4 ~ \text{Hf}(\chi^{(8)}_4)\,,\\[0.5em]
    \mathcal{A}[6^+,2^-] =  3\cdot 2^5 i\, \alpha^{(8)}_6 ~ \text{Hf}(\chi^{(8)}_6)\,,\\[0.5em]
    \mathcal{A}[8^+,0^-] =  2^6i\, \alpha^{(8)}_8 ~ \text{Hf}(\chi^{(8)}_8)\,, \label{eq:A8pt_irr}
\end{gather}
with Hafnians collected in App. \ref{app:hafnians}.

These amplitudes, together with the coefficients given in Table \ref{tab:EFT_coefs},  are consistent with the results obtained in \cite{Martin:2003gb} for the Euler-Heisenberg EFT and its scalar counterpart.

%% file: Sections/csw_expansion.tex
\section{Amplitude Factorization}
\label{se:CSW}

In this section, we outline an on-shell method for computing arbitrary helicity amplitudes in the EFT of electromagnetism. 

The fact that on-shell momenta  factorize into product of spinors  underlies on-shell  methods such as CSW \cite{Cachazo:2004kj} and BCFW \cite{Britto:2004ap, Britto:2005fq}.  These methods apply to pieces of the amplitudes that have poles. 
In our theory,  the $\langle + -\rangle$ self-dual propagator exhibits a pole (see  section \ref{se:SD_basis}), hence the  $+-$ lines can be cut. 

Consider for example the $[i|q\ket{j}^2$ term appearing 
in diagram \eqref{eqs:PropDiagramsLocal}, in which a  $\langle + -\rangle$  propagator is connected to two photon states. When the $q^{\dot a a}$ momentum becomes on-shell, we obtain 
\begin{equation}
    \frac{([i|_{\dot{a}}\,q^{\dot{a}a}\,\ket{j}_a)^2}{q^2}\Big|_{q^2\to 0 } = \frac{1}{q^2}([i|_{\dot{a}}\,|q]^{\dot{a}}\,\bra{q}^{a}\,\ket{j}_a)^2 = [iq]^2\frac{1}{q^2}\ev{qj}^2\,.
    \label{eq:facto_example1}
\end{equation}
On the r.h.s, the resulting    $[i q]^2$ and $\langle q j \rangle^2$ factors can be viewed as pieces of two independent subamplitudes, sewed together and continued off-shell. Factorizations like this are the backbone of recursion methods.  

{Unfortunately, no factorization is possible for the $\langle++\rangle$ and $\langle--\rangle$ propagators, as they have no poles. The resulting contact contributions seemingly obstruct a standard recursive construction of higher-point amplitudes. How, then,  can one formulate an efficient factorization-based method for  the EFT of electromagnetism?} The key insight is to identify and exploit an underlying structure  in the diagrams.

\subsection{Amplitude Substructure and a CSW-like Expansion}
\label{se:amp_substructure}

\begin{figure}
    \centering
    \includegraphics[width=0.8\linewidth]{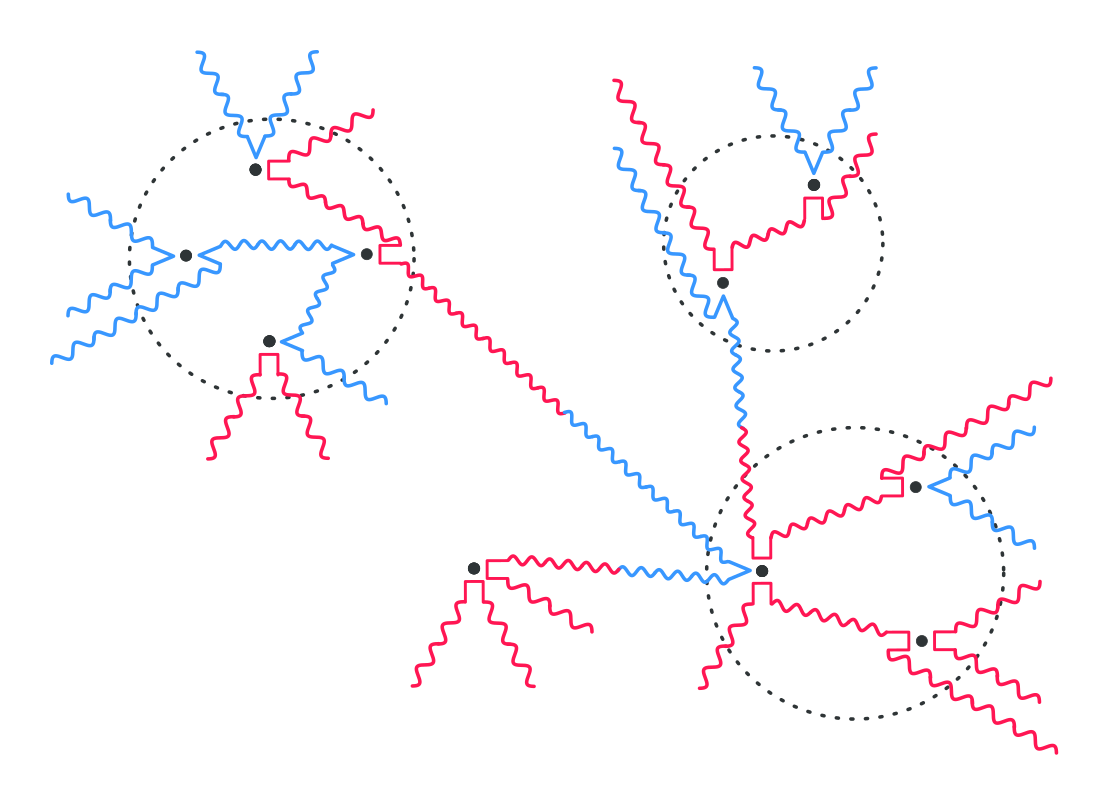}
    \caption{A contribution to the  $N=24$ photon amplitude with $K=13$ positive helicities, i.e. ${\cal A}[13^+, 11^-]$,  in the EFT of electromagnetism. This contribution can be viewed as  four distinct contact subamplitudes (one of which  being  a simple vertex) connected by $+-$ lines. }
    \label{fig:amplitude_factorization}
\end{figure}

We observe that any helicity amplitude can be decomposed into substructures built solely from $++$ and $--$ lines,  which are connected to one another through $+-$ lines. Using the fact that every $+-$ line factorizes, as exemplified in \eqref{eq:facto_example1}, each of these substructures is itself an helicity amplitude. Moreover, since they involve  only local propagators, these amplitudes  are purely contact. We thus refer to them as \textit{contact amplitudes}. {In short, any generic diagram can be viewed as a web of contact amplitudes connected by $+-$ lines. This structure is exemplified for a representative diagram in Fig.\,\ref{fig:amplitude_factorization} }.

The identified substructure provides a framework to recursively construct generic amplitudes. 
More precisely, a $N$-point amplitude splits into two pieces
\begin{equation}
{\cal A}={\cal A}_{\rm factorizable}+{\cal A}_{\rm contact}\,,
\end{equation}
with the factorizable part being expressed in terms of lower-point contact amplitudes, sewn together through $\langle+-\rangle $ propagators, similar to the CSW construction. The remaining non-factorizable part of the amplitude is simply the $N$-point contact amplitude.

{We have shown in section \ref{se:hafnians} that amplitudes originating from irreducible tree diagrams can be represented by a single Hafnian (see \eqref{eq:irred_amps}). Since the contact amplitudes behave as effective single vertices, the same calculation holds, and they are also given by a single Hafnian. Moreover, factorization implies that generic photon scattering amplitudes are expressed as combinations of Hafnians.}

In this section, the decomposition of amplitudes into a web of contact amplitudes is identified  at the diagrammatic level.
However, the functional formalism introduced in the next section provides a more formal proof  (see section  \ref{se:proof_substructure}).

\subsubsection{Contact amplitudes as building blocks}
\label{se:building_blocks}

For a contact amplitude with a given number of legs, only a finite number of tree diagrams contribute. We represent this finite sum diagrammatically by using an open circle, e.g.
\begin{gather}
\diagram{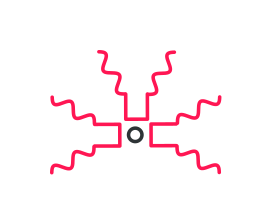} = \diagram{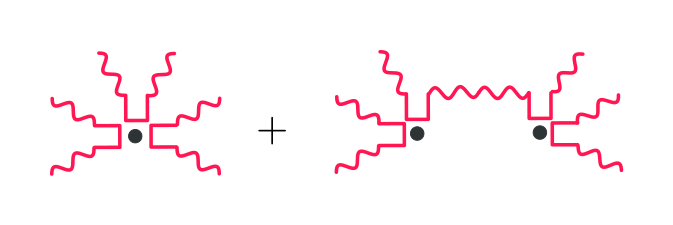}\,,\label{eq:6pt_contact1} \\
\diagram{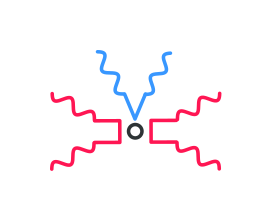} = \diagram{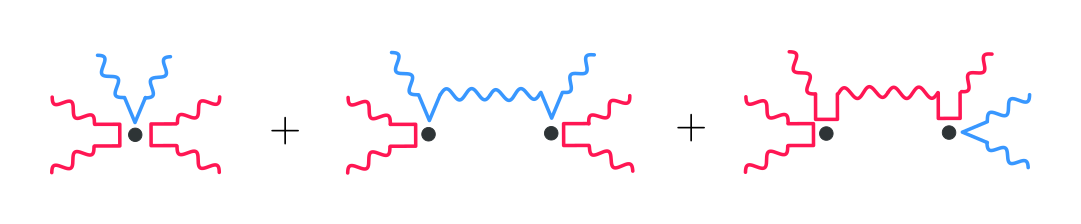}\,,\label{eq:6pt_contact2}
\end{gather}
for the 6-point contact amplitudes. Notice that the contact amplitudes inherit the property of the fundamental vertices of having even number of $+$ and $-$ legs.

Although a systematic computation method is presented in section \ref{se:Contact_lagrangian},  lower point contact amplitudes can also be evaluated manually.
Of course, a traditional computation of \eqref{eq:6pt_contact1} and \eqref{eq:6pt_contact2} requires  careful bookkeeping of the multiplicity factors coming from the various ways to connect vertices and external particles. Along these lines, it is worth mentioning that 
the  external particles multiplicity can still be captured in the $C^{(N)}_K$ from \eqref{eq:external_multiplicity}, while the internal vertices combinations have to be computed separately (using labelled-graph considerations, for example). 

 As an example, for the 6-point contact amplitudes (denoted ${\cal A}_c$), we find
\begin{align} \label{eq:Ac60}
    &\mathcal{A}_c[6^+,0^-] = -3\cdot 2^4\,i \,\qty[\alpha^{(6)}_6  + 2^3 \left(\alpha^{(4)}_4\right)^{2}] \, \text{Hf}(\chi^6_0)\,,  \\[0.5em]
    &\mathcal{A}_c[4^+,2^-] = -2^4\,i \,\qty[\alpha^{(6)}_4 + 2^2\left(\alpha^{(4)}_{2}\right)^{2} + 2^3\alpha^{(4)}_4\,\alpha^{(4)}_2] \, \text{Hf}(\chi^4_2)\,. \label{eq:Ac42}
\end{align}
 In this simple case, the contact amplitudes happen to correspond to the full amplitude, but this is generally not true for higher points.

\subsubsection{Sewing amplitudes: practical steps}
\label{se:practical_steps}

Assuming that the contact amplitudes are known, the factorizable part of a generic helicity amplitude can be obtained as follows. 

\paragraph{Finding topologies.}
As a first step, it is necessary to find {all} factorization topologies. These are conveniently determined by drawing all topologies connecting contact amplitudes through $+-$ lines. Allowed topologies are dictated by the choice of $N$ and $K$, and by the 
connections between $+$ and $-$ helicities carried by each contact amplitude.

\paragraph{Sewing representatives.}
The next step is to label external and internal photon legs on each topology,  thereby selecting  representative channels. One then writes each diagram as a product of contact amplitudes.  Each connection between internal states $q_i^\pm$ contributes a factor of $i/q_i^2$.  By the identity \eqref{eq:facto_example1}, the
 momenta  $q_i$ can be continued  off-shell, thereby sewing together the corresponding contact amplitudes.

\paragraph{Summing channels.}

Finally, one sums over every inequivalent channels associated to each topology. These channels correspond to permutations of external states among the \textit{different} Hafnians in a given topology. No further sums are required, as all remaining permutations are already encoded within the Hafnians. 
The result is the complete factorizable part of the helicity  amplitude.

\subsection{Example: Sewing 6-point Amplitudes}

\label{se:sewing_examples}

As an elementary example, let us compute the generic 6-point amplitudes through sewing. 

We start with the $\mathcal{A}[5^+,1^-]$ amplitude. In this case, there is only one possible topology: 
\\[-1em]
\begin{equation}
    \mathcal{A}[5^+,1^-] = \diagram{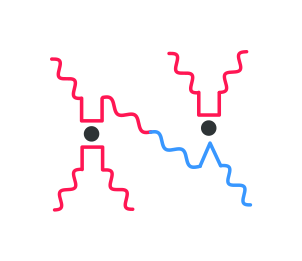} = \diagram{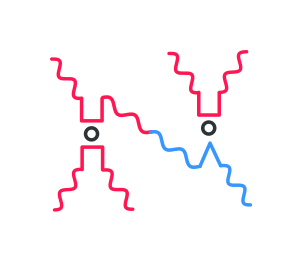} \,
\end{equation}
\\[-1em]
The only building blocks are the  4-pt contact amplitudes, which are just the 4-pt vertices.
A representative is obtained by assigning the set o momenta $\{1^+, 2^+, 3^+,q^+\}$ and $\{4^+, 5^+, 6^-,q^-\}$ to the left and right vertices, respectively, with $q$ being the momentum that goes off-shell. This yields the expression
\\[-1em]
\begin{equation}
    \begin{aligned}
    \hspace{-1em}\diagram{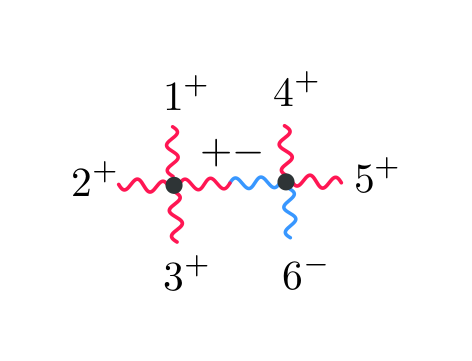} &= \qty[\alpha^{(4)}_4 ~\widehat{\text{Hf}}\qty(\chi^{(4)}_4(q^+))]\,\frac{i}{q^2}\,\qty[\alpha^{(4)}_2~\widehat{\text{Hf}}\qty(\chi^{(4)}_2(q^-))] \, \\[-2em] &= \Big[2^3i\,\alpha^{(4)}_4\,([12]^2[3q]^2+[23]^2[1q]^2+[13]^2[2q]^2)\Big] \frac{i}{q^2}\qty[2^2i\,\alpha^{(4)}_2 \,[45]^2\ev{6q}^2]\\[0.5em]
    &= ~ -2^5i\, \alpha^{(4)}_4\alpha^{(4)}_2 \,[12]^2[45]^2\frac{[3|q\ket{6}^2}{q^2} + (13) + (23)\,, \label{eq:A51_rep}
    \end{aligned}
\end{equation}
with $(ij)$ denoting the permutation of the external states $i$ and $j$. 
We remind that these  are  only a small subset of all  photon permutations.

\begin{figure}
    \centering
    \includegraphics[width=\linewidth]{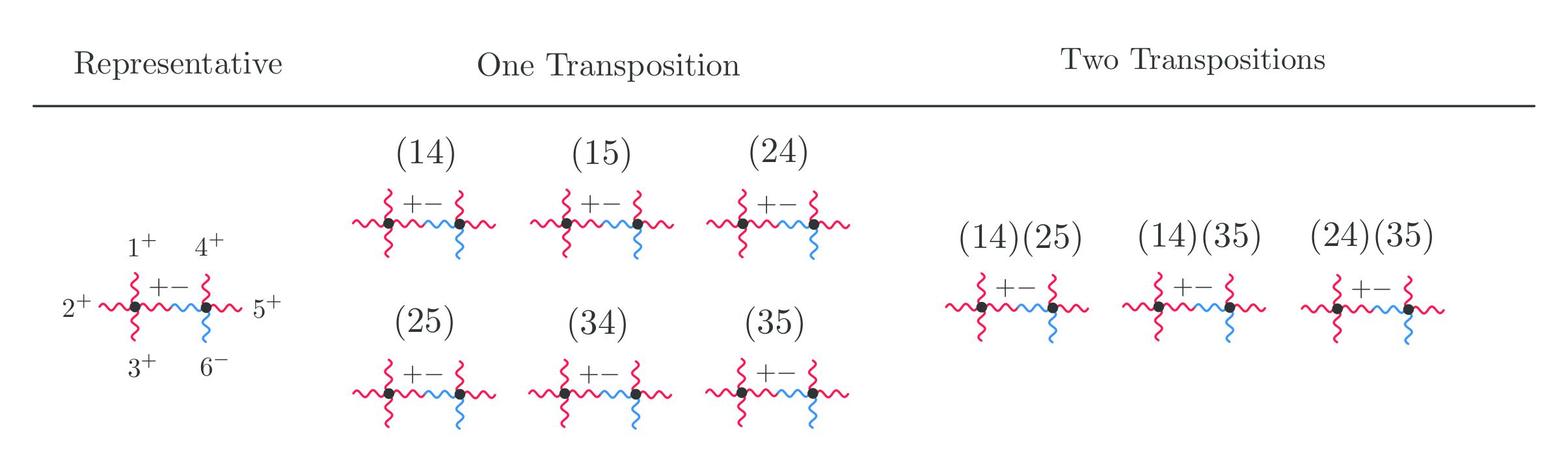}\,.
    \caption{All possible channels contributing to the $\mathcal{A}[5^+,1^-]$ helicity amplitude. The channels are generated by the $(ij)$ transpositions  applied to   the external states of the  representative diagram. }
    \label{fig:6pt_K5_channels}
\end{figure}

Finally, one sums over channels. The $q$'s in each channel are conveniently labelled as $q_{ijk\cdots}= p_i+p_j+p_k+\cdots$. For example, one has  $q=q_{123}$  in the representative \eqref{eq:A51_rep}. The 10 inequivalent channels are listed in Fig.\,\ref{fig:6pt_K5_channels}. {Summing over the 10 inequivalent channels  yields the factorizable part of $\mathcal{A}[5^+,1^-]$.
 Moreover, due to the odd number of polarizations, there is no contact contribution, i.e. $\mathcal{A}_c[5^+,1^-]=0 $. We have thus obtained the complete amplitude:
 \begin{equation}
    \mathcal{A}[5^+,1^-] =  -2^5i\,\alpha^{(4)}_4\alpha^{(4)}_2\, [12]^2[45]^2\frac{[3|q_{123}\ket{6}^2}{q_{123}^2} + (13) + (23) + (\text{other channels}) \,.
\end{equation}
}

We turn to the $\mathcal{A}[3^+,3^-]$ amplitude. In this case there are two inequivalent topologies:
\\[-1em]
\begin{equation}
    \begin{aligned}
    \mathcal{A}[3^+,3^-] &=\diagram{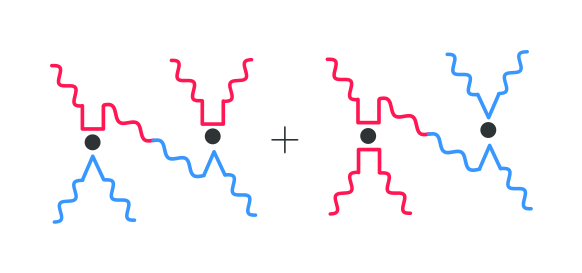}
    \end{aligned}
\end{equation}
We write the two representatives  in terms of Hafnians, then sum over all channels in each topology. The first topology has $9$ channels, while the second has only one. Since ${\cal A}_c[3^+ 3^-]=0$, the sewing directly provides the complete $\mathcal{A}[3^+,3^-] $ amplitude. The final result is
\begin{equation}
    \begin{aligned}
    \mathcal{A}[3^+,3^-] &=\qty{\diagram{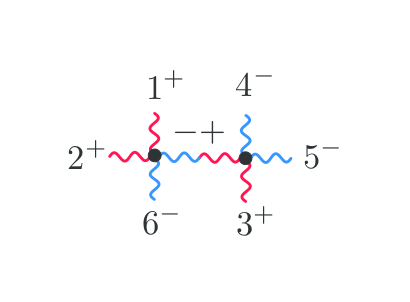} + \text{other channels}} + \qty{\diagram{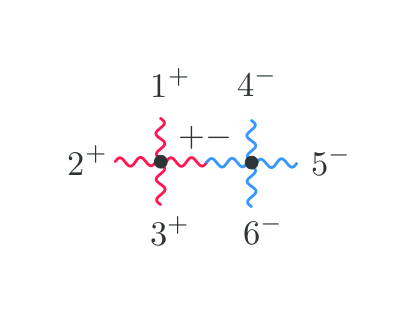}} \\
    &=  -2^4i\,\alpha^{(4)}_2\alpha^{(4)}_2 \left[ [12]^2\ev{45}^2 \frac{[3|q_{126}\ket{6}^2}{q_{126}^2} +\text{other channels} \right]\\ 
        &\quad -2^6i\,\alpha^{(4)}_4\alpha^{(4)}_0 \left[ [12]^2\ev{45}^2\frac{[3|q_{123}\ket{6}^2}{q_{123}^2} + \text{perms.} \right]\,
    \end{aligned}
    \label{eq:A33_final}
\end{equation}
In the contribution of the second topology (second line), the permutations indicated  are just those from the Hafnians, which appear as
$
~\widehat{\text{Hf}}\qty(\chi^{(4)}_4(q^+)) ~\widehat{\text{Hf}}\qty(\chi^{(4)}_0(q^-))
$.

As a sanity check, we have computed all 6-pt amplitudes through traditional Feynman rules methods. The results matche all  expressions obtained above,  but the computation requires a (very) careful combinatorial bookkeeping.

%% file: Sections/contact_eff_lag.tex
\section{The Effective Lagrangian of Contact Amplitudes}

\label{se:Contact_lagrangian}

We have identified the contact  amplitudes as the fundamental on-shell building blocks from which arbitrary  amplitudes can be  built. While low-point contact amplitudes such as those in \eqref{eq:Ac60}, \eqref{eq:Ac42} can be manually computed via standard methods, the task becomes exponentially difficult for higher number of legs.
In this section we present a functional method that systematically  computes these amplitudes.  

\subsection{The Effective Contact Lagrangian}

To compute the contact amplitudes, we introduce a reformulation of electromagnetism that  isolates the contact contributions of the propagator (see \eqref{eq:prop_matrix}). 
This reformulation trades the pair of Bianchi-constrained fields, $(F_+,F_-)$, for two pairs of \textit{unconstrained} ones, $(f_+, f_-)$ and $(\phi_+, \phi_-)$.

\subsubsection{An unconstrained formulation of electrodynamics}

We introduce the decomposition
\begin{equation}
    F_\pm = f_\pm + \phi_\pm\,,
    \label{eq:Fpm_dec}
\end{equation}
such that
\begin{equation}\label{eq:G_dec}
    \mqty( \ev{F_+ F_+} & \ev{F_+ F_-} \\ \ev{F_- F_+} & \ev{F_- F_-} ) = \mqty( 0 & \ev{f_+ f_-} \\ \ev{f_- f_+} & 0 ) + \mqty( \ev{\phi_+ \phi_+} & 0 \\ 0 & \ev{\phi_- \phi_-} )\,.
\end{equation}
As established in \eqref{eq:prop_matrix}, the non-diagonal entries are nonlocal while the diagonal entries are purely local. Therefore, while the $f_\pm$ fields propagate, the $\phi_\pm$ fields are purely static. The correlators $\ev{f\phi}$ vanish as a defining property of the above decomposition.\,\footnote{
The  decomposition defined by \eqref{eq:Fpm_dec}, \eqref{eq:G_dec} may be viewed as being in the spirit of the background field method, in which  the field is split into slow and fast modes, i.e. into background plus fluctuation. In our case, the background fields $\phi_\pm$ are truly constant, and the $f_\pm$ fields are fluctuations that vanish at zero momentum.
The  splitting of a field into slow and fast modes is for example encountered in the EFT of heavy particles  \cite{Hussain:1994zr} and of  black hole binaries  \cite{Goldberger:2004jt}. 
We mention that, while the goal in the background field method is usually to integrate out the fluctuation, here we  instead integrate out the static component.  
}

By definition, the decomposition \eqref{eq:Fpm_dec}
disentangles the  $\ev{\pm\pm}$ propagators from the $\ev{\pm\mp}$ ones, implying that $\pm\pm$ and $\pm\mp$ lines are solely generated by exchange of $\phi_\pm$ and $f_\pm$, respectively. Hence our decomposition 
makes  the amplitude substructure --- uncovered in section \ref{se:amp_substructure} --- 
manifest already at the level of fields.

Regarding  Bianchi, since the propagator \eqref{eq:prop_matrix} encodes the Bianchi constraint, our reformulation in terms of the $f$, $\phi$ fields must do so as well. It turns out that the Bianchi  identity \eqref{eq:Bianchi3}  reduces to a simple relation between the two  propagators of unconstrained fields,  $ Q_{+-}\ev{f_- f_+} = - \ev{\phi_+ \phi_+}$. 
Notice also that while the propagator matrix
 $G$ is singular,  its diagonal and anti-diagonal pieces are independently invertible, see section \ref{se:L_kin_discussion}.  
This means that the kinetic Lagrangian is uniquely determined, consistent with the fact that $f_\pm$ and $\phi_\pm$ are unconstrained.

We have therefore traded the usual formulation of electrodynamics, based on constrained fields, for an equivalent formulation in terms of {unconstrained} fields. 
This reformulation is very advantageous for 
functional calculations that follow, because integrations on unconstrained fields are  much easier to handle.

\subsubsection{Lagrangian for unconstrained fields}

Explicit evaluation of the inverse of our unconstrained field propagators, through  relations  \eqref{eq:OpsAlgebra}, determines  the kinetic Lagrangian. We find  
\begin{align}
\hat {\cal L}_{F,{\rm kin}}[f_\pm,\phi_\pm] 
& = \frac{1}{8}\left(
f_+ Q_{+-} f_- + + f_- Q_{-+} f_+ -\phi_+ P_+ \phi_+ - \phi_- P_- \phi_- 
\right)  \label{eq:hatLF_kin}
\\ & =  \frac{1}{2 q^2} f^{ab}_- q_{a \dot a} q_{b \dot b}  f^{\dot a \dot b}_+ -\frac{1}{4} \left( (\phi^+_{\dot{a}\dot{b}})^2+ (\phi^-_{ab})^2 \right)
\,  \label{eq:hatLF_kin_red}
\end{align}
 in terms of non-reduced (first line) and reduced (second line) self-dual fields. See definitions in sections \ref{se:propagators}, \ref{se:spinor_structures}.
Notice that the kinetic term for the $\phi$ fields is local, but the one for the $f$ fields is nonlocal. This non-locality was hidden in the Bianchi constraint in the original formulation.

The interactions in $\hat {\cal L}_F$ are simply the same as those of ${\cal L}_F$, expressed with the new fields, and are thus local.  The complete Lagrangian $\hat{\cal L}_F$ for the unconstrained variables is  given by
\be
\hat {\cal L}_F[f_\pm,\phi_\pm] = \hat {\cal L}_{F,\rm kin}[f_\pm,\phi_\pm]  +
{\cal L}_{F,{\rm int}}[f_\pm+\phi_\pm] \,. 
\label{eq:hatL_F_def}
\ee

\subsubsection{Generating functionals}

The self-dual correlators and the helicity amplitudes can be generated using a partition function with sources $J_\pm$ coupled to the original $F_\pm$  fields. It is
\be
Z[J_\pm]=\int_{\rm Bianchi}  {\cal D}F_\pm  e^{i S_F[F_\pm] + i\int d^4x J_\pm F_\pm }\,, \label{eq:ZF1}
\ee
where the integral on the self-dual variables is constrained by \eqref{eq:Bianchi2}.  The $S_F$ action is $S_F=\int d^4x {\cal L}_F$, with ${\cal L}_F$ the general EFT Lagrangian given in   \eqref{eq:Lag_SD}.

Introducing our decomposition into unconstrained variables \eqref{eq:Fpm_dec}, the partition function \eqref{eq:ZF1} takes the equivalent form
\be
Z[J_\pm]=\int  {\cal D}f_\pm {\cal D}\phi_\pm  e^{i \hat S_F[f_\pm,\,\phi_\pm] + i\int d^4x J_\pm (f_\pm+\phi_\pm) }\,,
\label{eq:ZF2}
\ee
with $\hat S_F =\int d^4 x \hat {\cal L}_F$ the Lagrangian defined in \eqref{eq:hatL_F_def}. 
Our focus being on generating the helicity amplitudes, we are only interested in sources probing the $f_\pm$ components. We thus choose the source to have no constant component, such that it does not couple to $\phi_\pm$. 
Given this restriction, the $\phi_\pm$ in \eqref{eq:ZF2} become internal variables that can  be integrated out. Doing so defines an effective action  for the $f_\pm$ fields, denoted $S_c[f_\pm]$, and given by
\be
Z[J_\pm]=
\int  {\cal D}f_\pm   e^{i S_c[f_\pm] + i\int d^4x J_\pm f_\pm }
\,,\quad\quad e^{iS_{c}[f_\pm]} = \int {\cal D} \phi_\pm e^{i \hat S_F[f_\pm,\phi_\pm] } \,. 
\label{eq:S_c_def}
\ee
The  effective action $S_c$ does not  exhibit any new nonlocality compared to 
$\hat S _F$,  since the  $\phi_\pm$ propagators  are purely local. We thus refer to $S_c$ as the \textit{contact effective action}, and to the corresponding Lagrangian ${\cal L}_c$ as the \textit{contact Lagrangian}.

The kinetic part of ${\cal L}_c$ is determined by \eqref{eq:hatLF_kin}. 
The general form of the interaction part of the contact Lagrangian must be a series of all possible monomials of $f_+$ and $f_-$ ---just like the original effective Lagrangian ${\cal L}_F$ of  \eqref{eq:L_F_def}. 
We thus write ${\cal L}_c$ as 
\be
{\cal L}_c = {\cal L}_{c,{\rm kin}} + {\cal L}_{c,{\rm int}}
\label{eq:L_c_def}
\ee
\be
{\cal L}_{c,{\rm kin}}[f_\pm] = 
\frac{1}{2 q^2} f^{ab}_- q_{a \dot a} q_{b \dot b}  f^{\dot a \dot b}_+\,,\quad\quad {\cal L}_{c,{\rm int}}[f_\pm] = 
\sum_{n=1}^\infty \sum_{k=0}^n \vhat^{(2n)}_{2k} f_+^{2n-2k}f^{2k}_- \,. 
\label{eq:L_c_kin_int}
\ee

The structure of ${\cal L}_{c,{\rm int}}$ is essentially the same as of \eqref{eq:L_F_def}, except that the coefficients  differ. In the second expression of \eqref{eq:L_c_kin_int}, our convention is that the fields have Lorentzian indexes, with $(f_\pm)^2=f_\pm^{\mu\nu} f_{\pm,\mu\nu}$,  matching the convention in \eqref{eq:L_F_def}. The translation to the reduced self-dual fields is simply given by 
$(f_-)^2 = \frac{1}{2} f^{ab}_-f_{-,ab}$ and  $(f_+)^2 = \frac{1}{2} f^{\dot a \dot b}_+f_{+, \dot a \dot b}$.

What have we achieved? Since all $\phi_\pm$ lines have been integrated out, the interacting contact Lagrangian \eqref{eq:L_c_def} encodes, by construction,  all  contact amplitudes. Each contact amplitude is in one-to-one correspondence with an operator in ${\cal L}_c$. Consequently, these amplitudes are   given directly by the Hafnian formula  \eqref{eq:irred_amps}, with the  $ \vhat_K^N$  as couplings.  

Our  definition of the contact effective action \eqref{eq:S_c_def} relies solely on the integration over $\phi_\pm$ and  thus 
holds  to all loop orders.
However, by construction,  all loop diagrams contributing to the $\vhat_K^N $s coefficients involve only the $\langle++\rangle$, $\langle--\rangle$ propagators, such that the loop integrands do not contain any loop momenta in their denominators.  As a result, all loop contributions to the contact Lagrangian  vanish identically upon using  dimensional regularization.  The contact Lagrangian therefore receives    contributions only at tree-level, 
\be
\vhat_K^N = \vhat_K^N \Big|_{\rm tree}\,. 
\ee

\subsubsection{Functional proof of amplitude substructure}

\label{se:proof_substructure}

Our functional formalism provides a simple proof of the contact substructure identified diagrammatically   at tree-level in section \ref{se:CSW} (see e.g. Fig.\,\ref{fig:amplitude_factorization}). 
Starting from the partition function $Z[J_\pm]$,  one  defines the generating functional $W[J_\pm]=i\log Z[J_\pm]$. By construction, the $W[J_\pm]$ generates the connected diagrams of the theory which, in the present formulation, are  composed of $\langle+-\rangle$ propagators connecting the vertices of the contact Lagrangian. This is precisely the decomposition into a web of contact amplitudes. The functional formalism shows  that this structure  holds beyond    tree-level, applying to general correlators, and thus also to scattering amplitudes.

\subsection{Evaluation of the Contact Lagrangian: Exact Formulas}

The contact Lagrangian   is conveniently computed by  functional methods. 
Throughout this section we work with the reduced self-dual fields, whose kinetic Lagrangian is given in \eqref{eq:hatLF_kin}. We will often omit the spinor indices to alleviate the notation. 

The contact Lagrangian is obtained by performing the $\phi_\pm$ integral in \eqref{eq:S_c_def}.  When evaluated at tree-level, this yields the relation
\be
S_c[f_\pm]=\hat S_F[f_\pm, \bar\phi_\pm] \,,
\label{eq:S_c_calc}
\ee
where $\bar\phi_\pm \equiv \langle \phi_\pm \rangle$ is the classical value of $\phi_\pm$ satisfying the corresponding classical equation of motion. 
Since the kinetic Lagrangian is  $  -\frac{1}{4} \left( (\phi^+)^2+ (\phi^-)^2 \right)$, the classical equation of motions  are
\be
\bar \phi_\pm  = 2  \frac{\partial }{\partial \phi_\pm} {\cal L}_{F,{\rm int}}[f_\pm+\phi_\pm] \Bigg|_{\phi_\pm=\bar \phi_\pm} \,. 
\label{eq:phibar_EOM1}
\ee

Plugging the equation of motion \eqref{eq:phibar_EOM1} into the $\hat {\cal L }_F$ Lagrangian expresses the contact Lagrangian as 
\be
{\cal L}_c[f_\pm] =  {\cal L}_{F,{\rm int}}[\bar\phi_\pm+ f_\pm] - \left[ 
\frac{\partial }{\partial \phi_+} {\cal L}_{F,{\rm int}}[f_\pm+\phi_\pm]  
\right]^2_{\phi_+=\bar \phi_+}
- \left[ 
\frac{\partial }{\partial \phi_-} {\cal L}_{F,{\rm int}}[f_\pm+\phi_\pm]  
\right]^2_{\phi_-=\bar \phi_-}
\,. \label{eq:Lc_2}
\ee 

An equivalent form is obtained by performing an appropriate shift of the $\bar \phi$ fields, as observed in \cite{Novotny:2018iph}.  This follows from the freedom to redefine $\phi_\pm$  since it is a dummy variable of the field integral, see \eqref{eq:S_c_def}. Defining $\phi^' = \phi + f $,  one obtains
\be
{\cal L}_c[f_\pm] =  {\cal L}_{F,{\rm int}}[{\bar\phi}^'_\pm] - \left[ 
\frac{\partial }{\partial \phi^'_+} {\cal L}_{F,{\rm int}}[\phi^'_\pm]  
\right]^2_{\phi^'_+={\bar \phi}^'_+} 
- \left[ 
\frac{\partial }{\partial \phi^'_-} {\cal L}_{F,{\rm int}}[\phi^'_\pm]  
\right]^2_{\phi^'_-={\bar \phi}^'_-} 
\label{eq:Lc_3}
\ee
with the equation of motion
\be
\bar \phi^'_\pm  = f_\pm +   2  \frac{\partial }{\partial \phi_\pm^'} {\cal L}_{F,{\rm int}}[\phi^'_\pm] \Bigg|_{\phi_\pm^'={\bar \phi}^'_\pm} \,. 
\label{eq:phibar_EOM2}
\ee
In this new form, the $f$ field only appears as an additive term in the equation of motion. 

Furthermore, it is convenient to define the Lorentz scalars  \cite{Novotny:2018iph}
\be
\x_\pm = {\bar\phi}^'_{\pm,a \dot a} {\bar\phi}_{\pm}^{'\, \dot a a }\,\quad\quad 
\y_\pm = f_{\pm,a \dot a} f_{\pm}^{\dot a a } \,. 
\label{eq:xy_def}
\ee
Using these scalar variables, the Lagrangian \eqref{eq:Lc_2} and equation of motion  \eqref{eq:phibar_EOM2} become 
\be
{\cal L}_c[\y_\pm] =  {\cal L}_{F,{\rm int}}[\x_\pm] - 4 x_+ \left(  
 \frac{\partial}{\partial x_+} {\cal L}_{F,{\rm int}}[\x_\pm]  
\right)^2
- 4 x_-  \left( \frac{\partial}{\partial x_-} {\cal L}_{F,{\rm int}}[\x_\pm]  
\right)^2
\,, \label{eq:Lc_4}
\ee
\be
y_\pm= x_\pm \left(1-4 \frac{\partial}{\partial x_\pm}{\cal L}_{F,{\rm int}}[\x_\pm]  \right)^2\,. \label{eq:phibar_EOM3}
\ee

\subsection{Evaluation of the Contact Lagrangian: Iterations and Tree Graphs}

The equation of motion being nonlinear, it generally cannot be solved exactly for $\bar \phi_\pm$. In the perturbative regime, however, it can be solved order by order using the fixed point iteration method around $\phi_\pm=0$. In fact, from the diagrammatic viewpoint, plugging iteratively the classical equation into the fundamental action produces all  connected tree graphs of the theory (see e.g. \cite{Skinner:lecture}). 

In our case, because the $\phi_\pm$ propagators are local,  each tree graph reduces to a single vertex. These  vertices correspond directly to the contact amplitudes, see definition in section \ref{se:amp_substructure}. Therefore, plugging iteratively  \eqref{eq:phibar_EOM1} into $\hat S_F[f_\pm, \bar\phi_\pm]$ produces monomials of $f_+$, $f_-$ that are precisely the contact operators  corresponding to each contact amplitude! 

The iterative process provides therefore a systematic  method for computing the contact Lagrangian. Furthermore, since the operators in $S_c$ are  generated only at tree-level, the definition of the contact action \eqref{eq:S_c_calc} is exact at \textit{all} loop order, hence the iterative process  produces the exact contact Lagrangian.  We emphasize that the  notion of iteration appearing in this section is unrelated to the notion of recursion underlying the CSW-like sewing of  contact amplitudes.  Both  are used in this work, but  at different stages: the former in the computation of the contact amplitudes, and the latter in  sewing them into full amplitudes.

\subsubsection{Branching depth and stopping criterion} 

In practice, performing the iterations  can be computationally intensive. An efficient approach to implement the $m$ iterations is to first perform the $m-1$ iterations at the level of the equation of motions, and only then substitute into the action as the final $m$-th iteration.

It is also very useful to know the stopping criterion for the number of iterations  needed to obtain all topologies for a target number of external legs $N$.   
The derivation of the sufficient number of iterations, denoted $m_\min$,   is achieved by considering the lowest-point vertices, which are here  the quartic ones. 

\begin{figure}
    \centering
    \includegraphics[width=1.\linewidth]{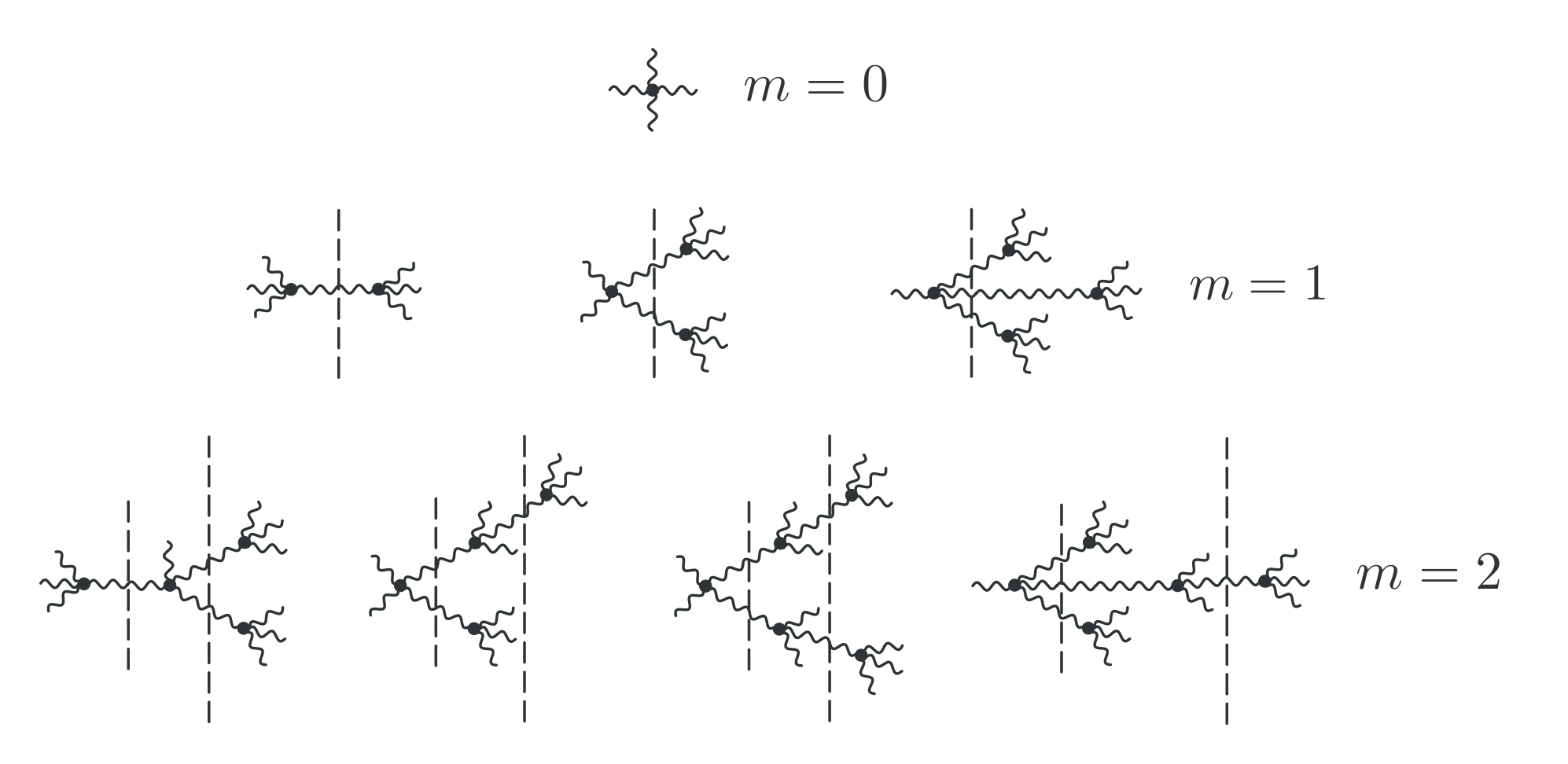}
    \caption{Examples of topologies produced with quartic vertices through fixed-point iteration, up to $m=2$ steps. The $m=2$ step produces all topologies for $N=10$ photons, while the $m=1$ step already saturates all possible topologies for $N=6$ and $N=8$ photons.}
    \label{fig:tree_gens}
\end{figure}

 The interaction Lagrangian  being a polynomial of $\bar \phi+f$ (using  the representation \eqref{eq:Lc_2}),  each iteration produces trees with various number of vertices and of branching points  (i.e. vertices adjacent to more than two other vertices). 
The stopping criterion to generate all topologies with $N$ photons is related to the number of branching points. More precisely, it is related to the \textit{branching depth}, denoted $b$, which is the greatest number of levels one must go through when following a chain of branches from the starting point to the furthest endpoint. The key idea is that, for a target number of photons, all tree topologies are generated when a sufficiently high branching depth $b_{\max}$ is reached. 
 This is illustrated in Fig.\,\ref{fig:tree_gens}. 

Let us compute $b_{\max}$, then $m_\min$, as a function of $N$. A tree with $v$ internal quartic vertices has $N = 4v - 2(v-1) = 2v+2$ external legs, giving $v = (N-2)/2$. 
The maximum branching depth  in such a tree is
\begin{equation}
    b_{\max} = \left\lfloor\frac{v-2}{2}\right\rfloor = \left\lfloor\frac{N-6}{4}\right\rfloor\,.
\end{equation}
  Running  the iteration  $ m\geq b_{\max}+1$ times guarantees to produce all possible topologies.\,\footnote{In practice, it is convenient to perform $b_{\max}$ iterations at the level of the equation of motion, then substitute into the action.} The stopping criterion on the number of iterations is therefore 
\be
m_\min=\left\lfloor\frac{N-2}{4}\right\rfloor \,.
\ee
This result is exemplified in Fig.\,\ref{fig:tree_gens}. 

\subsection{Contact Lagrangian of the General EFT of Electromagnetism}

Applying the fixed-point method, the contact Lagrangian in the general EFT of electromagnetism, expressed in terms of the original $\vv_i^{(N)}$ couplings and up to $N=8$ photons, is found to be
\begin{equation}
    \begin{aligned}
        \mathcal{L}_{c,{\rm int}} & = ~~(\vv^{(4)}_0-\vv^{(4)}_1)\qty[f_+^4+f_-^4] + 2(\vv^{(4)}_0+\vv^{(4)}_1)f_+^2f_-^2\\
        &\quad + (\vv^{(6)}_0-\vv^{(6)}_1 - 8(\vv^{(4)}_0-\vv^{(4)}_1)^2)\qty[f_+^6+f_-^6] \\
        &\quad+ (3\vv^{(6)}_0+\vv^{(6)}_1+8(2\vv^{(4)}_0\vv^{(4)}_1+3(\vv^{(4)}_0)^2-(\vv^{(4)}_1)^2))\qty[f_+^4f_-^2+f_+^2f_-^4] \\
        &\quad + (\vv^{(8)}_{1} - \vv^{(8)}_{2} + \vv^{(8)}_{3} + 96 \left(\vv^{(4)}_{1} - \vv^{(4)}_{2}\right)^{3}+ 24 \left(\vv^{(4)}_{1} - \vv^{(4)}_{2}\right) \left(\vv^{(6)}_{1} - \vv^{(6)}_{2}\right))\qty[f_+^8+f_-^8] \\
        &\quad + 4(96 (\vv^{(4)}_0)^{3} - 64 \vv^{(4)}_{1} (\vv^{(4)}_{2})^{2} + 24 \vv^{(4)}_{1} \vv^{(6)}_{1} + 32 (\vv^{(4)}_{2})^{3} - 8 \vv^{(4)}_{2} \vv^{(6)}_{2}+ \vv^{(8)}_{1} - \vv^{(8)}_{3})\qty[f_+^6f_-^2 + f_+^2f_-^6]\\
        &\quad\, \begin{aligned}+ \, 2(&3 \vv^{(8)}_0 + \vv^{(8)}_1 + 3 \vv^{(8)}_2 - 32 \vv^{(4)}_0 (\vv^{(4)}_1)^2 + 288 (\vv^{(4)}_0)^2 \vv^{(4)}_1 + 72 \vv^{(6)}_0 \vv^{(4)}_0\\ & \quad\quad + 24 \vv^{(6)}_0 \vv^{(4)}_1 + 24 \vv^{(6)}_1 \vv^{(4)}_0 + 8 \vv^{(6)}_1 \vv^{(4)}_1 + 288 (\vv^{(4)}_0)^3 - 32 (\vv^{(4)}_1)^3)f_+^4 f_-^4\end{aligned}   \\
        &\quad +\mathcal{O}(f^{10}) \,. \label{eq:L_c_EFT}
    \end{aligned}
\end{equation}
We remind that the fields have Lorentz indices, with $(f_\pm)^2= (f_\pm^{\mu\nu}f_{\pm,\mu\nu})$. 
The expression up to $N=14$ photons is given in App.\,\ref{app:contact_lag}.

%% file: Sections/BI_bootstrap.tex
\section{Contact Lagrangian in Helicity-Conserving and  Born-Infeld Electromagnetism}

\label{se:hel_conserving_lagrangian}

\subsection{Helicity Conservation and Duality }

Certain  theories of electromagnetism exhibit helicity conservation at the level of scattering amplitudes. In such theories, any amplitude ${\cal A}[K^+, (N-K)^-]$ vanishes  when $N\neq 2K$.  

Helicity conservation is in particular expected for BI 
electromagnetism, at least at tree-level.  A tree-level proof based of  the on-shell invariance under electromagnetic duality was presented in  \cite{Rosly:2002jt}, and a diagrammatic argument was given in \cite{Boels:2008fc}. A more evolved proof, valid also at loop-level, was later proposed in \cite{Novotny:2018iph}. We will see below that our functional approach produces a considerably simpler version of the latter proof.

Is there an EFT that captures \textit{all} and \textit{only}  helicity-conserving theories? While addressing this question directly with the conventional EFT Lagrangian appears rather challenging, the contact Lagrangian provides a natural framework. Indeed, if all amplitudes conserve helicity, then every contact amplitude must conserve helicity as well.  Moreover, since the $+$ and $-$ helicities each appear an even number of times in the contact amplitudes,  the only non-vanishing contact amplitudes are of the form
${\cal A}_c[P^+, P^-]$ with $P\in 2\mathbb{N}$. 
The general helicity-conserving  contact Lagrangian is thus built from monomials of $f^2_+f^2_-$, namely
\be
{\cal L}_{c, {\rm int}}^{{\rm hel. \,cons. }}  = \sum_{P=1}^\infty \hat \alpha^{(2P)}_{P} (f^2_+f^2_-)^{P} \,. 
\label{eq:L_c_BI_def}
\ee

Helicity conservation is  manifest at the level of the contact Lagrangian \eqref{eq:L_c_BI_def}, in sharp contrast to the original EFT Lagrangian. For example, when the BI contact amplitudes are computed from the original ${\cal L}_F$ Lagrangian, the vanishing of the helicity-violating amplitudes  occurs only after intricate combinations of diagrams with different topologies.
By contrast, all these combinations are already encoded into the vertices of  the contact Lagrangian. In fact, substituting the BI coefficients of Tab.\,\ref{tab:EFT_coefs} into the generic contact Lagrangian of \eqref{eq:L_c_EFT}, one does find that all helicity-violating operators cancel.

So far we have argued that any helicity-conserving theory admits a contact Lagrangian of the form \eqref{eq:L_c_BI_def}. The converse, however,  is also true.
As established in section \ref{se:proof_substructure} (see also section \ref{se:CSW}), any amplitude can be decomposed into contact vertices connected by $\langle+- \rangle$ propagators. Since each propagator connects a $+$ leg to a $-$ leg,   helicity conservation is preserved in any  amplitude of the theory. It follows that the contact Lagrangian \eqref{eq:L_c_BI_def} provides an EFT framework to capture every  helicity-conserving theories.

Finally, one  observes that \eqref{eq:L_c_BI_def} is invariant under the electromagnetic duality transformation \be
f_+\to f_+ e^{i\theta}\,\quad\quad
f_-\to f_- e^{-i\theta} \,. 
\label{eq:duality_transfo}
\ee
The statement that a  theory is  invariant under electromagnetic duality---even off-shell---is thus equivalent to requiring its  contact Lagrangian to take  the form \eqref{eq:L_c_BI_def}. 
This establishes, both at tree and loop level, that a theory is invariant under (off-shell) electromagnetic duality if and only if its amplitudes conserve helicity.  A similar conclusion has been obtained in \cite{Novotny:2018iph}.

In a nutshell, the contact Lagrangian reveals the hidden simplicity of helicity-conserving theories. In particular, it allows us to establish the equivalence between electromagnetic duality and helicity conservation in a remarkably simple way.

\subsection{Constraint Equation}

The electromagnetic duality transformation acts on the scalar variables 
$y_\pm$ defined in \eqref{eq:xy_def} as 
\be
y_+ \to y_+ e^{2 i \theta} \,\quad\quad 
y_- \to y_- e^{- 2 i \theta} \,.  
\label{eq:duality_transfo2}
\ee
We have shown above that helicity conservation is equivalent to the contact Lagrangian being invariant under these transformations. 

Even when the contact Lagrangian is invariant under \eqref{eq:duality_transfo2}, it is \textit{not} guaranteed that 
 the $x_\pm$ variables transform simply under duality transformation.  Because the relation between the $x_+$, $x_-$ and $y_+$, $y_-$ variables is nonlinear, see \eqref{eq:phibar_EOM3}, the $y_\pm$ are generally not irreps of the $U(1)$ transformation.    
However,  for both technical and conceptual purposes, it is very advantageous to have variables that transform as irreps of the $U(1)$ group. Finding such variables amounts to solving the following problem. 

We introduce the variables $z_\pm $ that are irreps of the $U(1)$ group \textit{by  assumption}.  We further introduce the invertible  functions $\h_\pm$ such that 
\be
x_+= \h_+(z_+, z_-)\,,\quad x_-= \h_-(z_+,z_-)\,.
\label{eq:h_def}
\quad
\ee
Substituting \eqref{eq:h_def} into the  equations of motion  \eqref{eq:phibar_EOM3}
yields
\be
\y_\pm= \y_\pm(z_+,z_-)= \h_\pm(z_+,z_-) \left(1-4 \frac{\partial}{\partial h_\pm} {\cal L}_{F,{\rm int}}[\h_+(z_+,z_-),\h_-(z_+,z_-)]  \right)^2\,. \label{eq:phibar_EOM4}
\ee

Viewing the $x_\pm$, $y_\pm$ and $z_\pm$ variables as complex scalars, we can notice that  $y_-=y_+^*$ together with  \eqref{eq:phibar_EOM3} imply that $x_-=x_+^*$. 
Combined with \eqref{eq:h_def} and $z_-=z_+^*$, this implies that the $h_+$ and $h_-$ functions are related as 
\be
  h_+(z_+,z_-) = h_-(z_-,z_+)\,. \label{eq:hp_hm_rel}
\ee
There is thus a single function of two variables to determine. For clarity it is still convenient to use the $h_+$, $h_-$ notation in the following.

Consider a generic smooth function  $g$ that is  invariant under the duality transformation  \eqref{eq:duality_transfo2}, i.e. $g(\y_+, \y_-)=g(\y_+ \y_-)$. Upon using the nonlinear equations of motion \eqref{eq:phibar_EOM4}, the invariance of $g$  can be used to constrain  the $h_\pm$ functions. 
To this end, we use  the infinitesimal form of the transformation in $z_\pm$. The  invariance of $g$ provides the relation
\be
z_+\frac{\partial}{\partial z_+} g\left(y_+(z_+,z_-) y_-(z_+,z_-) \right) = z_-\frac{\partial}{\partial z_-} g\left(y_+(z_+,z_-) y_-(z_+,z_-) \right) \,.\label{eq:gen_condition}
\ee
Any theory that has duality invariance must satisfy \eqref{eq:gen_condition}, for any  $g$.  The $g$ function  could be anything, including the identity. It turns out that Eq.\,\eqref{eq:gen_condition} takes a fairly simple form when $g$ is taken to be the contact Lagrangian itself, using the form in \eqref{eq:Lc_4}.

\subsection{Bootstrapping the  Born-Infeld Contact Lagrangian}

We apply the above approach to BI electromagnetism.
Setting $g\equiv {\cal L}_c$, we find that  the constraint equation \eqref{eq:gen_condition} takes the form 
\be
\frac{  h_+ \partial_- h_- \left(8\Lambda^2+h_+^2-h_-^2 \right) +
 h_- \partial_- h_+ \left(8\Lambda^2+h_-^2-h_- \right)
}{
 h_+ \partial_+ h_- \left(8\Lambda^2+h_+^2-h_-^2 \right) +
 h_- \partial_+ h_+ \left(8\Lambda^2+h_-^2-h_- \right)
} = \frac{z_+}{z_-} \,, \label{eq:constraint_BI}
\ee
where $\partial_\pm = \frac{\partial }{\partial z_\pm}$.
Noticing that the constraint can be expressed  in terms of $h_\pm^2$ only, we search solutions using the ansatz 
\be
h_+^2(z_+, z_-) = p_+(z_+)p_-(z_-) r(z_+ z_-)\,.
\ee
We take the $p_+$, $p_-$ functions as polynomials, and then derive  $r$ through the constraint \eqref{eq:gen_condition}. 
We find the consistent solutions
\be
h_+^2(z_+, z_-) = 8 a \Lambda^4 \frac{z_- (1+a z_+)^2}{(1-a^2 z_+z_-)^2}\, \label{eq:h2}
\ee
with arbitrary $a\in \mathbb{R}^*$. For $a=\frac{1}{2}$, our result \eqref{eq:h2} matches the change of variable introduced in \cite{Bellucci:2000bd} in the context of D-brane models. For $a=1$, it matches the one used in \cite{Novotny:2018iph, Ferrara:2016crd}.   Below we set $a=1$. 

We can see that, as planned,  our bootstrap method has identified the functions $h_\pm$ that expresses the static field $x_\pm = (\phi^'_{\pm})^2$ in terms of variables  $z_\pm$ that are irreps of the $U(1)$ duality group. 
With the new variable, the equation of motion becomes
\be
y_+ = \frac{8\Lambda^4 z_+}{(1-z_+ z_-)^2}\,,\quad\quad 
y_- = \frac{8\Lambda^4 z_-}{(1-z_+ z_-)^2}
\ee
and 
the interaction part of the original EFT Lagrangian and of the contact Lagrangian  
become respectively
\begin{align}
{\cal L}_{F,{\rm int}}= 2 \Lambda^4 z_+ z_-\frac{(1+z_+)(1+z_-)}{(1-z_+z_-)^2} \,,\quad\quad 
{\cal L}_{c,{\rm int}}= 2 \Lambda^4 z_+ z_-\frac{1+z_+ z_-}{(1-z_+z_-)^2}\,. \quad\quad
\label{eq:Lint_BI1}
\end{align}
These results are essentially consistent with \cite{Novotny:2018iph}.\,\footnote{Our result for  ${\cal L}_{c,{\rm int}}$ differs from (6.31) of  \cite{Novotny:2018iph} by a factor of $-2$, which is most likely a typo.   }

The key properties arising through our method is that ${\cal L}_{c,{\rm int}}$ depends only on the product $z_+ z_-$, and that $z_\pm$ is proportional to some power of $y_\pm$ since both are irreps of the $U(1)$. The latter implies that $y_+ y_-$ is a function of $z_+z_-$. Namely, we have 
\be
y_+ y_- =\frac{64\Lambda^8 z_+z_-}{(1-z_+z_-)^4}
\,.
\label{eq:EOM_5}
\ee

\subsubsection*{Inversion formula}

The remaining task  to obtain the   contact Lagrangian is to substitute $z_+z_-$ for $y_+ y_-$ in   \eqref{eq:Lint_BI1} using the equation of motion \eqref{eq:EOM_5}. This equation can be  inverted thanks to the \textit{Lagrange inversion theorem} (see e.g. \cite{Whittaker_Watson_1996, Wilf}).  

We write the equation of motion \eqref{eq:EOM_5} as
\be
Y= \frac{Z}{(1 - Z)^4}\,, \quad \quad Y=\frac{y_+ y_-}{64\Lambda^8}\,,\quad \quad Z=z_+z_-\,.
\ee
It takes the form $Y\equiv f(Z)\equiv\frac{Z}{\varphi(Z)}$ with $\varphi$ analytical in $Z$. Furthermore, the contact Lagrangian is also analytical in $Z$ when written as ${\cal L}_{c,{\rm int}}= 2\Lambda^4 Y (1+Z)(1-Z)^2 $. 

These are the conditions needed to apply the    Lagrange-Burmann inversion formula, which  is stated as follows,  
\be
[Y^n]H(f^{-1}(Y))= \frac{1}{n}[Z^{n-1}](H'(Z) \varphi^n(Z))\label{eq:LB_formula} \,.
\ee
Here  $[x^n]$ is an operator that selects the coefficient of the $x^n$ monomial in a Taylor series. In our case we have  $\varphi(Z)=(1-Z)^4$, $H(Z)= (1+Z)(1-Z)^2$. 

Using $H'(Z)=-(1-Z)(1+3Z)$, we obtain 
\begin{align}
[Y^n]H(f^{-1}(Y))& = \frac{1}{n} [Z^{n-1}] (-(1+3Z)(1-Z)^{4n+1} )= \frac{1}{n} [Z^{n-1}] (1+3Z)  \sum_{k=0}^{4n+1}  {4n+1 \choose k} (-Z)^{k+1} \nn 
 \\ 
 & = \frac{1}{n} \left( (-1)^{n} {4n+1 \choose n-1} + 3  (-1)^{n-1} {4n+1 \choose n-2} \right)= (-1)^{n}\frac{6(4n+1)!}{n!(3n+3)!} \,. \label{eq:coef_calculation}
\end{align}

\subsubsection*{The complete BI contact Lagrangian}

The coefficients obtained in \eqref{eq:coef_calculation} completely determine  the  Taylor series representation for the contact Lagrangian, that was introduced in generic form  in \eqref{eq:L_c_BI_def}.  The series can be identified as a hypergeometric $_2F_3$ function. 
Coming back to fields with Lorentz indices via $y_+ y_- = 4 (f^{\mu\nu}_+)^2(f^{\mu\nu}_-)^2$, our final result for   the contact Lagrangian of BI electromagnetism is
\begin{align}
{\cal L}_{c,{\rm int}} & = \frac{1}{8 \Lambda^4} f_+^2 f_-^2\sum_{n=0}^\infty \frac{6(4n+1)!}{n!(3n+3)!} \left(- \frac{f_+^2 f_-^2}{16 \Lambda^8}\right)^n  \\
& = 6\Lambda^4 \left[1-~_3F_2\left(-\frac{1}{2},-\frac{1}{4}, \frac{1}{4} ; \frac{1}{3}, \frac{2}{3} ; - \frac{4}{27}  \frac{f_+^2 f_-^2}{16 \Lambda^8}\right) \right] \,.
\end{align}
This is  consistent with the result presented in \cite{Novotny:2018iph}. 

%% file: Sections/applications.tex
\section{Further Results}

\label{se:further_results}

As a final application of the methods and results developed in this work, we compute the 8-point amplitudes of the general EFT of electromagnetism and the 10-point amplitudes of BI theory. 
We do so by combining the practical sewing recipe developed in Sec.~\ref{se:practical_steps} with the contact Lagrangian derived in \eqref{eq:L_c_EFT}. The results are expressed in terms of the Hafnians introduced in Sec.~\ref{se:hafnians}.

\subsection{8-pt Generic EFT Amplitudes}

The first step is to determine all possible topologies. We find 
\begin{align} \label{eq:A8pt_diag80}
    &\mathcal{A}[8^+,0^-] = ~\diagram{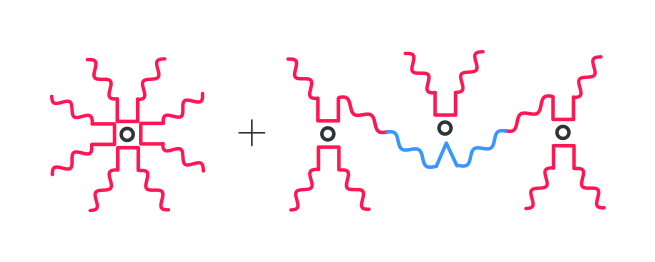} \\[-1.5em]
    \label{eq:A8pt_diag71}
    &\mathcal{A}[7^+,1^-] = ~\diagram{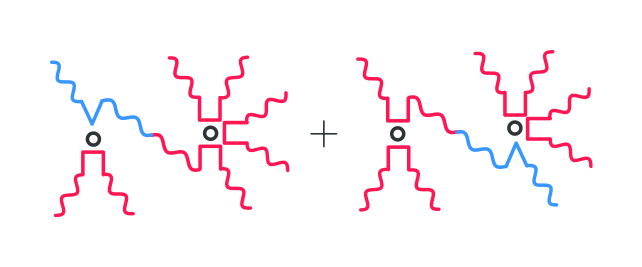} \\[-1.5em]
    \label{eq:A8pt_diag62}
    &\mathcal{A}[6^+,2^-] = ~\diagram{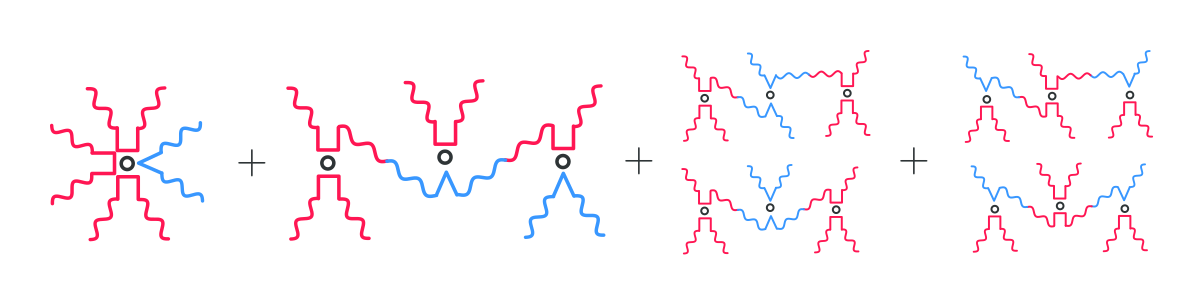} \\[-1.5em]
    \label{eq:A8pt_diag53}
    &\mathcal{A}[5^+,3^-] = ~\diagram{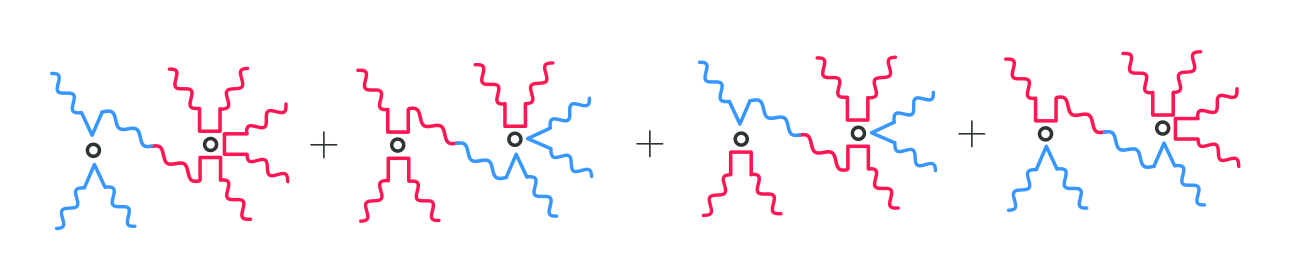} \\[-1.5em]
    \label{eq:A8pt_diag44}
    &\mathcal{A}[4^+,4^-] = ~\diagram[0.2]{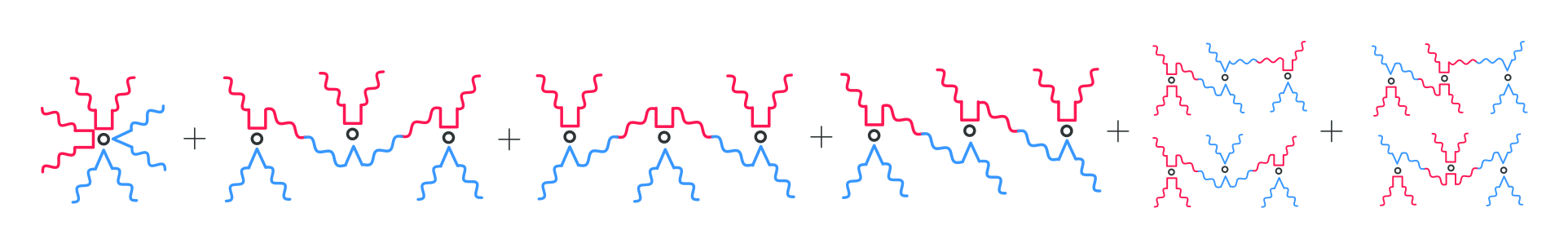} 
\end{align}
We remind that the circle vertices represent the contact amplitudes, see definition in section \ref{se:building_blocks}. 
In these diagrams, we have specified the substructure of the vertices (see section \ref{se:diags}), but we emphasize again that this is extra information and is not required to determine the topologies. 
For instance, the stacked diagrams have same topology and are encoded by the same products of Hafnians. 

Following our sewing recipe, each of the amplitudes is then written as a sum over  
products of Hafnian corresponding to each of the topologies drawn above.  Finally we sum over all existing channels, which are simply given by permutations of external states between the  different Hafnians (see examples in section \ref{se:sewing_examples}).

 The complete results for the 8-pt helicity amplitudes generated by the  EFT of electromagnetism are found to be
\begin{align}
    &\begin{aligned}
        \mathcal{A}[8^+,0^-] = &\qty[\vhat_8^{(8)} \widehat{\text{Hf}}\qty(\chi_8^{(8)})] \\[0.5em] 
        &+ \qty[\sum_{\text{channels}}\frac{\vhat_4^{(4)} \vhat_2^{(4)} \vhat_4^{(4)}}{q_1^2 q_2^2}~ \widehat{\text{Hf}}\qty(\chi_4^{(4)}(q_1^+))  \widehat{\text{Hf}}\qty(\chi_2^{(4)}(q_1^-,q_2^-)) \widehat{\text{Hf}}\qty(\chi_4^{(4)}(q_2^+))]
    \end{aligned} \\[2em]
    &\begin{aligned}
        \mathcal{A}[7^+,1^-] = &\qty[\sum_{\text{channels}}\frac{\vhat_2^{(4)}\vhat_6^{(6)}}{q^2} ~\widehat{\text{Hf}}\qty(\chi_2^{(4)}(q^-))\widehat{\text{Hf}}\qty(\chi_6^{(6)}(q^+))] \\[0.5em] 
        &+ \qty[\sum_{\text{channels}}\frac{\vhat_4^{(4)}\vhat_4^{(6)}}{q^2} ~\widehat{\text{Hf}}\qty(\chi_2^{(4)}(q^+))\widehat{\text{Hf}}\qty(\chi_4^{(6)}(q^-))]
    \end{aligned} \\[2em]
    &\begin{aligned}
        \mathcal{A}[6^+,2^-] = &\qty[\vhat_6^{(8)} \widehat{\text{Hf}}\qty(\chi_6^{(8)})] \\[0.5em] 
        &+ \qty[\sum_{\text{channels}}\frac{\vhat_4^{(4)} \vhat_2^{(4)} \vhat_2^{(4)}}{q_1^2 q_2^2}~ \widehat{\text{Hf}}\qty(\chi_4^{(4)}(q_1^+))  \widehat{\text{Hf}}\qty(\chi_2^{(4)}(q_1^-,q_2^-)) \widehat{\text{Hf}}\qty(\chi_2^{(4)}(q_2^+))] \\[0.5em]
        &+\qty[\sum_{\text{channels}}\frac{\vhat_4^{(4)} \vhat_0^{(4)} \vhat_4^{(4)}}{q_1^2 q_2^2}~ \widehat{\text{Hf}}\qty(\chi_4^{(4)}(q_1^+))  \widehat{\text{Hf}}\qty(\chi_0^{(4)}(q_1^-,q_2^-)) \widehat{\text{Hf}}\qty(\chi_4^{(4)}(q_2^+))] \\[0.5em]
        &+\qty[\sum_{\text{channels}}\frac{\vhat_2^{(4)} \vhat_4^{(4)} \vhat_2^{(4)}}{q_1^2 q_2^2}~ \widehat{\text{Hf}}\qty(\chi_2^{(4)}(q_1^-))  \widehat{\text{Hf}}\qty(\chi_4^{(4)}(q_1^+,q_2^+)) \widehat{\text{Hf}}\qty(\chi_2^{(4)}(q_2^-))]
    \end{aligned} \\[2em]
    &\begin{aligned}
        \mathcal{A}[5^+,3^-] = &\qty[\sum_{\text{channels}}\frac{\vhat_0^{(4)}\vhat_6^{(6)}}{q^2} ~\widehat{\text{Hf}}\qty(\chi_0^{(4)}(q^-))\widehat{\text{Hf}}\qty(\chi_6^{(6)}(q^+))] \\[0.5em]
        &+ \qty[\sum_{\text{channels}}\frac{\vhat_4^{(4)}\vhat_2^{(6)}}{q^2} ~\widehat{\text{Hf}}\qty(\chi_4^{(4)}(q^+))\widehat{\text{Hf}}\qty(\chi_2^{(6)}(q^-))] \\[0.5em]
        &+\qty[\sum_{\text{channels}}\frac{\vhat_2^{(4)}\vhat_4^{(6)}}{q^2} ~\widehat{\text{Hf}}\qty(\chi_2^{(4)}(q^-))\widehat{\text{Hf}}\qty(\chi_4^{(6)}(q^+))] \\[0.5em] 
        &+ \qty[\sum_{\text{channels}}\frac{\vhat_2^{(4)}\vhat_4^{(6)}}{q^2} ~\widehat{\text{Hf}}\qty(\chi_2^{(4)}(q^+))\widehat{\text{Hf}}\qty(\chi_4^{(6)}(q^-))]
    \end{aligned}  
\end{align}
\begin{align}
\begin{aligned}
        \mathcal{A}[4^+,4^-] = &\qty[\vhat_4^{(8)} \widehat{\text{Hf}}\qty(\chi_4^{(8)})] \\[0.5em]
        &+ \qty[\sum_{\text{channels}}\frac{\vhat_2^{(4)} \vhat_2^{(4)} \vhat_2^{(4)}}{q_1^2 q_2^2}~ \widehat{\text{Hf}}\qty(\chi_2^{(4)}(q_1^+))  \widehat{\text{Hf}}\qty(\chi_2^{(4)}(q_1^-,q_2^-)) \widehat{\text{Hf}}\qty(\chi_2^{(4)}(q_2^+))] \\[0.5em]
        &+\qty[\sum_{\text{channels}}\frac{\vhat_2^{(4)} \vhat_2^{(4)} \vhat_2^{(4)}}{q_1^2 q_2^2}~ \widehat{\text{Hf}}\qty(\chi_2^{(4)}(q_1^-))  \widehat{\text{Hf}}\qty(\chi_2^{(4)}(q_1^+,q_2^+)) \widehat{\text{Hf}}\qty(\chi_2^{(4)}(q_2^-))] \\[0.5em]
        &+\qty[\sum_{\text{channels}}\frac{\vhat_2^{(4)} \vhat_2^{(4)} \vhat_2^{(4)}}{q_1^2 q_2^2}~ \widehat{\text{Hf}}\qty(\chi_2^{(4)}(q_1^+))  \widehat{\text{Hf}}\qty(\chi_2^{(4)}(q_1^-,q_2^+)) \widehat{\text{Hf}}\qty(\chi_2^{(4)}(q_2^-))]  \\[0.5em]
        &+\qty[\sum_{\text{channels}}\frac{\vhat_4^{(4)} \vhat_0^{(4)} \vhat_2^{(4)}}{q_1^2 q_2^2}~ \widehat{\text{Hf}}\qty(\chi_4^{(4)}(q_1^+))  \widehat{\text{Hf}}\qty(\chi_0^{(4)}(q_1^-,q_2^-)) \widehat{\text{Hf}}\qty(\chi_2^{(4)}(q_2^+))] \\[0.5em]
        &+\qty[\sum_{\text{channels}}\frac{\vhat_2^{(4)} \vhat_4^{(4)} \vhat_0^{(4)}}{q_1^2 q_2^2}~ \widehat{\text{Hf}}\qty(\chi_2^{(4)}(q_1^-))  \widehat{\text{Hf}}\qty(\chi_4^{(4)}(q_1^+,q_2^+)) \widehat{\text{Hf}}\qty(\chi_0^{(4)}(q_2^-))]
    \end{aligned}
\end{align}
In these expressions, each bracket corresponds to the contribution of a topology, following the ordering of the diagrams drawn in \eqref{eq:A8pt_diag80}--\eqref{eq:A8pt_diag44}.  

The contact 6-pt and 8-pt amplitudes   contribute to build these 8-pt amplitudes. Interestingly, they do not contribute to every set of polarizations. 
The 8-pt contact amplitude contributes only as a local term in  $ \mathcal{A}[8^+,0^-]$, $ \mathcal{A}[6^+,2^-]$, $ \mathcal{A}[4^+,4^-]$, whereas  the 6-pt amplitude contributes only in $\mathcal{A}[7^+,1^-]$ and $\mathcal{A}[5^+,3^-]$.

\subsection{10-pt BI Amplitudes}

Finally we close with the  10-pt amplitude of BI theory of electromagnetism.  We have established that the only contact amplitudes have $4p$ photons where  $p\geq 1$ is integer (see section \ref{se:hel_conserving_lagrangian}).   Therefore only the 4-pt and 8-pt contact amplitudes are available.   
Furthermore,  due to helicity conservation the only nonzero 10-pt amplitude is $\mathcal{A}^{\rm BI}[5^+,5^-]$. Consequently, it turns out that 
the 10-pt amplitude is only built from  4-pt contact amplitudes. 

Following our recipe, we start by drawing all possible topologies,
\begin{equation}
    \begin{aligned}
        \mathcal{A}[5^+,5^-] = \diagram[0.2]{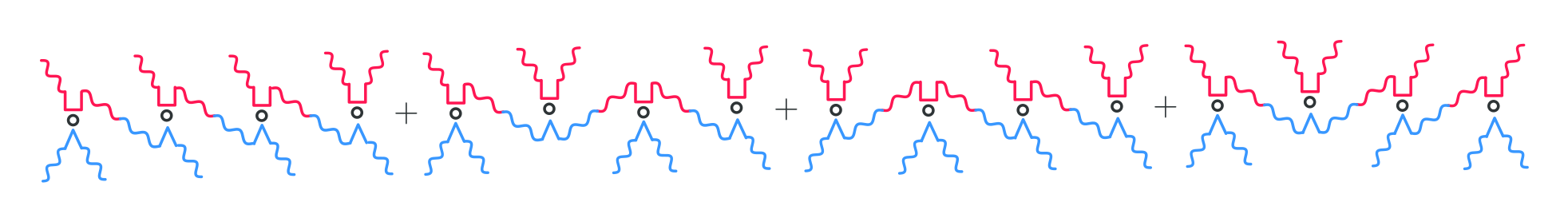} \\[-1em]
        \diagram[0.2]{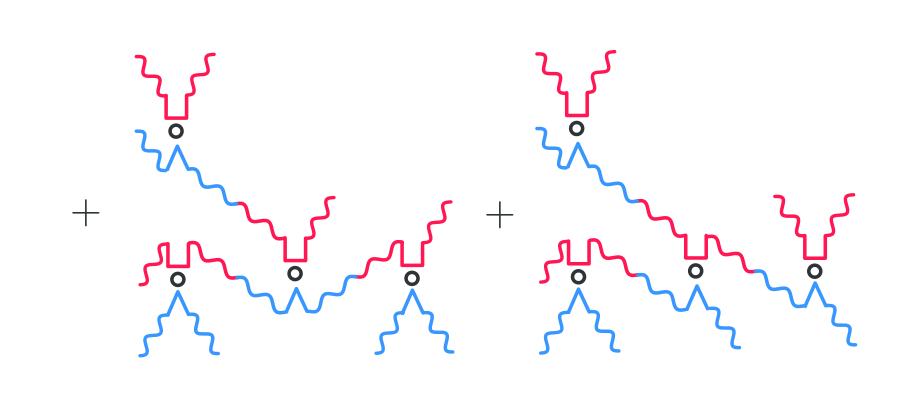}
    \end{aligned}
\end{equation}
We then write the corresponding expression in terms of Hafnians and sum over channels. The final result for $\mathcal{A}^{\rm BI}[5^+,5^-]$ is \begin{equation}
    \begin{aligned}
        \mathcal{A}^{\rm BI} & [5^+,5^-]  = \\[0.5em] &\qty[\sum_{\text{channels}}\frac{(2^4 \Lambda^2)^{-4}}{q_1^2 q_2^2 q_3^2}~ \widehat{\text{Hf}}\qty(\chi_2^{(4)}(q_1^+)) \widehat{\text{Hf}}\qty(\chi_2^{(4)}(q_1^-, q_2^+))  \widehat{\text{Hf}}\qty(\chi_2^{(4)}(q_2^-,q_3^+)) \widehat{\text{Hf}}\qty(\chi_2^{(4)}(q_3^-))] \\[0.5em]
        +&\qty[\sum_{\text{channels}}\frac{(2^4 \Lambda^2)^{-4}}{q_1^2 q_2^2 q_3^2}~ \widehat{\text{Hf}}\qty(\chi_2^{(4)}(q_1^+)) \widehat{\text{Hf}}\qty(\chi_2^{(4)}(q_1^-, q_2^-)) \widehat{\text{Hf}}\qty(\chi_2^{(4)}(q_2^+,q_3^+)) \widehat{\text{Hf}}\qty(\chi_2^{(4)}(q_3^-))] \\[0.5em]
        +&\qty[\sum_{\text{channels}}\frac{(2^4 \Lambda^2)^{-4}}{q_1^2 q_2^2 q_3^2}~ \widehat{\text{Hf}}\qty(\chi_2^{(4)}(q_1^-)) \widehat{\text{Hf}}\qty(\chi_2^{(4)}(q_1^+,q_2^+)) \widehat{\text{Hf}}\qty(\chi_2^{(4)}(q_2^-,q_3^+)) \widehat{\text{Hf}}\qty(\chi_2^{(4)}(q_3^-))]  \\[0.5em]
        +&\qty[\sum_{\text{channels}}\frac{(2^4 \Lambda^2)^{-4}}{q_1^2 q_2^2 q_3^2}~ \widehat{\text{Hf}}\qty(\chi_2^{(4)}(q_1^+)) \widehat{\text{Hf}}\qty(\chi_2^{(4)}(q_1^-,q_2^-)) \widehat{\text{Hf}}\qty(\chi_2^{(4)}(q_2^+,q_3^-)) \widehat{\text{Hf}}\qty(\chi_2^{(4)}(q_3^+))]  \\[0.5em]
        +&\qty[\sum_{\text{channels}}\frac{(2^4 \Lambda^2)^{-4}}{q_1^2 q_2^2 q_3^2}~ \widehat{\text{Hf}}\qty(\chi_2^{(4)}(q_1^+)) \widehat{\text{Hf}}\qty(\chi_2^{(4)}(q_2^-)) \widehat{\text{Hf}}\qty(\chi_2^{(4)}(q_1^-,q_2^+,q_3^+)) \widehat{\text{Hf}}\qty(\chi_2^{(4)}(q_3^-))] \\[0.5em]
        +&\qty[\sum_{\text{channels}}\frac{(2^4 \Lambda^2)^{-4}}{q_1^2 q_2^2 q_3^2}~ \widehat{\text{Hf}}\qty(\chi_2^{(4)}(q_1^+)) \widehat{\text{Hf}}\qty(\chi_2^{(4)}(q_2^-)) \widehat{\text{Hf}}\qty(\chi_2^{(4)}(q_1^-,q_2^+,q_3^-)) \widehat{\text{Hf}}\qty(\chi_2^{(4)}(q_3^+))]
    \end{aligned}
\end{equation}

%% file: Sections/Summary.tex
\section{Summary}
\label{se:Summary}

We have presented a recursive framework for computing arbitrary tree-level helicity amplitudes in the generic effective field theory of electromagnetism. The framework has two complementary components, combining both on-shell and functional methods.
First, in a construction reminiscent of the CSW method, a generic helicity amplitude is expressed as a web of on-shell building blocks, the contact amplitudes. Second, these contact amplitudes are themselves computed efficiently through a functional approach, which naturally encodes them into a contact Lagrangian.

As a preliminary step, we define the most general EFT of electromagnetism, making clear that the number of independent operators grows as $\lfloor N/2\rfloor$, with $N$ the number of photons.  Throughout the whole paper, we work solely at the level of the photon field strength, so that the entire formulation is gauge invariant. The price of this choice is that the field integrals are constrained by the Bianchi identity.

We express the EFT in the self-dual basis $(F_+, F_-)$. 
We show that the $\langle F_\pm F_\pm\rangle $ propagators are purely local, whereas $\langle F_+ F_- \rangle $ is nonlocal. Although these features can already be found in the Lorentz-tensor formalism, the simplicity of the photon propagator in the self-dual basis becomes fully manifest in the spinor formalism. As an aside, we argue that the Bianchi identity fixes  the nonlocal structure of $\langle F_+ F_- \rangle $ and is, in a sense, what allows the photon propagates. 
 We have also shown in details that the inversion of the propagator is defined with respect to the identity of the subpace of Bianchi-constrained fields.

Working at the diagrammatic level, every leg of every vertex carries either a self-dual ($+$) or anti-self dual ($-$) label, even off shell, matching  the  photon helicities when on-shell. The EFT vertices  naturally exhibit a pairing substructure. We introduce a  diagrammatic representation that  provides an intuitive understanding of the spinor brackets appearing in the amplitudes. 

We further show that, as a consequence of the pairing structure of the vertices, all contact amplitudes are expressed in terms of Hafnians,  the combinatorial object associated with perfect matchings.  
More precisely they are given by weighted Hafnians that encode the full combinatorics, thereby greatly simplifying subsequent calculations. 

Turning to factorization, we identify a key feature:  any  tree amplitude can be viewed as a web of subamplitudes involving only  $++$, $--$ lines,  connected  through $+-$ lines, as illustrated in Fig.\,\ref{fig:amplitude_factorization}. Since the $++$ and $--$ propagators are local, the subamplitudes are themselves local.  We refer to them  as contact  amplitudes.

This amplitude substructure can be leveraged to define a CSW-like construction. Indeed, using that the $+-$ lines are simple poles and the factorization  of spinor brackets, 
the non-contact piece of any tree-level amplitude factorizes, and can thus    be reconstructed by appropriately sewing together lower-point contact amplitudes. For a fixed multiplicity, the knowledge of a finite number of contact amplitudes is needed. 

We outline a  practical recipe to construct arbitrary helicity amplitudes from the contact amplitudes. 
The procedure consists of \textit{i)} determining the available topologies, \textit{ii)} writing the amplitudes in terms of Hafnians and \textit{iii)} summing over all  channels. 
 We illustrate the method by computing all 6-pt helicity amplitudes and verify that the results agree with those obtained using the traditional Feynman-rules approach,
which requires tricky combinatorial bookkeeping. In contrast, our sewing method does not require combinatorial bookkeeping at all.

 We turn to a functional formalism to develop a systematic computation method for the contact amplitudes.
We first introduce a reformulation of electromagnetism that trades the Bianchi-constrained fields for  unconstrained fields,  that separately generate the $+-$ and $\pm \pm$ propagators. The fields producing the latter do not propagate, and can  be integrated out exactly. 
The integration gives rise to an effective Lagrangian, referred to as the contact Lagrangian, whose operators are in one-to-one correspondence with the contact amplitudes. 

We show how to evaluate the contact Lagrangian using a fixed-point method and establish the stopping criterion for the iterations.
We present the contact Lagrangian up to  $N=14$ photons in the general EFT of electromagnetism. We also argue that the contact Lagrangian only receives tree-level contributions, implying that the approach remains valid to all loop orders.

The use of the contact Lagrangian is especially enlightening when applied to helicity-conserving theories such as BI electromagnetism. 
We argue that, while the vanishing of helicity-violating amplitudes requires intricate cancellations among the coefficients of the original EFT Lagrangian, all cancellations become manifest in the contact Lagrangian.  The contact Lagrangian therefore reveals the hidden simplicity of helicity-conserving theories. In particular, the contact Lagrangian  provides a simple proof that a theory is invariant under off-shell electromagnetic duality if and only if its amplitudes conserve helicity.

We further argue that it is possible to use the invariance  under $U(1)$ electromagnetic duality to bootstrap the contact Lagrangian in order to identify variables that are irreps of the $U(1)$ duality group. We apply this approach to Born-Infeld electromagnetism and, using the Lagrange inversion theorem, derive the complete contact BI Lagrangian  in closed form. We verify that our result is in agreement with \cite{Novotny:2018iph}.

As a final application, bringing together all the methods and results developed in this work, we compute all 8-pt amplitudes in the general EFT of electromagnetism, as well as the unique nonvanishing 10-pt amplitude of Born-Infeld electromagnetism.

 \section*{Acknowledgements}

SF was supported  by the São Paulo Research Foundation (FAPESP) under Grants \#2025/27083-0 and \#2021/10128-0.

%% file: Appendices/shf_review.tex
\section{Spinor Helicity Formalism: Conventions and Identities}\label{app:SHF}

This appendix complements the essential aspects of  the spinor helicity formalism  introduced in sections \ref{se:Spinor_Rep} and \ref{se:SHF}. 

\subsection{Spinor Tensors}

We work in  four-dimensional Minkowski spacetime, with mostly-minus metric defining the line element
\begin{equation}\label{eq:minkowski_interval}
    \dd{s}^2 = \eta_{\mu\nu} \dd{x^\mu} \dd{x^\nu} = \dd{x^0} - \dd{x^1} - \dd{x^2} - \dd{x^3}\,.
\end{equation}

The map from the Lorentz to the $SL(2,\mathbb{C})$ group given in  \eqref{eq:SL2_map}
 is expressed in terms of the Pauli matrices $\sigma^\mu$, $\bar \sigma^\mu$,  where  
\begin{equation}
    \sigma^1 = \mqty(0 & 1\\ 1&0 ) \,, \quad \sigma^2 = \mqty(0 &-i \\ i& 0) \,, \quad \sigma^3 = \mqty(1 &0 \\ 0& -1) \,,
\end{equation}
and $(\sigma^\mu)_{a\dot{a}} = (\mathbb{1},\sigma^i)$ and $(\bar{\sigma}^\mu)^{\dot{a}a} = (\mathbb{1},-\sigma^i)$. 
For example, a Lorentz vector is expressed under the map \eqref{eq:SL2_map} as a bispinor with  
\begin{equation}
    \begin{aligned}
         & v_{a\dot{a}} \equiv v_\mu (\sigma^\mu)_{a\dot{a}} = \mqty( v^0-v^3         & -v^1+iv^2 \\ -\qty(v^1+iv^2) & v^0+v^3 ) \,, \\[0.5em]
         & v^{\dot{a}a} \equiv v_\mu (\bar{\sigma}^\mu)^{\dot{a}a} = \mqty( v^0+v^3 & v^1-iv^2  \\ v^1+iv^2 & v^0-v^3 )\,.
    \end{aligned}
\end{equation}
The Minkowski metric maps onto the skew-symmetric spinor metrics,
\begin{equation}
    \eta_{\mu\nu} \, (\sigma_{a\dot{a}})^\mu (\sigma_{b\dot{b}})^\nu =  2\varepsilon_{ab} \varepsilon_{\dot{a}\dot{b}} \,, \quad (\sigma^\mu)_{a\dot{a}} \,(\bar{\sigma}^\nu)^{\dot{a}a} = 2\eta^{\mu\nu}
\end{equation}
with
\begin{equation}\label{eq:spinor_metrics}
    \varepsilon^{ab}=\varepsilon^{\dot{a}\dot{b}} = \mqty(0 & 1 \\ -1 & 0) = -\varepsilon_{ab} = -\varepsilon_{\dot{a}\dot{b}}
\end{equation}
satisfying  $\varepsilon^{ab} \varepsilon_{bc}= \delta^a_c$ and likewise for dotted indices.

From the above definitions, the norm of a Lorentz vector takes the form
\begin{equation}
   v^\mu v_\mu = \det({v^{\dot{a}a}}) =   \frac{1}{2}v^{\dot{a}a} v_{a\dot{a}} \,.
\end{equation}
We also get the useful relation
\be
     q^{a}_{\dot{b}} \, k_{a\dot{a}}  = \varepsilon_{\dot{b}\dot{a}} \,q^\mu k_\mu  \,. 
     \label{eq:scalar_product_identity}
\ee
Along similar lines, a full contraction of rank-$n$ Lorentz tensors becomes 
\begin{equation}\label{eq:tensor_contraction_map}
    T^{\mu\nu\,\cdots}\ W_{\mu\nu\,\cdots} = \frac{1}{2n} T^{(\dot{a}a)(\dot{b}b)\,\cdots} \ W_{(a\dot{a})(b\dot{b})\,\cdots}\,
\end{equation}
in terms of the corresponding spinor tensors.

\subsection{Massless Particles }

As presented in section \ref{se:SHF}, if $v$ is a null vector, i.e. $v^\mu v_\mu = 0$, its determinant vanishes, and it is possible to decompose the  corresponding bispinor as 
\begin{equation}
    v^{\dot{a}a} = v^{\dot{a}}v^a\,.    
\end{equation}
This can be recast in the square/angle braket notation,
\begin{equation}
    \begin{aligned}
        &v^{\dot{a}} = |v]\,,\quad &&v_{\dot{a}} = [v| \,,\\[0.5em]
        &v^{a} = \bra{v}\,,\quad &&v_a = \ket{v} \,.
    \end{aligned}
\end{equation}
The metric of \eqref{eq:spinor_metrics} naturally defines the $SL(2,\mathbb{C})$-invariant inner products 
\begin{equation}
    \begin{aligned}
        [v\,w] &\equiv \varepsilon_{\dot{b}\dot{a}}\,v^{\dot{a}} w^{\dot{b}} = v_{\dot{a}} w^{\dot{a}}\, \\[0.5em]
        \ev{v\,w} &\equiv \varepsilon_{ab}\,v^{a} w^{b} = v^{a} w_{a}\,.
    \end{aligned}
\end{equation}
Some care should be taken about the conventions  for raising and lowering spinor indices. We have 
\begin{equation}
    \ev{v \, w}= \varepsilon_{ab}\,v^{a} w^{b} = v^{a} w_{a} = -\varepsilon_{ba}\,v^{a} w^{b} = -v_{a} w^{a} = -\ev{w \, v}\,,
\end{equation}
and similarly for square brakets.

The inner products  satisfy the following general properties:
\begin{enumerate}
    \item $[v\, w] = -[w\, v]\,, \quad \ev{v\, w} = -\ev{w\, v}$
    \item If $v \in \mathbb{R}\,, \ [v| =  \ket{v}^*\,, \ \bra{v} = |v]^*$
    \label{rel:conjugation}
    \item  $[w\,v]\ev{v\,w} = 2v^\mu w_\mu  $
    \item If $k^\mu k_\mu=0\,, \ \bra{v}^a \,k_{a\dot{a}}\,{}^{\dot{a}}|w] = \ev{v\,k} [k\,w]$
    \item $\ket{u}\langle vw \rangle + \ket{v}\langle wu \rangle +\ket{w}\langle uv \rangle = 0$ \quad \quad(Schouten Relation)
    \item $\bra{v}\sigma^\mu|w]\bra{u}\sigma_\mu|r] = -2\ev{v\,u}[w\,r]$  \quad \quad (Fierz Identity) 
\end{enumerate}

Additionally, at the level of scattering amplitudes for massless particles, we also have 
\begin{enumerate}[resume]
    \item $s_{ij} = [ji]\ev{ij} = (p_i + p_j)^2$ \quad\quad(Mandelstam Variables)\label{rel:mandelstam}
    \item $\sum_{j=1}^N \ket{j}[j| = 0$  \quad\quad (Momentum Conservation)\label{rel:momentum_conservation}
\end{enumerate}

\subsection{Weyl Spinors and Phase Conventions}\label{app:phase_conventions}

It is sometimes necessary to write the  momentum bispinors in a more explicit notation in order to compare results from different formalisms. 

 For a massless particle with momentum
$p^\mu = E(1,\sin{\phi}\cos{\theta},\sin{\phi}\sin{\theta},\cos{\phi})$, we  write
\begin{equation}
    \begin{aligned}
         & p^{\dot{a}a} = 2E\mqty( \cos^2{\tfrac{\phi}{2}} & \tfrac{1}{2}\sin{\phi}\,e^{-i\theta}  \\ \tfrac{1}{2}\sin{\phi}\,e^{i\theta} & \sin^2{\tfrac{\phi}{2}} ) \\[0.5em]
         & p_{a\dot{a}} = 2E\mqty( \sin^2{\tfrac{\phi}{2}} & -\tfrac{1}{2}\sin{\phi}\,e^{-i\theta} \\ -\tfrac{1}{2}\sin{\phi}\,e^{i\theta} & \cos^2{\tfrac{\phi}{2}} )
    \end{aligned} 
\end{equation}
from which we obtain the spinors
\begin{equation}\label{eq:spinors}
    \begin{gathered}
        |p]_a = \sqrt{2E}\mqty(\sin{\tfrac{\phi}{2}} \\[0.5em] -\cos{\tfrac{\phi}{2}}e^{i\theta})\,, \quad \ket{p}^{\dot{a}} = \sqrt{2E}\mqty(\cos{\tfrac{\phi}{2}} \\[0.5em] \sin{\tfrac{\phi}{2}}e^{i\theta})\,, \\[0.5em]
        [p|^a = \sqrt{2E}(\cos{\tfrac{\phi}{2}},~\sin{\tfrac{\phi}{2}}e^{-i\theta})\,, \quad \bra{p}_{\dot{a}} = \sqrt{2E}(\sin{\tfrac{\phi}{2}},~ -\cos{\tfrac{\phi}{2}}e^{-i\theta})\,.
    \end{gathered}
\end{equation}

Finally, using relations \ref{rel:conjugation} and \ref{rel:mandelstam}, we express the inner products as
\begin{equation}
    \langle ij \rangle = -\sqrt{|s_{ij}|} e^{i\varphi_{ij}}\,, \quad [ij] = \sqrt{|s_{ij}|} e^{-i\varphi_{ij}}\,.
\end{equation}
The $\varphi_{ij}$ phases can be fixed  upon specification of the kinematics. This will be relevant when comparing  our results
with those from  traditional computations.

%% file: Appendices/comparing_amps.tex
\section{Symmetries and Polarization Bases}
\label{app:symmetries}

The scattering amplitudes from our general  EFT of electromagnetism transform under some fundamental symmetries: {spatial parity} ($\mathcal{P}$), {time-reversal} ($\mathcal{T}$) and {boson-exchange} ($\mathcal{E}$). 
In this appendix we provide a summary of how these symmetries act on helicity amplitudes. We also show how these connect to amplitudes of linearly polarized amplitudes.

\subsection{Symmetries and Representative Amplitudes}

For a given $N$-photons amplitude, there are $2^N$ possible helicity assignments, which are not all independent.
Parity mirrors space coordinates, whereas time-reversal reverses time direction; both flip helicity assignments. Boson-exchange works at the level of relabelling of particles.  This should not be  confused with {crossing relations}, which relates different scattering channels by changing incoming/outgoing assignments of external legs.

For $\gamma\gamma \to \gamma\gamma$ scattering, we can write the following relations between the $2^4=16$ amplitudes,
\begin{center}
    \begin{tikzpicture}

        \node (eq1) at (-2.5,0) {$\mathcal{A}[--\to--] = \mathcal{A}[++\to++]\,,$};
        \node (eq2) at (-2.5,-0.6) {$\mathcal{A}[+-\to+-] = \mathcal{A}[-+\to-+]\,,$};
        \node (eq3) at (-2.5,-1.2) {$\mathcal{A}[+-\to-+] = \mathcal{A}[-+\to+-]\,,$};
        \node (eq4) at (-2.5,-1.8) {$\mathcal{A}[++\to--] = \mathcal{A}[--\to++]\,,$};
        \draw[->, thick] (-3.8,-2.2) -- (-1.,-2.2) node[midway, below] {$\mathcal{P}$ or $\mathcal{T}$};

        \node[anchor=west] (eq5) at (0.5,0)     {  $ \hspace{1.1em} \mathcal{A}[++\to+-] = \mathcal{A}[++\to-+]$};
        \node[anchor=west] (eq6) at (0.5,-0.6)  {$= \mathcal{A}[--\to-+] = \mathcal{A}[--\to+-]$};
        \node[anchor=west] (eq7) at (0.5,-1.2)  {$= \mathcal{A}[+-\to--] = \mathcal{A}[-+\to--]$};
        \node[anchor=west] (eq8) at (0.5,-1.8)  {$= \mathcal{A}[-+\to++] = \mathcal{A}[+-\to++]\,.$};

        \draw[->, thick] (6.5,0.) -- (6.5,-0.6) node[midway,right] {$\mathcal{P}$};
        \draw[->, thick] (7.4,0.) -- (7.4,-1.2) node[midway,right] {$\mathcal{T}$};
        \draw[->, thick] (8.4,0.) -- (8.4,-1.8) node[midway,right] {$\mathcal{P}+\mathcal{T}$};
        \draw[->, thick] (2.,-2.2) -- (5.2,-2.2) node[midway, below] {$\mathcal{E}$};

    \end{tikzpicture}\,
\end{center}
Crossing further relates the remaining amplitudes through
\begin{equation}\label{eq:crossing_relations}
    \begin{cases} \mathcal{A}[+-\to-+](s,t,u) = \mathcal{A}[++\to--](u,t,s) \\[0.5em] \mathcal{A}[+-\to+-](s,t,u) = \mathcal{A}[++\to--](t,s,u)
    \end{cases}\ 
\end{equation}
This leaves only three independent amplitudes. 

A similar drop  happens for 6-photons scattering, where only four from the $2^6 = 64$ possibilities are independent. For 8-photons, the drop
is even more drastic, as only five from the $2^8=256$ amplitudes need to be considered.
Clearly, there are $\frac{N}{2}+1$ independent representatives, which  can be simply labelled by the number of $+$ and $-$ photons. 
These are the representative amplitudes ${\cal A}[K^+, (N-K)^-]$ used throughout the main text, see section \ref{se:SHF}. 

Finally, it is worth mentioning  that in the all-incoming (or all-outgoing) convention,  the crossing relations are accounted for by the larger boson-exchange symmetry which permutes between \textit{all} states. Yet, in the recipe of section \ref{se:practical_steps},  they stand out as special permutations that require  
a specific step.

\subsection{Mapping Linear and Helicity Bases}

The amplitudes evaluated in App.\,\ref{app:traditional_amps} 
are written in terms of generic linear polarizations. Linear polarizations can be decomposed into helicity eigenstates as
\begin{equation}\label{eq:linear_pol}
    \epsilon_{i}^{(\pm)} = \frac{1}{\sqrt{2}}(\epsilon^{(x)}_i \pm i\epsilon^{(y)}_i)\,.
\end{equation}
This identity defines the mapping  between the helicity and the linear bases. 
  For example, we find
\begin{equation}\label{eq:F4_lin_hel}
    \begin{aligned}
        \mathcal{A}_{xx\to xx} & =\frac{1}{2}\qty(\mathcal{A}[2^+,2^-]+4\mathcal{A}^{(34)}[3^+,1^-]+\mathcal{A}[4^+,0^-]+\mathcal{A}^{(24)}[2^+,2^-]+\mathcal{A}^{(23)}[2^+,2^-]) \,, \\
        \mathcal{A}_{xy\to xy} & =\frac{1}{2}\qty(-\mathcal{A}[2^+,2^-]+\mathcal{A}[4^+,0^-]-\mathcal{A}^{(24)}[2^+,2^-]+\mathcal{A}^{(23)}[2^+,2^-]) \,,                    
    \end{aligned}
\end{equation}
for some of the 4-photons amplitudes. For 6 photons we have for instance
\begin{equation}
    \begin{aligned} \label{eq:F6_lin_hel}
         & \begin{aligned} \mathcal{A}_{xxx\to xxx} = \tfrac{1}{4}( &\mathcal{A}[3^+,3^-] + 6\mathcal{A}^{(46)}[4^+,2^-] + 6\mathcal{A}^{(46)}[5^+,1^-] + \mathcal{A}[6^+,0^-] + 3\mathcal{A}^{(36)}[3^+,3^-] \\[0.5em]&+ 6\mathcal{A}^{(35)}[3^+,3^-] + 6\mathcal{A}^{(35)(46)}[4^+,2^-] + 3\mathcal{A}^{(35)}[4^+,2^-])\,,\end{aligned} \\[0.5em]
         & \begin{aligned} \mathcal{A}_{xyx\to xyx} = \tfrac{1}{4}( &-\mathcal{A}[3^+,3^-] -2\mathcal{A}^{(46)}[4^+,2^-] + 2\mathcal{A}^{(46)}[5^+,1^-] + \mathcal{A}[6^+,0^-] - 3\mathcal{A}^{(36)}[3^+,3^-] \\[0.5em]&+ 2\mathcal{A}^{(35)}[3^+,3^-] - 2\mathcal{A}^{(35)(46)}[4^+,2^-] + 3\mathcal{A}^{(35)}[4^+,2^-])\,\end{aligned}\,
    \end{aligned}
\end{equation}
Here the $(ij)$ denotes a  permutation between legs $i$ and $j$, as usual. 

We emphasize  that the left- and right-hand sides of \eqref{eq:F4_lin_hel} and \eqref{eq:F6_lin_hel} are written in different conventions. 
On the l.h.s, we employ a notation with half of the photons taken as incoming and half as outgoing, which is more transparent for linear polarizations. 
On the r.h.s, the all-incoming convention adopted throughout the main text is used. The two descriptions are, of course, equivalent and can be related by crossing prescriptions.

%% file: Appendices/traditional_amps.tex
\section{Consistency Check: Forward 6pt Amplitudes from Lorentzian Formalism}

\label{app:traditional_amps}

In order to perform sanity checks, we independently compute  the  6-photon (and also 4-pt)  scattering amplitudes in the standard ``Lorentzian'' formalism. In this formalism, it is convenient to choose that half of the states is ingoing  and the other half is outgoing.   
We take the forward limit,  for which the initial ket states remain unchanged after scattering,  $\ket{i} \to \ket{i}$. 
We define 
\begin{equation}
    s_{ij} \equiv 2 (p_i \cdot p_j)\,,\quad \pi_{ij} \equiv \epsilon_i \cdot \epsilon_j\,, \quad a_{ij} \equiv p_i \cdot \epsilon_j = 0\,: ~\forall\,i,j\,.
\end{equation}

In this section we used the  basis of cyclic operators,
\begin{equation}\label{eq:EFTlagrangianCyclic}
    \begin{aligned}
        \mathcal{L}_{F} = &-\tfrac{1}{4}\Tr(F^2)
        + \beta_0^{(4)}\Tr(F^2)^2+ \beta_1^{(4)}\Tr(F^4) 
        + \beta_0^{(6)}\Tr(F^2)^3 + \beta_1^{(6)}\Tr(F^2)\Tr(F^4) \\[0.5em]
        &+ \beta_0^{(8)}\Tr(F^2)^4 + \beta_1^{(8)}\Tr(F^2)^2\Tr(F^4) + \beta_2^{(8)}\Tr(F^4)^3 
        + \mathcal{O}(F^{10}) \,.
    \end{aligned}
\end{equation}
Using  identity \eqref{eq:cyclic_to_dual}, the $\beta_i$ coefficients are related to the $\kappa$ coefficients of \eqref{eq:LF_standardbasis} as follows, 
\begin{equation}
\begin{gathered}
    \beta_0^{(4)} = \vv_0^{(4)} - 2 \vv_1^{(4)}\,, \quad \beta_1^{(4)} = 4 \vv_1^{(4)} \\[0.5em]
    \beta_0^{(6)} = \vv_0^{(6)} - 2 \vv_1^{(6)}\,, \quad \beta_1^{(6)} = 4 \vv_1^{(6)} \\[0.5em]
    \beta_0^{(8)} = \vv_0^{(8)} - 2 \vv_1^{(8)} + 4 \vv_2^{(8)}\,, \quad \beta_1^{(8)} = 4 \vv_1^{(8)} - 16 \vv_2^{(8)}\,, \quad \beta_2^{(8)} = 16 \vv_2^{(8)}\,.
\end{gathered}
\end{equation}

\subsection{Forward 
$\gamma\gamma\to\gamma\gamma$  Amplitude}

The states satisfy  $p_1=p_3$, $p_2=p_4$, $\epsilon_1=\epsilon_3$, $\epsilon_2=\epsilon_4$, and
one has $a_{ij}=0$. We define the angle between the two polarizations as $\epsilon_1 \cdot \epsilon_2 = \cos \theta$. 
We obtain 
\begin{align} \label{eq:F4-Amp}
    \mathcal{A}^{(4)}_{\ket{i} \rightarrow \ket{i}}(\theta)= 16\,\beta^{(4)}_0i~s_{12}^2~\cos^2{\theta} + 4\,\beta^{(4)}_1i~s_{12}^2~(1+\cos^2{\theta})\,.
\end{align}

\subsection{ Forward 
$\gamma\gamma\gamma\to\gamma\gamma\gamma$  Amplitude}

In that case we have $p_1=p_4$, $p_2=p_5$, $p_3=p_6$, $\epsilon_1=\epsilon_4$, $\epsilon_2=\epsilon_5$, $\epsilon_3=\epsilon_6$. 
We take  the center of momentum frame of two initial photons to be on  the $z$-axis, and append the third photon with an azimutal angle $\phi$ in the $xz$-plane. The momenta and polarizations read
\begin{equation}\label{eq:3-photon-kinematics}
    \begin{gathered}
        \epsilon_1 = (0,\cos{\varphi_1},\sin{\varphi_1},0), \quad p_1 = E_1(1,0,0,1),\\[0.5em]
        \epsilon_2 =  (0,\cos{\varphi_2},\sin{\varphi_2},0), \quad p_2 = E_2(1,0,0,-1),\\[0.5em]
        \epsilon_3 = (0,\cos{\varphi_3}\cos{\phi},\sin{\varphi_3},-\cos{\varphi_3}\sin{\phi}), \quad p_3 = E_3(1,\sin{\phi},0,\cos{\phi})\,.
    \end{gathered}
\end{equation}
We  write the products of non-vanishing $a_{ij}$ as functions of $s_{ij}$'s and the relevant angles,
\begin{equation}
    \begin{gathered}
        a_{13} \ a_{31} = -\frac{s_{13}\cos{\varphi_3}\cos{\varphi_1}\sin^2{\phi}}{2(1-\cos{\phi})} \ , \qquad a_{23} \ a_{31} = \frac{s_{23}\cos{\varphi_3}\cos{\varphi_1}\sin^2{\phi}}{2(1+\cos{\phi})}\,, \\[0.5em]
        a_{13} \ a_{32} = -\frac{s_{13}\cos{\varphi_3}\cos{\varphi_2}\sin^2{\phi}}{2(1-\cos{\phi})} \ , \qquad  a_{23} \ a_{32} = \frac{s_{23}\cos{\varphi_3}\cos{\varphi_2}\sin^2{\phi}}{2(1+\cos{\phi})}\,, \\[0.5em]
        a_{31} \ a_{32} = \frac{s_{13}\ s_{23}}{s_{12}}\cos{\varphi_1}\cos{\varphi_2}\,.
    \end{gathered}
\end{equation}
The final amplitude is found to be
\begin{equation} \label{eq:F6-Amp}
    \begin{aligned}
        \mathcal{A}^{(6)}_{\ket{i} \rightarrow \ket{i}}(\varphi_1,\varphi_2,\varphi_3) & =-8i\Big[(12 \beta^{(6)}_0+3\beta^{(6)}_1)\cos{(2 (\varphi_1-\varphi_2+\varphi_3}))+\\[0.5em]
        &\hspace{3.5em}+(12 \beta^{(6)}_0+7\beta^{(6)}_1)\sum_i^3 \cos (2\varphi_i)\Big]s_{12} s_{13} s_{23}
    \end{aligned}
\end{equation}
The expression is surprisingly simple. Remarkably, it is 
 independent of the $\phi$ angle.\,\footnote{However, the Mandelstam variables $s_{ij}$ generally depend on $\phi$.} In fact, the only relevant parameters here are the angles related to the projections of polarizations in the $xy$-plane.

\subsection{Consistency Checks}

Using \eqref{eq:F4_lin_hel} and \eqref{eq:F6_lin_hel}, we can compare the amplitudes computed via  Lorentzian formalism  to the  amplitudes computed via spinor helicity formalism in section \ref{se:hafnians}.  
To this end, we need to express the helicity amplitudes, given  in terms of square/angle spinor brackets, as functions
of the Mandelstam variables, $s_{ij}$. 

Taking into account the phase discussed in App.\,\ref{app:phase_conventions}, alongside the kinematical conventions used for results \eqref{eq:F4-Amp}, \eqref{eq:F6-Amp}, we find
\begin{equation}
    \gamma\gamma\to\gamma\gamma \ :\quad \begin{gathered}\  [12] = \sqrt{s}\,, \quad [13] = \sqrt{t}\,, \quad [14] = \sqrt{u}\,, \\[0.5em]
                                                        \varphi_{12}=\varphi_{13}=\varphi_{14}=0\,,
                                    \end{gathered}
\end{equation}
and
\begin{equation}
    \gamma\gamma\gamma\to\gamma\gamma\gamma \ :\quad \begin{gathered}
         [12] =\sqrt{s_{12}}\,, \quad [13] = \sqrt{s_{13}}\,, \quad[23] =-\sqrt{s_{23}}\,, \\[0.5em]
         \varphi_{12} = 0\,, \quad \varphi_{13} = 0\,, \quad \varphi_{23} = \pi\,.
     \end{gathered}
\end{equation}

Using the $\kappa^{(N)}_i$  coefficients from the original EFT basis, the helicity amplitudes found in section \ref{se:hafnians}, once taken in the forward limit,  are
\begin{align}
\label{eq:forward_helicity_amps_40}
     &\mathcal{A}_{\ket{i} \rightarrow \ket{i}}[4^+,0^-](s) =2^4i\,(\vv^{(4)}_0-\vv^{(4)}_1)s^2\,, \\[0.5em]
     \label{eq:forward_helicity_amps_22}
     &\mathcal{A}_{\ket{i} \rightarrow \ket{i}}[2^+,2^-]= \mathcal{A}^{(24)}_{\ket{i} \rightarrow \ket{i}}[2^+,2^-]  = 2^3i\,(\vv^{(4)}_0+\vv^{(4)}_1)s^2 \,, \\[0.5em] &\mathcal{A}^{(23)}_{\ket{i} \rightarrow \ket{i}}[2^+,2^-]=0\,,\\[0.5em]
     \label{eq:forward_helicity_amps_22b}
     &\mathcal{A}_{\ket{i} \rightarrow \ket{i}}[6^+,0^-] =-3\cdot2^7i\,(\vv^{(6)}_0-\vv^{(6)}_1)s_{12}s_{13}s_{23}\,, \\[0.5em]
     \label{eq:forward_helicity_amps_60}
     &\mathcal{A}_{\ket{i} \rightarrow \ket{i}}[4^+,2^-]=\mathcal{A}^{(46)}_{\ket{i} \rightarrow \ket{i}}[4^+,2^-]=\mathcal{A}^{(35)(46)}_{\ket{i} \rightarrow \ket{i}}[4^+,2^-] = -2^5i\,(3\vv^{(6)}_0+\vv^{(6)}_1)s_{12}s_{13}s_{23} \,, \\[0.5em] \label{eq:forward_helicity_amps_42} &\mathcal{A}^{(35)}_{\ket{i} \rightarrow \ket{i}}[4^+,2^-] = 0\,.
\end{align}
On the other hand, the amplitudes with linear polarizations \eqref{eq:F4-Amp} and \eqref{eq:F6-Amp} yield
\begin{gather}
\label{eq:forward_linear_amps_4}
    \mathcal{A}_{xx\to xx} = 2^4 i\,\vv^{(4)}_0\, s^2\,,\quad  \mathcal{A}_{xy\to xy} = 2^4 i\,\vv^{(4)}_1 \,s^2\,, \\[0.5em]
    \mathcal{A}_{xxx\to xxx} = -3\cdot2^7i\, \vv^{(6)}_0 \, s_{12}s_{13}s_{23} \,,\quad  \mathcal{A}_{xyx\to xyx} = -2^7i\, \vv^{(6)}_1 \, s_{12}s_{13}s_{23}\,,\label{eq:forward_linear_amps_6}
\end{gather}
respectively. 
 We now substitute the helicity amplitudes \eqref{eq:forward_helicity_amps_40}-\eqref{eq:forward_helicity_amps_22b} and \eqref{eq:forward_helicity_amps_60}-\eqref{eq:forward_helicity_amps_42}
into \eqref{eq:F4_lin_hel} and \eqref{eq:F6_lin_hel}. The outcome matches exactly the linearly polarized amplitudes \eqref{eq:forward_linear_amps_4} and 
\eqref{eq:forward_linear_amps_6}, hence providing  an overall consistency check of our computations.

%% file: Appendices/interactions.tex
\section{8-point Hafnians}
\label{app:hafnians}
In this appendix we expand the  expressions of the Hafnians appearing in the $8$-point amplitudes computed in \eqref{eq:A8pt_irr}. 

\subparagraph{$N=8$, $K=4$ Hafnian} 
\begin{dmath*}
\text{Hf}(\chi^{(8)}_{4}) = \, \left(\left[1 2\right]^2\left[3 4\right]^2 + \left[1 3\right]^2\left[2 4\right]^2 + \left[1 4\right]^2\left[2 3\right]^2\right) \left(\langle5 6\rangle^2\langle7 8\rangle^2 + \langle5 7\rangle^2\langle6 8\rangle^2 + \langle5 8\rangle^2\langle6 7\rangle^2 \right)\,
\end{dmath*}

\subparagraph{$N=8$, $K=6$ Hafnian}
\begin{dmath*}
\text{Hf}(\chi^{(8)}_{6}) = \, \left(\left[1 2\right]^2\left[3 4\right]^2\left[5 6\right]^2 + \left[1 2\right]^2\left[3 5\right]^2\left[4 6\right]^2 + \left[1 2\right]^2\left[3 6\right]^2\left[4 5\right]^2 + \left[1 3\right]^2\left[2 4\right]^2\left[5 6\right]^2 + \left[1 3\right]^2\left[2 5\right]^2\left[4 6\right]^2 + \left[1 3\right]^2\left[2 6\right]^2\left[4 5\right]^2 + \left[1 4\right]^2\left[2 3\right]^2\left[5 6\right]^2 + \left[1 4\right]^2\left[2 5\right]^2\left[3 6\right]^2 + \left[1 4\right]^2\left[2 6\right]^2\left[3 5\right]^2 + \left[1 5\right]^2\left[2 3\right]^2\left[4 6\right]^2 + \left[1 5\right]^2\left[2 4\right]^2\left[3 6\right]^2 + \left[1 5\right]^2\left[2 6\right]^2\left[3 4\right]^2 + \left[1 6\right]^2\left[2 3\right]^2\left[4 5\right]^2 + \left[1 6\right]^2\left[2 4\right]^2\left[3 5\right]^2 + \left[1 6\right]^2\left[2 5\right]^2\left[3 4\right]^2\right)\langle7 8\rangle^2\,
\end{dmath*}

\subparagraph{$N=8$, $K=8$ Hafnian}
\begin{dmath*}
\text{Hf}(\chi^{(8)}_{8}) = \, \left[1 2\right]^2\left[3 4\right]^2\left[5 6\right]^2\left[7 8\right]^2 + \left[1 2\right]^2\left[3 4\right]^2\left[5 7\right]^2\left[6 8\right]^2 + \left[1 2\right]^2\left[3 4\right]^2\left[5 8\right]^2\left[6 7\right]^2 + \left[1 2\right]^2\left[3 5\right]^2\left[4 6\right]^2\left[7 8\right]^2 + \left[1 2\right]^2\left[3 5\right]^2\left[4 7\right]^2\left[6 8\right]^2 + \left[1 2\right]^2\left[3 5\right]^2\left[4 8\right]^2\left[6 7\right]^2 + \left[1 2\right]^2\left[3 6\right]^2\left[4 5\right]^2\left[7 8\right]^2 + \left[1 2\right]^2\left[3 6\right]^2\left[4 7\right]^2\left[5 8\right]^2 + \left[1 2\right]^2\left[3 6\right]^2\left[4 8\right]^2\left[5 7\right]^2 + \left[1 2\right]^2\left[3 7\right]^2\left[4 5\right]^2\left[6 8\right]^2 + \left[1 2\right]^2\left[3 7\right]^2\left[4 6\right]^2\left[5 8\right]^2 + \left[1 2\right]^2\left[3 7\right]^2\left[4 8\right]^2\left[5 6\right]^2 + \left[1 2\right]^2\left[3 8\right]^2\left[4 5\right]^2\left[6 7\right]^2 + \left[1 2\right]^2\left[3 8\right]^2\left[4 6\right]^2\left[5 7\right]^2 + \left[1 2\right]^2\left[3 8\right]^2\left[4 7\right]^2\left[5 6\right]^2 + \left[1 3\right]^2\left[2 4\right]^2\left[5 6\right]^2\left[7 8\right]^2 + \left[1 3\right]^2\left[2 4\right]^2\left[5 7\right]^2\left[6 8\right]^2 + \left[1 3\right]^2\left[2 4\right]^2\left[5 8\right]^2\left[6 7\right]^2 + \left[1 3\right]^2\left[2 5\right]^2\left[4 6\right]^2\left[7 8\right]^2 + \left[1 3\right]^2\left[2 5\right]^2\left[4 7\right]^2\left[6 8\right]^2 + \left[1 3\right]^2\left[2 5\right]^2\left[4 8\right]^2\left[6 7\right]^2 + \left[1 3\right]^2\left[2 6\right]^2\left[4 5\right]^2\left[7 8\right]^2 + \left[1 3\right]^2\left[2 6\right]^2\left[4 7\right]^2\left[5 8\right]^2 + \left[1 3\right]^2\left[2 6\right]^2\left[4 8\right]^2\left[5 7\right]^2 + \left[1 3\right]^2\left[2 7\right]^2\left[4 5\right]^2\left[6 8\right]^2 + \left[1 3\right]^2\left[2 7\right]^2\left[4 6\right]^2\left[5 8\right]^2 + \left[1 3\right]^2\left[2 7\right]^2\left[4 8\right]^2\left[5 6\right]^2 + \left[1 3\right]^2\left[2 8\right]^2\left[4 5\right]^2\left[6 7\right]^2 + \left[1 3\right]^2\left[2 8\right]^2\left[4 6\right]^2\left[5 7\right]^2 + \left[1 3\right]^2\left[2 8\right]^2\left[4 7\right]^2\left[5 6\right]^2 + \left[1 4\right]^2\left[2 3\right]^2\left[5 6\right]^2\left[7 8\right]^2 + \left[1 4\right]^2\left[2 3\right]^2\left[5 7\right]^2\left[6 8\right]^2 + \left[1 4\right]^2\left[2 3\right]^2\left[5 8\right]^2\left[6 7\right]^2 + \left[1 4\right]^2\left[2 5\right]^2\left[3 6\right]^2\left[7 8\right]^2 + \left[1 4\right]^2\left[2 5\right]^2\left[3 7\right]^2\left[6 8\right]^2 + \left[1 4\right]^2\left[2 5\right]^2\left[3 8\right]^2\left[6 7\right]^2 + \left[1 4\right]^2\left[2 6\right]^2\left[3 5\right]^2\left[7 8\right]^2 + \left[1 4\right]^2\left[2 6\right]^2\left[3 7\right]^2\left[5 8\right]^2 + \left[1 4\right]^2\left[2 6\right]^2\left[3 8\right]^2\left[5 7\right]^2 + \left[1 4\right]^2\left[2 7\right]^2\left[3 5\right]^2\left[6 8\right]^2 + \left[1 4\right]^2\left[2 7\right]^2\left[3 6\right]^2\left[5 8\right]^2 + \left[1 4\right]^2\left[2 7\right]^2\left[3 8\right]^2\left[5 6\right]^2 + \left[1 4\right]^2\left[2 8\right]^2\left[3 5\right]^2\left[6 7\right]^2 + \left[1 4\right]^2\left[2 8\right]^2\left[3 6\right]^2\left[5 7\right]^2 + \left[1 4\right]^2\left[2 8\right]^2\left[3 7\right]^2\left[5 6\right]^2 + \left[1 5\right]^2\left[2 3\right]^2\left[4 6\right]^2\left[7 8\right]^2 + \left[1 5\right]^2\left[2 3\right]^2\left[4 7\right]^2\left[6 8\right]^2 + \left[1 5\right]^2\left[2 3\right]^2\left[4 8\right]^2\left[6 7\right]^2 + \left[1 5\right]^2\left[2 4\right]^2\left[3 6\right]^2\left[7 8\right]^2 + \left[1 5\right]^2\left[2 4\right]^2\left[3 7\right]^2\left[6 8\right]^2 + \left[1 5\right]^2\left[2 4\right]^2\left[3 8\right]^2\left[6 7\right]^2 + \left[1 5\right]^2\left[2 6\right]^2\left[3 4\right]^2\left[7 8\right]^2 + \left[1 5\right]^2\left[2 6\right]^2\left[3 7\right]^2\left[4 8\right]^2 + \left[1 5\right]^2\left[2 6\right]^2\left[3 8\right]^2\left[4 7\right]^2 + \left[1 5\right]^2\left[2 7\right]^2\left[3 4\right]^2\left[6 8\right]^2 + \left[1 5\right]^2\left[2 7\right]^2\left[3 6\right]^2\left[4 8\right]^2 + \left[1 5\right]^2\left[2 7\right]^2\left[3 8\right]^2\left[4 6\right]^2 + \left[1 5\right]^2\left[2 8\right]^2\left[3 4\right]^2\left[6 7\right]^2 + \left[1 5\right]^2\left[2 8\right]^2\left[3 6\right]^2\left[4 7\right]^2 + \left[1 5\right]^2\left[2 8\right]^2\left[3 7\right]^2\left[4 6\right]^2 + \left[1 6\right]^2\left[2 3\right]^2\left[4 5\right]^2\left[7 8\right]^2 + \left[1 6\right]^2\left[2 3\right]^2\left[4 7\right]^2\left[5 8\right]^2 + \left[1 6\right]^2\left[2 3\right]^2\left[4 8\right]^2\left[5 7\right]^2 + \left[1 6\right]^2\left[2 4\right]^2\left[3 5\right]^2\left[7 8\right]^2 + \left[1 6\right]^2\left[2 4\right]^2\left[3 7\right]^2\left[5 8\right]^2 + \left[1 6\right]^2\left[2 4\right]^2\left[3 8\right]^2\left[5 7\right]^2 + \left[1 6\right]^2\left[2 5\right]^2\left[3 4\right]^2\left[7 8\right]^2 + \left[1 6\right]^2\left[2 5\right]^2\left[3 7\right]^2\left[4 8\right]^2 + \left[1 6\right]^2\left[2 5\right]^2\left[3 8\right]^2\left[4 7\right]^2 + \left[1 6\right]^2\left[2 7\right]^2\left[3 4\right]^2\left[5 8\right]^2 + \left[1 6\right]^2\left[2 7\right]^2\left[3 5\right]^2\left[4 8\right]^2 + \left[1 6\right]^2\left[2 7\right]^2\left[3 8\right]^2\left[4 5\right]^2 + \left[1 6\right]^2\left[2 8\right]^2\left[3 4\right]^2\left[5 7\right]^2 + \left[1 6\right]^2\left[2 8\right]^2\left[3 5\right]^2\left[4 7\right]^2 + \left[1 6\right]^2\left[2 8\right]^2\left[3 7\right]^2\left[4 5\right]^2 + \left[1 7\right]^2\left[2 3\right]^2\left[4 5\right]^2\left[6 8\right]^2 + \left[1 7\right]^2\left[2 3\right]^2\left[4 6\right]^2\left[5 8\right]^2 + \left[1 7\right]^2\left[2 3\right]^2\left[4 8\right]^2\left[5 6\right]^2 + \left[1 7\right]^2\left[2 4\right]^2\left[3 5\right]^2\left[6 8\right]^2 + \left[1 7\right]^2\left[2 4\right]^2\left[3 6\right]^2\left[5 8\right]^2 + \left[1 7\right]^2\left[2 4\right]^2\left[3 8\right]^2\left[5 6\right]^2 + \left[1 7\right]^2\left[2 5\right]^2\left[3 4\right]^2\left[6 8\right]^2 + \left[1 7\right]^2\left[2 5\right]^2\left[3 6\right]^2\left[4 8\right]^2 + \left[1 7\right]^2\left[2 5\right]^2\left[3 8\right]^2\left[4 6\right]^2 + \left[1 7\right]^2\left[2 6\right]^2\left[3 4\right]^2\left[5 8\right]^2 + \left[1 7\right]^2\left[2 6\right]^2\left[3 5\right]^2\left[4 8\right]^2 + \left[1 7\right]^2\left[2 6\right]^2\left[3 8\right]^2\left[4 5\right]^2 + \left[1 7\right]^2\left[2 8\right]^2\left[3 4\right]^2\left[5 6\right]^2 + \left[1 7\right]^2\left[2 8\right]^2\left[3 5\right]^2\left[4 6\right]^2 + \left[1 7\right]^2\left[2 8\right]^2\left[3 6\right]^2\left[4 5\right]^2 + \left[1 8\right]^2\left[2 3\right]^2\left[4 5\right]^2\left[6 7\right]^2 + \left[1 8\right]^2\left[2 3\right]^2\left[4 6\right]^2\left[5 7\right]^2 + \left[1 8\right]^2\left[2 3\right]^2\left[4 7\right]^2\left[5 6\right]^2 + \left[1 8\right]^2\left[2 4\right]^2\left[3 5\right]^2\left[6 7\right]^2 + \left[1 8\right]^2\left[2 4\right]^2\left[3 6\right]^2\left[5 7\right]^2 + \left[1 8\right]^2\left[2 4\right]^2\left[3 7\right]^2\left[5 6\right]^2 + \left[1 8\right]^2\left[2 5\right]^2\left[3 4\right]^2\left[6 7\right]^2 + \left[1 8\right]^2\left[2 5\right]^2\left[3 6\right]^2\left[4 7\right]^2 + \left[1 8\right]^2\left[2 5\right]^2\left[3 7\right]^2\left[4 6\right]^2 + \left[1 8\right]^2\left[2 6\right]^2\left[3 4\right]^2\left[5 7\right]^2 + \left[1 8\right]^2\left[2 6\right]^2\left[3 5\right]^2\left[4 7\right]^2 + \left[1 8\right]^2\left[2 6\right]^2\left[3 7\right]^2\left[4 5\right]^2 + \left[1 8\right]^2\left[2 7\right]^2\left[3 4\right]^2\left[5 6\right]^2 + \left[1 8\right]^2\left[2 7\right]^2\left[3 5\right]^2\left[4 6\right]^2 + \left[1 8\right]^2\left[2 7\right]^2\left[3 6\right]^2\left[4 5\right]^2 \,.
\end{dmath*}

\section{Contact Lagrangian up to $O(F^{14})$}
\label{app:contact_lag}

In this appendix, we write our results for the effective contact Lagrangian up to operators of order $N=14$, in terms of the couplings $\vv^{(N)}_i$ of the original basis, defined in  \eqref{eq:LF_standardbasis}.  
 The couplings for each $N$ are organized from lowest to highest $K$.

{\setlength{\intereqskip}{4ex}%

\subsubsection*{$F^4$ Operators}
\begin{dgroup*}

\begin{dmath*}
  \vhat^{(4)}_2 = 2 \vv^{(4)}_0 + 2 \vv^{(4)}_1
\end{dmath*}

\begin{dmath*}
  \vhat^{(4)}_4 = \vv^{(4)}_0 - \vv^{(4)}_1
\end{dmath*}

\end{dgroup*}

\subsubsection*{$F^6$ Operators}

\begin{dgroup*}

\begin{dmath*}
  \vhat^{(6)}_4 = 3 \vv^{(6)}_0 + \vv^{(6)}_1 + 16 \vv^{(4)}_0 \vv^{(4)}_1 + 24 (\vv^{(4)}_0)^2 - 8 (\vv^{(4)}_1)^2
\end{dmath*}

\begin{dmath*}
  \vhat^{(6)}_6 = \vv^{(6)}_0 - \vv^{(6)}_1 - 16 \vv^{(4)}_0 \vv^{(4)}_1 + 8 (\vv^{(4)}_0)^2 + 8 (\vv^{(4)}_1)^2
\end{dmath*}

\end{dgroup*}

\subsubsection*{$F^8$ Operators}
\begin{dgroup*}[spread={1.5ex}]

    \begin{dmath*}
      \vhat^{(8)}_4 = 6 \vv^{(8)}_0 + 2 \vv^{(8)}_1 + 6 \vv^{(8)}_2 - 64 \vv^{(4)}_0 (\vv^{(4)}_1)^2 + 576 (\vv^{(4)}_0)^2 \vv^{(4)}_1 + 144 \vv^{(6)}_0 \vv^{(4)}_0 + 48 \vv^{(6)}_0 \vv^{(4)}_1 + 48 \vv^{(6)}_1 \vv^{(4)}_0 + 16 \vv^{(6)}_1 \vv^{(4)}_1 + 576 (\vv^{(4)}_0)^3 - 64 (\vv^{(4)}_1)^3
    \end{dmath*}
    
    \begin{dmath*}
      \vhat^{(8)}_6 = 4 \vv^{(8)}_0 - 4 \vv^{(8)}_2 - 256 \vv^{(4)}_0 (\vv^{(4)}_1)^2 + 96 \vv^{(6)}_0 \vv^{(4)}_0 - 32 \vv^{(6)}_1 \vv^{(4)}_1 + 384 (\vv^{(4)}_0)^3 + 128 (\vv^{(4)}_1)^3
    \end{dmath*}
    
    \begin{dmath*}
      \vhat^{(8)}_8 = \vv^{(8)}_0 - \vv^{(8)}_1 + \vv^{(8)}_2 + 288 \vv^{(4)}_0 (\vv^{(4)}_1)^2 - 288 (\vv^{(4)}_0)^2 \vv^{(4)}_1 + 24 \vv^{(6)}_0 \vv^{(4)}_0 - 24 \vv^{(6)}_0 \vv^{(4)}_1 - 24 \vv^{(6)}_1 \vv^{(4)}_0 + 24 \vv^{(6)}_1 \vv^{(4)}_1 + 96 (\vv^{(4)}_0)^3 - 96 (\vv^{(4)}_1)^3
    \end{dmath*}
    
\end{dgroup*}

\subsubsection*{$F^{10}$ Operators}
\begin{dgroup*}[spread={1.5ex}]

\begin{dmath*}
  \vhat^{(10)}_6 =  10 \vv^{(10)}_0 + 2 \vv^{(10)}_1 + 2 \vv^{(10)}_2 - 1024 \vv^{(4)}_0 (\vv^{(4)}_1)^3 + 320 \vv^{(4)}_0 \vv^{(8)}_0 + 64 \vv^{(4)}_0 \vv^{(8)}_1 + 64 \vv^{(4)}_0 \vv^{(8)}_2 - 4608 (\vv^{(4)}_0)^2 (\vv^{(4)}_1)^2 + 11264 (\vv^{(4)}_0)^3 \vv^{(4)}_1 + 64 \vv^{(4)}_1 \vv^{(8)}_0 - 64 \vv^{(4)}_1 \vv^{(8)}_2 + 5280 \vv^{(6)}_0 (\vv^{(4)}_0)^2 - 288 \vv^{(6)}_0 (\vv^{(4)}_1)^2 + 1056 \vv^{(6)}_1 (\vv^{(4)}_0)^2 - 288 \vv^{(6)}_1 (\vv^{(4)}_1)^2 + 72 \vv^{(6)}_1 \vv^{(6)}_0 + 2112 \vv^{(6)}_0 \vv^{(4)}_0 \vv^{(4)}_1 - 192 \vv^{(6)}_1 \vv^{(4)}_0 \vv^{(4)}_1 + 14080 (\vv^{(4)}_0)^4 + 768 (\vv^{(4)}_1)^4 + 180 (\vv^{(6)}_0)^2 + 4 (\vv^{(6)}_1)^2
\end{dmath*}

\begin{dmath*}
  \vhat^{(10)}_8 =  5 \vv^{(10)}_0 - \vv^{(10)}_1 - 3 \vv^{(10)}_2 + 6656 \vv^{(4)}_0 (\vv^{(4)}_1)^3 + 160 \vv^{(4)}_0 \vv^{(8)}_0 - 32 \vv^{(4)}_0 \vv^{(8)}_1 - 96 \vv^{(4)}_0 \vv^{(8)}_2 - 3840 (\vv^{(4)}_0)^2 (\vv^{(4)}_1)^2 - 5632 (\vv^{(4)}_0)^3 \vv^{(4)}_1 - 32 \vv^{(4)}_1 \vv^{(8)}_0 - 32 \vv^{(4)}_1 \vv^{(8)}_1 + 96 \vv^{(4)}_1 \vv^{(8)}_2 + 2640 \vv^{(6)}_0 (\vv^{(4)}_0)^2 - 240 \vv^{(6)}_0 (\vv^{(4)}_1)^2 - 528 \vv^{(6)}_1 (\vv^{(4)}_0)^2 + 816 \vv^{(6)}_1 (\vv^{(4)}_1)^2 - 36 \vv^{(6)}_1 \vv^{(6)}_0 - 1056 \vv^{(6)}_0 \vv^{(4)}_0 \vv^{(4)}_1 - 864 \vv^{(6)}_1 \vv^{(4)}_0 \vv^{(4)}_1 + 7040 (\vv^{(4)}_0)^4 - 2176 (\vv^{(4)}_1)^4 + 90 (\vv^{(6)}_0)^2 - 22 (\vv^{(6)}_1)^2
\end{dmath*}

\begin{dmath*}
  \vhat^{(10)}_{10} =  \vv^{(10)}_0 - \vv^{(10)}_1 + \vv^{(10)}_2 - 5632 \vv^{(4)}_0 (\vv^{(4)}_1)^3 + 32 \vv^{(4)}_0 \vv^{(8)}_0 - 32 \vv^{(4)}_0 \vv^{(8)}_1 + 32 \vv^{(4)}_0 \vv^{(8)}_2 + 8448 (\vv^{(4)}_0)^2 (\vv^{(4)}_1)^2 - 5632 (\vv^{(4)}_0)^3 \vv^{(4)}_1 - 32 \vv^{(4)}_1 \vv^{(8)}_0 + 32 \vv^{(4)}_1 \vv^{(8)}_1 - 32 \vv^{(4)}_1 \vv^{(8)}_2 + 528 \vv^{(6)}_0 (\vv^{(4)}_0)^2 + 528 \vv^{(6)}_0 (\vv^{(4)}_1)^2 - 528 \vv^{(6)}_1 (\vv^{(4)}_0)^2 - 528 \vv^{(6)}_1 (\vv^{(4)}_1)^2 - 36 \vv^{(6)}_1 \vv^{(6)}_0 - 1056 \vv^{(6)}_0 \vv^{(4)}_0 \vv^{(4)}_1 + 1056 \vv^{(6)}_1 \vv^{(4)}_0 \vv^{(4)}_1 + 1408 (\vv^{(4)}_0)^4 + 1408 (\vv^{(4)}_1)^4 + 18 (\vv^{(6)}_0)^2 + 18 (\vv^{(6)}_1)^2
\end{dmath*}

\end{dgroup*}

\subsubsection*{$F^{12}$ Operators}
\begin{dgroup*}[spread={1.5ex}]

\begin{dmath*}
  \vhat^{(12)}_{6} = 20 \vv^{(12)}_0 + 4 \vv^{(12)}_1 + 4 \vv^{(12)}_2 + 20 \vv^{(12)}_3 + 7168 \vv^{(4)}_0 (\vv^{(4)}_1)^4 + 800 \vv^{(4)}_0 \vv^{(10)}_0 + 160 \vv^{(4)}_0 \vv^{(10)}_1 + 160 \vv^{(4)}_0 \vv^{(10)}_2 - 71680 (\vv^{(4)}_0)^2 (\vv^{(4)}_1)^3 + 16640 (\vv^{(4)}_0)^2 \vv^{(8)}_0 + 3328 (\vv^{(4)}_0)^2 \vv^{(8)}_1 + 3328 (\vv^{(4)}_0)^2 \vv^{(8)}_2 - 71680 (\vv^{(4)}_0)^3 (\vv^{(4)}_1)^2 + 465920 (\vv^{(4)}_0)^4 \vv^{(4)}_1 + 160 \vv^{(4)}_1 \vv^{(10)}_0 + 32 \vv^{(4)}_1 \vv^{(10)}_1 + 32 \vv^{(4)}_1 \vv^{(10)}_2 - 256 (\vv^{(4)}_1)^2 \vv^{(8)}_0 - 256 (\vv^{(4)}_1)^2 \vv^{(8)}_1 - 1280 (\vv^{(4)}_1)^2 \vv^{(8)}_2 + 232960 \vv^{(6)}_0 (\vv^{(4)}_0)^3 - 3584 \vv^{(6)}_0 (\vv^{(4)}_1)^3 + 960 \vv^{(6)}_0 \vv^{(8)}_0 + 192 \vv^{(6)}_0 \vv^{(8)}_1 + 192 \vv^{(6)}_0 \vv^{(8)}_2 + 18720 (\vv^{(6)}_0)^2 \vv^{(4)}_0 + 3744 (\vv^{(6)}_0)^2 \vv^{(4)}_1 + 46592 \vv^{(6)}_1 (\vv^{(4)}_0)^3 - 3584 \vv^{(6)}_1 (\vv^{(4)}_1)^3 + 192 \vv^{(6)}_1 \vv^{(8)}_0 + 64 \vv^{(6)}_1 \vv^{(8)}_1 + 192 \vv^{(6)}_1 \vv^{(8)}_2 + 928 (\vv^{(6)}_1)^2 \vv^{(4)}_0 + 288 (\vv^{(6)}_1)^2 \vv^{(4)}_1 + 6656 \vv^{(4)}_0 \vv^{(4)}_1 \vv^{(8)}_0 + 1024 \vv^{(4)}_0 \vv^{(4)}_1 \vv^{(8)}_1 - 512 \vv^{(4)}_0 \vv^{(4)}_1 \vv^{(8)}_2 - 10752 \vv^{(6)}_0 \vv^{(4)}_0 (\vv^{(4)}_1)^2 + 139776 \vv^{(6)}_0 (\vv^{(4)}_0)^2 \vv^{(4)}_1 - 10752 \vv^{(6)}_1 \vv^{(4)}_0 (\vv^{(4)}_1)^2 + 10752 \vv^{(6)}_1 (\vv^{(4)}_0)^2 \vv^{(4)}_1 + 7488 \vv^{(6)}_1 \vv^{(6)}_0 \vv^{(4)}_0 + 576 \vv^{(6)}_1 \vv^{(6)}_0 \vv^{(4)}_1 + 465920 (\vv^{(4)}_0)^5 + 7168 (\vv^{(4)}_1)^5
\end{dmath*}

\begin{dmath*}
  \vhat^{(12)}_{8} = 15 \vv^{(12)}_0 + \vv^{(12)}_1 - \vv^{(12)}_2 - 15 \vv^{(12)}_3 + 47360 \vv^{(4)}_0 (\vv^{(4)}_1)^4 + 600 \vv^{(4)}_0 \vv^{(10)}_0 + 40 \vv^{(4)}_0 \vv^{(10)}_1 - 40 \vv^{(4)}_0 \vv^{(10)}_2 + 69120 (\vv^{(4)}_0)^2 (\vv^{(4)}_1)^3 + 12480 (\vv^{(4)}_0)^2 \vv^{(8)}_0 + 832 (\vv^{(4)}_0)^2 \vv^{(8)}_1 - 832 (\vv^{(4)}_0)^2 \vv^{(8)}_2 - 232960 (\vv^{(4)}_0)^3 (\vv^{(4)}_1)^2 + 116480 (\vv^{(4)}_0)^4 \vv^{(4)}_1 + 40 \vv^{(4)}_1 \vv^{(10)}_0 - 40 \vv^{(4)}_1 \vv^{(10)}_1 - 88 \vv^{(4)}_1 \vv^{(10)}_2 - 832 (\vv^{(4)}_1)^2 \vv^{(8)}_0 - 192 (\vv^{(4)}_1)^2 \vv^{(8)}_1 + 2240 (\vv^{(4)}_1)^2 \vv^{(8)}_2 + 174720 \vv^{(6)}_0 (\vv^{(4)}_0)^3 + 3456 \vv^{(6)}_0 (\vv^{(4)}_1)^3 + 720 \vv^{(6)}_0 \vv^{(8)}_0 + 48 \vv^{(6)}_0 \vv^{(8)}_1 - 48 \vv^{(6)}_0 \vv^{(8)}_2 + 14040 (\vv^{(6)}_0)^2 \vv^{(4)}_0 + 936 (\vv^{(6)}_0)^2 \vv^{(4)}_1 + 11648 \vv^{(6)}_1 (\vv^{(4)}_0)^3 + 10880 \vv^{(6)}_1 (\vv^{(4)}_1)^3 + 48 \vv^{(6)}_1 \vv^{(8)}_0 - 48 \vv^{(6)}_1 \vv^{(8)}_1 - 208 \vv^{(6)}_1 \vv^{(8)}_2 - 936 (\vv^{(6)}_1)^2 \vv^{(4)}_0 - 728 (\vv^{(6)}_1)^2 \vv^{(4)}_1 + 1664 \vv^{(4)}_0 \vv^{(4)}_1 \vv^{(8)}_0 - 1664 \vv^{(4)}_0 \vv^{(4)}_1 \vv^{(8)}_1 - 2432 \vv^{(4)}_0 \vv^{(4)}_1 \vv^{(8)}_2 - 34944 \vv^{(6)}_0 \vv^{(4)}_0 (\vv^{(4)}_1)^2 + 34944 \vv^{(6)}_0 (\vv^{(4)}_0)^2 \vv^{(4)}_1 - 1920 \vv^{(6)}_1 \vv^{(4)}_0 (\vv^{(4)}_1)^2 - 34944 \vv^{(6)}_1 (\vv^{(4)}_0)^2 \vv^{(4)}_1 + 1872 \vv^{(6)}_1 \vv^{(6)}_0 \vv^{(4)}_0 - 1872 \vv^{(6)}_1 \vv^{(6)}_0 \vv^{(4)}_1 + 349440 (\vv^{(4)}_0)^5 - 21760 (\vv^{(4)}_1)^5
\end{dmath*}

\begin{dmath*}
  \vhat^{(12)}_{10} = 6 \vv^{(12)}_0 - 2 \vv^{(12)}_1 - 2 \vv^{(12)}_2 + 6 \vv^{(12)}_3 - 167424 \vv^{(4)}_0 (\vv^{(4)}_1)^4 + 240 \vv^{(4)}_0 \vv^{(10)}_0 - 80 \vv^{(4)}_0 \vv^{(10)}_1 - 80 \vv^{(4)}_0 \vv^{(10)}_2 + 199680 (\vv^{(4)}_0)^2 (\vv^{(4)}_1)^3 + 4992 (\vv^{(4)}_0)^2 \vv^{(8)}_0 - 1664 (\vv^{(4)}_0)^2 \vv^{(8)}_1 - 1664 (\vv^{(4)}_0)^2 \vv^{(8)}_2 + 35840 (\vv^{(4)}_0)^3 (\vv^{(4)}_1)^2 - 232960 (\vv^{(4)}_0)^4 \vv^{(4)}_1 - 80 \vv^{(4)}_1 \vv^{(10)}_0 - 16 \vv^{(4)}_1 \vv^{(10)}_1 + 112 \vv^{(4)}_1 \vv^{(10)}_2 + 128 (\vv^{(4)}_1)^2 \vv^{(8)}_0 + 1152 (\vv^{(4)}_1)^2 \vv^{(8)}_1 - 2432 (\vv^{(4)}_1)^2 \vv^{(8)}_2 + 69888 \vv^{(6)}_0 (\vv^{(4)}_0)^3 + 9984 \vv^{(6)}_0 (\vv^{(4)}_1)^3 + 288 \vv^{(6)}_0 \vv^{(8)}_0 - 96 \vv^{(6)}_0 \vv^{(8)}_1 - 96 \vv^{(6)}_0 \vv^{(8)}_2 + 5616 (\vv^{(6)}_0)^2 \vv^{(4)}_0 - 1872 (\vv^{(6)}_0)^2 \vv^{(4)}_1 - 23296 \vv^{(6)}_1 (\vv^{(4)}_0)^3 - 20736 \vv^{(6)}_1 (\vv^{(4)}_1)^3 - 96 \vv^{(6)}_1 \vv^{(8)}_0 - 32 \vv^{(6)}_1 \vv^{(8)}_1 + 160 \vv^{(6)}_1 \vv^{(8)}_2 - 464 (\vv^{(6)}_1)^2 \vv^{(4)}_0 + 1520 (\vv^{(6)}_1)^2 \vv^{(4)}_1 - 3328 \vv^{(4)}_0 \vv^{(4)}_1 \vv^{(8)}_0 - 512 \vv^{(4)}_0 \vv^{(4)}_1 \vv^{(8)}_1 + 4352 \vv^{(4)}_0 \vv^{(4)}_1 \vv^{(8)}_2 + 5376 \vv^{(6)}_0 \vv^{(4)}_0 (\vv^{(4)}_1)^2 - 69888 \vv^{(6)}_0 (\vv^{(4)}_0)^2 \vv^{(4)}_1 + 42240 \vv^{(6)}_1 \vv^{(4)}_0 (\vv^{(4)}_1)^2 - 5376 \vv^{(6)}_1 (\vv^{(4)}_0)^2 \vv^{(4)}_1 - 3744 \vv^{(6)}_1 \vv^{(6)}_0 \vv^{(4)}_0 - 288 \vv^{(6)}_1 \vv^{(6)}_0 \vv^{(4)}_1 + 139776 (\vv^{(4)}_0)^5 + 41472 (\vv^{(4)}_1)^5
\end{dmath*}

\begin{dmath*}
  \vhat^{(12)}_{12} = \vv^{(12)}_0 - \vv^{(12)}_1 + \vv^{(12)}_2 - \vv^{(12)}_3 + 116480 \vv^{(4)}_0 (\vv^{(4)}_1)^4 + 40 \vv^{(4)}_0 \vv^{(10)}_0 - 40 \vv^{(4)}_0 \vv^{(10)}_1 + 40 \vv^{(4)}_0 \vv^{(10)}_2 - 232960 (\vv^{(4)}_0)^2 (\vv^{(4)}_1)^3 + 832 (\vv^{(4)}_0)^2 \vv^{(8)}_0 - 832 (\vv^{(4)}_0)^2 \vv^{(8)}_1 + 832 (\vv^{(4)}_0)^2 \vv^{(8)}_2 + 232960 (\vv^{(4)}_0)^3 (\vv^{(4)}_1)^2 - 116480 (\vv^{(4)}_0)^4 \vv^{(4)}_1 - 40 \vv^{(4)}_1 \vv^{(10)}_0 + 40 \vv^{(4)}_1 \vv^{(10)}_1 - 40 \vv^{(4)}_1 \vv^{(10)}_2 + 832 (\vv^{(4)}_1)^2 \vv^{(8)}_0 - 832 (\vv^{(4)}_1)^2 \vv^{(8)}_1 + 832 (\vv^{(4)}_1)^2 \vv^{(8)}_2 + 11648 \vv^{(6)}_0 (\vv^{(4)}_0)^3 - 11648 \vv^{(6)}_0 (\vv^{(4)}_1)^3 + 48 \vv^{(6)}_0 \vv^{(8)}_0 - 48 \vv^{(6)}_0 \vv^{(8)}_1 + 48 \vv^{(6)}_0 \vv^{(8)}_2 + 936 (\vv^{(6)}_0)^2 \vv^{(4)}_0 - 936 (\vv^{(6)}_0)^2 \vv^{(4)}_1 - 11648 \vv^{(6)}_1 (\vv^{(4)}_0)^3 + 11648 \vv^{(6)}_1 (\vv^{(4)}_1)^3 - 48 \vv^{(6)}_1 \vv^{(8)}_0 + 48 \vv^{(6)}_1 \vv^{(8)}_1 - 48 \vv^{(6)}_1 \vv^{(8)}_2 + 936 (\vv^{(6)}_1)^2 \vv^{(4)}_0 - 936 (\vv^{(6)}_1)^2 \vv^{(4)}_1 - 1664 \vv^{(4)}_0 \vv^{(4)}_1 \vv^{(8)}_0 + 1664 \vv^{(4)}_0 \vv^{(4)}_1 \vv^{(8)}_1 - 1664 \vv^{(4)}_0 \vv^{(4)}_1 \vv^{(8)}_2 + 34944 \vv^{(6)}_0 \vv^{(4)}_0 (\vv^{(4)}_1)^2 - 34944 \vv^{(6)}_0 (\vv^{(4)}_0)^2 \vv^{(4)}_1 - 34944 \vv^{(6)}_1 \vv^{(4)}_0 (\vv^{(4)}_1)^2 + 34944 \vv^{(6)}_1 (\vv^{(4)}_0)^2 \vv^{(4)}_1 - 1872 \vv^{(6)}_1 \vv^{(6)}_0 \vv^{(4)}_0 + 1872 \vv^{(6)}_1 \vv^{(6)}_0 \vv^{(4)}_1 + 23296 (\vv^{(4)}_0)^5 - 23296 (\vv^{(4)}_1)^5
\end{dmath*}

\end{dgroup*}

\subsubsection*{$F^{14}$ Operators}

\begin{dgroup*}[spread={1.5ex}]

\begin{dmath*}
  \vhat^{(14)}_{8} =  35 \vv^{(14)}_0 + 5 \vv^{(14)}_1 + 3 \vv^{(14)}_2 + 5 \vv^{(14)}_3 + 147456 \vv^{(4)}_0 (\vv^{(4)}_1)^5 + 1680 \vv^{(4)}_0 \vv^{(12)}_0 + 240 \vv^{(4)}_0 \vv^{(12)}_1 + 144 \vv^{(4)}_0 \vv^{(12)}_2 + 240 \vv^{(4)}_0 \vv^{(12)}_3 + 1105920 (\vv^{(4)}_0)^2 (\vv^{(4)}_1)^4 + 42000 (\vv^{(4)}_0)^2 \vv^{(10)}_0 + 6000 (\vv^{(4)}_0)^2 \vv^{(10)}_1 + 3600 (\vv^{(4)}_0)^2 \vv^{(10)}_2 - 1671168 (\vv^{(4)}_0)^3 (\vv^{(4)}_1)^3 + 716800 (\vv^{(4)}_0)^3 \vv^{(8)}_0 + 102400 (\vv^{(4)}_0)^3 \vv^{(8)}_1 + 61440 (\vv^{(4)}_0)^3 \vv^{(8)}_2 - 4595712 (\vv^{(4)}_0)^4 (\vv^{(4)}_1)^2 + 12533760 (\vv^{(4)}_0)^5 \vv^{(4)}_1 + 240 \vv^{(4)}_1 \vv^{(12)}_0 + 16 \vv^{(4)}_1 \vv^{(12)}_1 - 16 \vv^{(4)}_1 \vv^{(12)}_2 - 240 \vv^{(4)}_1 \vv^{(12)}_3 - 880 (\vv^{(4)}_1)^2 \vv^{(10)}_0 - 528 (\vv^{(4)}_1)^2 \vv^{(10)}_1 - 880 (\vv^{(4)}_1)^2 \vv^{(10)}_2 - 4096 (\vv^{(4)}_1)^3 \vv^{(8)}_0 + 20480 (\vv^{(4)}_1)^3 \vv^{(8)}_2 + 9139200 \vv^{(6)}_0 (\vv^{(4)}_0)^4 + 46080 \vv^{(6)}_0 (\vv^{(4)}_1)^4 + 2100 \vv^{(6)}_0 \vv^{(10)}_0 + 300 \vv^{(6)}_0 \vv^{(10)}_1 + 180 \vv^{(6)}_0 \vv^{(10)}_2 + 1209600 (\vv^{(6)}_0)^2 (\vv^{(4)}_0)^2 - 25344 (\vv^{(6)}_0)^2 (\vv^{(4)}_1)^2 + 1305600 \vv^{(6)}_1 (\vv^{(4)}_0)^4 + 76800 \vv^{(6)}_1 (\vv^{(4)}_1)^4 + 8100 \vv^{(6)}_1 (\vv^{(6)}_0)^2 + 300 \vv^{(6)}_1 \vv^{(10)}_0 + 52 \vv^{(6)}_1 \vv^{(10)}_1 + 44 \vv^{(6)}_1 \vv^{(10)}_2 + 13568 (\vv^{(6)}_1)^2 (\vv^{(4)}_0)^2 - 8960 (\vv^{(6)}_1)^2 (\vv^{(4)}_1)^2 + 636 (\vv^{(6)}_1)^2 \vv^{(6)}_0 + 320 \vv^{(8)}_0 \vv^{(8)}_1 + 192 \vv^{(8)}_0 \vv^{(8)}_2 + 64 \vv^{(8)}_2 \vv^{(8)}_1 + 12000 \vv^{(4)}_0 \vv^{(4)}_1 \vv^{(10)}_0 + 544 \vv^{(4)}_0 \vv^{(4)}_1 \vv^{(10)}_1 - 1312 \vv^{(4)}_0 \vv^{(4)}_1 \vv^{(10)}_2 - 45056 \vv^{(4)}_0 (\vv^{(4)}_1)^2 \vv^{(8)}_0 - 24576 \vv^{(4)}_0 (\vv^{(4)}_1)^2 \vv^{(8)}_1 - 20480 \vv^{(4)}_0 (\vv^{(4)}_1)^2 \vv^{(8)}_2 + 307200 (\vv^{(4)}_0)^2 \vv^{(4)}_1 \vv^{(8)}_0 + 4096 (\vv^{(4)}_0)^2 \vv^{(4)}_1 \vv^{(8)}_1 - 53248 (\vv^{(4)}_0)^2 \vv^{(4)}_1 \vv^{(8)}_2 - 208896 \vv^{(6)}_0 \vv^{(4)}_0 (\vv^{(4)}_1)^3 + 100800 \vv^{(6)}_0 \vv^{(4)}_0 \vv^{(8)}_0 + 14400 \vv^{(6)}_0 \vv^{(4)}_0 \vv^{(8)}_1 + 8640 \vv^{(6)}_0 \vv^{(4)}_0 \vv^{(8)}_2 - 1148928 \vv^{(6)}_0 (\vv^{(4)}_0)^2 (\vv^{(4)}_1)^2 + 5222400 \vv^{(6)}_0 (\vv^{(4)}_0)^3 \vv^{(4)}_1 + 14400 \vv^{(6)}_0 \vv^{(4)}_1 \vv^{(8)}_0 + 192 \vv^{(6)}_0 \vv^{(4)}_1 \vv^{(8)}_1 - 2496 \vv^{(6)}_0 \vv^{(4)}_1 \vv^{(8)}_2 + 345600 (\vv^{(6)}_0)^2 \vv^{(4)}_0 \vv^{(4)}_1 + 61440 \vv^{(6)}_1 \vv^{(4)}_0 (\vv^{(4)}_1)^3 + 14400 \vv^{(6)}_1 \vv^{(4)}_0 \vv^{(8)}_0 + 1984 \vv^{(6)}_1 \vv^{(4)}_0 \vv^{(8)}_1 + 1088 \vv^{(6)}_1 \vv^{(4)}_0 \vv^{(8)}_2 - 522240 \vv^{(6)}_1 (\vv^{(4)}_0)^2 (\vv^{(4)}_1)^2 - 208896 \vv^{(6)}_1 (\vv^{(4)}_0)^3 \vv^{(4)}_1 - 576 \vv^{(6)}_1 \vv^{(4)}_1 \vv^{(8)}_0 - 704 \vv^{(6)}_1 \vv^{(4)}_1 \vv^{(8)}_1 - 4160 \vv^{(6)}_1 \vv^{(4)}_1 \vv^{(8)}_2 + 345600 \vv^{(6)}_1 \vv^{(6)}_0 (\vv^{(4)}_0)^2 - 23040 \vv^{(6)}_1 \vv^{(6)}_0 (\vv^{(4)}_1)^2 - 18944 (\vv^{(6)}_1)^2 \vv^{(4)}_0 \vv^{(4)}_1 - 27648 \vv^{(6)}_1 \vv^{(6)}_0 \vv^{(4)}_0 \vv^{(4)}_1 + 14622720 (\vv^{(4)}_0)^6 - 122880 (\vv^{(4)}_1)^6 + 18900 (\vv^{(6)}_0)^3 + 12 (\vv^{(6)}_1)^3 + 1120 (\vv^{(8)}_0)^2 + 32 (\vv^{(8)}_1)^2 - 160 (\vv^{(8)}_2)^2
\end{dmath*}

\begin{dmath*}
 \vhat^{(14)}_{10} =  21 \vv^{(14)}_0 - \vv^{(14)}_1 - 3 \vv^{(14)}_2 - 9 \vv^{(14)}_3 - 1884160 \vv^{(4)}_0 (\vv^{(4)}_1)^5 + 1008 \vv^{(4)}_0 \vv^{(12)}_0 - 48 \vv^{(4)}_0 \vv^{(12)}_1 - 144 \vv^{(4)}_0 \vv^{(12)}_2 - 432 \vv^{(4)}_0 \vv^{(12)}_3 + 73728 (\vv^{(4)}_0)^2 (\vv^{(4)}_1)^4 + 25200 (\vv^{(4)}_0)^2 \vv^{(10)}_0 - 1200 (\vv^{(4)}_0)^2 \vv^{(10)}_1 - 3600 (\vv^{(4)}_0)^2 \vv^{(10)}_2 + 6684672 (\vv^{(4)}_0)^3 (\vv^{(4)}_1)^3 + 430080 (\vv^{(4)}_0)^3 \vv^{(8)}_0 - 20480 (\vv^{(4)}_0)^3 \vv^{(8)}_1 - 61440 (\vv^{(4)}_0)^3 \vv^{(8)}_2 - 7102464 (\vv^{(4)}_0)^4 (\vv^{(4)}_1)^2 - 2506752 (\vv^{(4)}_0)^5 \vv^{(4)}_1 - 48 \vv^{(4)}_1 \vv^{(12)}_0 - 80 \vv^{(4)}_1 \vv^{(12)}_1 - 48 \vv^{(4)}_1 \vv^{(12)}_2 + 432 \vv^{(4)}_1 \vv^{(12)}_3 - 1360 (\vv^{(4)}_1)^2 \vv^{(10)}_0 + 528 (\vv^{(4)}_1)^2 \vv^{(10)}_1 + 3120 (\vv^{(4)}_1)^2 \vv^{(10)}_2 + 16384 (\vv^{(4)}_1)^3 \vv^{(8)}_0 + 14336 (\vv^{(4)}_1)^3 \vv^{(8)}_1 - 61440 (\vv^{(4)}_1)^3 \vv^{(8)}_2 + 5483520 \vv^{(6)}_0 (\vv^{(4)}_0)^4 + 3072 \vv^{(6)}_0 (\vv^{(4)}_1)^4 + 1260 \vv^{(6)}_0 \vv^{(10)}_0 - 60 \vv^{(6)}_0 \vv^{(10)}_1 - 180 \vv^{(6)}_0 \vv^{(10)}_2 + 725760 (\vv^{(6)}_0)^2 (\vv^{(4)}_0)^2 - 39168 (\vv^{(6)}_0)^2 (\vv^{(4)}_1)^2 - 261120 \vv^{(6)}_1 (\vv^{(4)}_0)^4 - 343040 \vv^{(6)}_1 (\vv^{(4)}_1)^4 - 1620 \vv^{(6)}_1 (\vv^{(6)}_0)^2 - 60 \vv^{(6)}_1 \vv^{(10)}_0 - 116 \vv^{(6)}_1 \vv^{(10)}_1 - 156 \vv^{(6)}_1 \vv^{(10)}_2 - 58624 (\vv^{(6)}_1)^2 (\vv^{(4)}_0)^2 + 40704 (\vv^{(6)}_1)^2 (\vv^{(4)}_1)^2 - 2748 (\vv^{(6)}_1)^2 \vv^{(6)}_0 - 64 \vv^{(8)}_0 \vv^{(8)}_1 - 192 \vv^{(8)}_0 \vv^{(8)}_2 - 192 \vv^{(8)}_2 \vv^{(8)}_1 - 2400 \vv^{(4)}_0 \vv^{(4)}_1 \vv^{(10)}_0 - 3872 \vv^{(4)}_0 \vv^{(4)}_1 \vv^{(10)}_1 - 1632 \vv^{(4)}_0 \vv^{(4)}_1 \vv^{(10)}_2 - 69632 \vv^{(4)}_0 (\vv^{(4)}_1)^2 \vv^{(8)}_0 + 30720 \vv^{(4)}_0 (\vv^{(4)}_1)^2 \vv^{(8)}_1 + 122880 \vv^{(4)}_0 (\vv^{(4)}_1)^2 \vv^{(8)}_2 - 61440 (\vv^{(4)}_0)^2 \vv^{(4)}_1 \vv^{(8)}_0 - 94208 (\vv^{(4)}_0)^2 \vv^{(4)}_1 \vv^{(8)}_1 - 12288 (\vv^{(4)}_0)^2 \vv^{(4)}_1 \vv^{(8)}_2 + 835584 \vv^{(6)}_0 \vv^{(4)}_0 (\vv^{(4)}_1)^3 + 60480 \vv^{(6)}_0 \vv^{(4)}_0 \vv^{(8)}_0 - 2880 \vv^{(6)}_0 \vv^{(4)}_0 \vv^{(8)}_1 - 8640 \vv^{(6)}_0 \vv^{(4)}_0 \vv^{(8)}_2 - 1775616 \vv^{(6)}_0 (\vv^{(4)}_0)^2 (\vv^{(4)}_1)^2 - 1044480 \vv^{(6)}_0 (\vv^{(4)}_0)^3 \vv^{(4)}_1 - 2880 \vv^{(6)}_0 \vv^{(4)}_1 \vv^{(8)}_0 - 4416 \vv^{(6)}_0 \vv^{(4)}_1 \vv^{(8)}_1 - 576 \vv^{(6)}_0 \vv^{(4)}_1 \vv^{(8)}_2 - 69120 (\vv^{(6)}_0)^2 \vv^{(4)}_0 \vv^{(4)}_1 + 491520 \vv^{(6)}_1 \vv^{(4)}_0 (\vv^{(4)}_1)^3 - 2880 \vv^{(6)}_1 \vv^{(4)}_0 \vv^{(8)}_0 - 5312 \vv^{(6)}_1 \vv^{(4)}_0 \vv^{(8)}_1 - 5952 \vv^{(6)}_1 \vv^{(4)}_0 \vv^{(8)}_2 + 940032 \vv^{(6)}_1 (\vv^{(4)}_0)^2 (\vv^{(4)}_1)^2 - 1462272 \vv^{(6)}_1 (\vv^{(4)}_0)^3 \vv^{(4)}_1 - 4032 \vv^{(6)}_1 \vv^{(4)}_1 \vv^{(8)}_0 - 192 \vv^{(6)}_1 \vv^{(4)}_1 \vv^{(8)}_1 + 10560 \vv^{(6)}_1 \vv^{(4)}_1 \vv^{(8)}_2 - 69120 \vv^{(6)}_1 \vv^{(6)}_0 (\vv^{(4)}_0)^2 + 41472 \vv^{(6)}_1 \vv^{(6)}_0 (\vv^{(4)}_1)^2 + 10752 (\vv^{(6)}_1)^2 \vv^{(4)}_0 \vv^{(4)}_1 - 193536 \vv^{(6)}_1 \vv^{(6)}_0 \vv^{(4)}_0 \vv^{(4)}_1 + 8773632 (\vv^{(4)}_0)^6 + 548864 (\vv^{(4)}_1)^6 + 11340 (\vv^{(6)}_0)^3 - 316 (\vv^{(6)}_1)^3 + 672 (\vv^{(8)}_0)^2 - 64 (\vv^{(8)}_1)^2 + 288 (\vv^{(8)}_2)^2
\end{dmath*}

\begin{dmath*}
  \vhat^{(14)}_{12} =  7 \vv^{(14)}_0 - 3 \vv^{(14)}_1 - \vv^{(14)}_2 + 5 \vv^{(14)}_3 + 4243456 \vv^{(4)}_0 (\vv^{(4)}_1)^5 + 336 \vv^{(4)}_0 \vv^{(12)}_0 - 144 \vv^{(4)}_0 \vv^{(12)}_1 - 48 \vv^{(4)}_0 \vv^{(12)}_2 + 240 \vv^{(4)}_0 \vv^{(12)}_3 - 7446528 (\vv^{(4)}_0)^2 (\vv^{(4)}_1)^4 + 8400 (\vv^{(4)}_0)^2 \vv^{(10)}_0 - 3600 (\vv^{(4)}_0)^2 \vv^{(10)}_1 - 1200 (\vv^{(4)}_0)^2 \vv^{(10)}_2 + 3342336 (\vv^{(4)}_0)^3 (\vv^{(4)}_1)^3 + 143360 (\vv^{(4)}_0)^3 \vv^{(8)}_0 - 61440 (\vv^{(4)}_0)^3 \vv^{(8)}_1 - 20480 (\vv^{(4)}_0)^3 \vv^{(8)}_2 + 5431296 (\vv^{(4)}_0)^4 (\vv^{(4)}_1)^2 - 7520256 (\vv^{(4)}_0)^5 \vv^{(4)}_1 - 144 \vv^{(4)}_1 \vv^{(12)}_0 + 16 \vv^{(4)}_1 \vv^{(12)}_1 + 112 \vv^{(4)}_1 \vv^{(12)}_2 - 240 \vv^{(4)}_1 \vv^{(12)}_3 + 1040 (\vv^{(4)}_1)^2 \vv^{(10)}_0 + 1200 (\vv^{(4)}_1)^2 \vv^{(10)}_1 - 3440 (\vv^{(4)}_1)^2 \vv^{(10)}_2 + 8192 (\vv^{(4)}_1)^3 \vv^{(8)}_0 - 34816 (\vv^{(4)}_1)^3 \vv^{(8)}_1 + 61440 (\vv^{(4)}_1)^3 \vv^{(8)}_2 + 1827840 \vv^{(6)}_0 (\vv^{(4)}_0)^4 - 310272 \vv^{(6)}_0 (\vv^{(4)}_1)^4 + 420 \vv^{(6)}_0 \vv^{(10)}_0 - 180 \vv^{(6)}_0 \vv^{(10)}_1 - 60 \vv^{(6)}_0 \vv^{(10)}_2 + 241920 (\vv^{(6)}_0)^2 (\vv^{(4)}_0)^2 + 29952 (\vv^{(6)}_0)^2 (\vv^{(4)}_1)^2 - 783360 \vv^{(6)}_1 (\vv^{(4)}_0)^4 + 527360 \vv^{(6)}_1 (\vv^{(4)}_1)^4 - 4860 \vv^{(6)}_1 (\vv^{(6)}_0)^2 - 180 \vv^{(6)}_1 \vv^{(10)}_0 + 4 \vv^{(6)}_1 \vv^{(10)}_1 + 172 \vv^{(6)}_1 \vv^{(10)}_2 + 10496 (\vv^{(6)}_1)^2 (\vv^{(4)}_0)^2 - 66304 (\vv^{(6)}_1)^2 (\vv^{(4)}_1)^2 + 492 (\vv^{(6)}_1)^2 \vv^{(6)}_0 - 192 \vv^{(8)}_0 \vv^{(8)}_1 - 64 \vv^{(8)}_0 \vv^{(8)}_2 + 192 \vv^{(8)}_2 \vv^{(8)}_1 - 7200 \vv^{(4)}_0 \vv^{(4)}_1 \vv^{(10)}_0 + 928 \vv^{(4)}_0 \vv^{(4)}_1 \vv^{(10)}_1 + 5344 \vv^{(4)}_0 \vv^{(4)}_1 \vv^{(10)}_2 + 53248 \vv^{(4)}_0 (\vv^{(4)}_1)^2 \vv^{(8)}_0 + 55296 \vv^{(4)}_0 (\vv^{(4)}_1)^2 \vv^{(8)}_1 - 163840 \vv^{(4)}_0 (\vv^{(4)}_1)^2 \vv^{(8)}_2 - 184320 (\vv^{(4)}_0)^2 \vv^{(4)}_1 \vv^{(8)}_0 + 28672 (\vv^{(4)}_0)^2 \vv^{(4)}_1 \vv^{(8)}_1 + 126976 (\vv^{(4)}_0)^2 \vv^{(4)}_1 \vv^{(8)}_2 + 417792 \vv^{(6)}_0 \vv^{(4)}_0 (\vv^{(4)}_1)^3 + 20160 \vv^{(6)}_0 \vv^{(4)}_0 \vv^{(8)}_0 - 8640 \vv^{(6)}_0 \vv^{(4)}_0 \vv^{(8)}_1 - 2880 \vv^{(6)}_0 \vv^{(4)}_0 \vv^{(8)}_2 + 1357824 \vv^{(6)}_0 (\vv^{(4)}_0)^2 (\vv^{(4)}_1)^2 - 3133440 \vv^{(6)}_0 (\vv^{(4)}_0)^3 \vv^{(4)}_1 - 8640 \vv^{(6)}_0 \vv^{(4)}_1 \vv^{(8)}_0 + 1344 \vv^{(6)}_0 \vv^{(4)}_1 \vv^{(8)}_1 + 5952 \vv^{(6)}_0 \vv^{(4)}_1 \vv^{(8)}_2 - 207360 (\vv^{(6)}_0)^2 \vv^{(4)}_0 \vv^{(4)}_1 - 1597440 \vv^{(6)}_1 \vv^{(4)}_0 (\vv^{(4)}_1)^3 - 8640 \vv^{(6)}_1 \vv^{(4)}_0 \vv^{(8)}_0 + 448 \vv^{(6)}_1 \vv^{(4)}_0 \vv^{(8)}_1 + 7744 \vv^{(6)}_1 \vv^{(4)}_0 \vv^{(8)}_2 + 1148928 \vv^{(6)}_1 (\vv^{(4)}_0)^2 (\vv^{(4)}_1)^2 + 626688 \vv^{(6)}_1 (\vv^{(4)}_0)^3 \vv^{(4)}_1 + 1728 \vv^{(6)}_1 \vv^{(4)}_1 \vv^{(8)}_0 + 3776 \vv^{(6)}_1 \vv^{(4)}_1 \vv^{(8)}_1 - 9280 \vv^{(6)}_1 \vv^{(4)}_1 \vv^{(8)}_2 - 207360 \vv^{(6)}_1 \vv^{(6)}_0 (\vv^{(4)}_0)^2 + 50688 \vv^{(6)}_1 \vv^{(6)}_0 (\vv^{(4)}_1)^2 + 77312 (\vv^{(6)}_1)^2 \vv^{(4)}_0 \vv^{(4)}_1 + 82944 \vv^{(6)}_1 \vv^{(6)}_0 \vv^{(4)}_0 \vv^{(4)}_1 + 2924544 (\vv^{(4)}_0)^6 - 843776 (\vv^{(4)}_1)^6 + 3780 (\vv^{(6)}_0)^3 + 844 (\vv^{(6)}_1)^3 + 224 (\vv^{(8)}_0)^2 - 160 (\vv^{(8)}_2)^2
\end{dmath*}

\begin{dmath*}
  \vhat^{(14)}_{14} =  \vv^{(14)}_0 - \vv^{(14)}_1 + \vv^{(14)}_2 - \vv^{(14)}_3 - 2506752 \vv^{(4)}_0 (\vv^{(4)}_1)^5 + 48 \vv^{(4)}_0 \vv^{(12)}_0 - 48 \vv^{(4)}_0 \vv^{(12)}_1 + 48 \vv^{(4)}_0 \vv^{(12)}_2 - 48 \vv^{(4)}_0 \vv^{(12)}_3 + 6266880 (\vv^{(4)}_0)^2 (\vv^{(4)}_1)^4 + 1200 (\vv^{(4)}_0)^2 \vv^{(10)}_0 - 1200 (\vv^{(4)}_0)^2 \vv^{(10)}_1 + 1200 (\vv^{(4)}_0)^2 \vv^{(10)}_2 - 8355840 (\vv^{(4)}_0)^3 (\vv^{(4)}_1)^3 + 20480 (\vv^{(4)}_0)^3 \vv^{(8)}_0 - 20480 (\vv^{(4)}_0)^3 \vv^{(8)}_1 + 20480 (\vv^{(4)}_0)^3 \vv^{(8)}_2 + 6266880 (\vv^{(4)}_0)^4 (\vv^{(4)}_1)^2 - 2506752 (\vv^{(4)}_0)^5 \vv^{(4)}_1 - 48 \vv^{(4)}_1 \vv^{(12)}_0 + 48 \vv^{(4)}_1 \vv^{(12)}_1 - 48 \vv^{(4)}_1 \vv^{(12)}_2 + 48 \vv^{(4)}_1 \vv^{(12)}_3 + 1200 (\vv^{(4)}_1)^2 \vv^{(10)}_0 - 1200 (\vv^{(4)}_1)^2 \vv^{(10)}_1 + 1200 (\vv^{(4)}_1)^2 \vv^{(10)}_2 - 20480 (\vv^{(4)}_1)^3 \vv^{(8)}_0 + 20480 (\vv^{(4)}_1)^3 \vv^{(8)}_1 - 20480 (\vv^{(4)}_1)^3 \vv^{(8)}_2 + 261120 \vv^{(6)}_0 (\vv^{(4)}_0)^4 + 261120 \vv^{(6)}_0 (\vv^{(4)}_1)^4 + 60 \vv^{(6)}_0 \vv^{(10)}_0 - 60 \vv^{(6)}_0 \vv^{(10)}_1 + 60 \vv^{(6)}_0 \vv^{(10)}_2 + 34560 (\vv^{(6)}_0)^2 (\vv^{(4)}_0)^2 + 34560 (\vv^{(6)}_0)^2 (\vv^{(4)}_1)^2 - 261120 \vv^{(6)}_1 (\vv^{(4)}_0)^4 - 261120 \vv^{(6)}_1 (\vv^{(4)}_1)^4 - 1620 \vv^{(6)}_1 (\vv^{(6)}_0)^2 - 60 \vv^{(6)}_1 \vv^{(10)}_0 + 60 \vv^{(6)}_1 \vv^{(10)}_1 - 60 \vv^{(6)}_1 \vv^{(10)}_2 + 34560 (\vv^{(6)}_1)^2 (\vv^{(4)}_0)^2 + 34560 (\vv^{(6)}_1)^2 (\vv^{(4)}_1)^2 + 1620 (\vv^{(6)}_1)^2 \vv^{(6)}_0 - 64 \vv^{(8)}_0 \vv^{(8)}_1 + 64 \vv^{(8)}_0 \vv^{(8)}_2 - 64 \vv^{(8)}_2 \vv^{(8)}_1 - 2400 \vv^{(4)}_0 \vv^{(4)}_1 \vv^{(10)}_0 + 2400 \vv^{(4)}_0 \vv^{(4)}_1 \vv^{(10)}_1 - 2400 \vv^{(4)}_0 \vv^{(4)}_1 \vv^{(10)}_2 + 61440 \vv^{(4)}_0 (\vv^{(4)}_1)^2 \vv^{(8)}_0 - 61440 \vv^{(4)}_0 (\vv^{(4)}_1)^2 \vv^{(8)}_1 + 61440 \vv^{(4)}_0 (\vv^{(4)}_1)^2 \vv^{(8)}_2 - 61440 (\vv^{(4)}_0)^2 \vv^{(4)}_1 \vv^{(8)}_0 + 61440 (\vv^{(4)}_0)^2 \vv^{(4)}_1 \vv^{(8)}_1 - 61440 (\vv^{(4)}_0)^2 \vv^{(4)}_1 \vv^{(8)}_2 - 1044480 \vv^{(6)}_0 \vv^{(4)}_0 (\vv^{(4)}_1)^3 + 2880 \vv^{(6)}_0 \vv^{(4)}_0 \vv^{(8)}_0 - 2880 \vv^{(6)}_0 \vv^{(4)}_0 \vv^{(8)}_1 + 2880 \vv^{(6)}_0 \vv^{(4)}_0 \vv^{(8)}_2 + 1566720 \vv^{(6)}_0 (\vv^{(4)}_0)^2 (\vv^{(4)}_1)^2 - 1044480 \vv^{(6)}_0 (\vv^{(4)}_0)^3 \vv^{(4)}_1 - 2880 \vv^{(6)}_0 \vv^{(4)}_1 \vv^{(8)}_0 + 2880 \vv^{(6)}_0 \vv^{(4)}_1 \vv^{(8)}_1 - 2880 \vv^{(6)}_0 \vv^{(4)}_1 \vv^{(8)}_2 - 69120 (\vv^{(6)}_0)^2 \vv^{(4)}_0 \vv^{(4)}_1 + 1044480 \vv^{(6)}_1 \vv^{(4)}_0 (\vv^{(4)}_1)^3 - 2880 \vv^{(6)}_1 \vv^{(4)}_0 \vv^{(8)}_0 + 2880 \vv^{(6)}_1 \vv^{(4)}_0 \vv^{(8)}_1 - 2880 \vv^{(6)}_1 \vv^{(4)}_0 \vv^{(8)}_2 - 1566720 \vv^{(6)}_1 (\vv^{(4)}_0)^2 (\vv^{(4)}_1)^2 + 1044480 \vv^{(6)}_1 (\vv^{(4)}_0)^3 \vv^{(4)}_1 + 2880 \vv^{(6)}_1 \vv^{(4)}_1 \vv^{(8)}_0 - 2880 \vv^{(6)}_1 \vv^{(4)}_1 \vv^{(8)}_1 + 2880 \vv^{(6)}_1 \vv^{(4)}_1 \vv^{(8)}_2 - 69120 \vv^{(6)}_1 \vv^{(6)}_0 (\vv^{(4)}_0)^2 - 69120 \vv^{(6)}_1 \vv^{(6)}_0 (\vv^{(4)}_1)^2 - 69120 (\vv^{(6)}_1)^2 \vv^{(4)}_0 \vv^{(4)}_1 + 138240 \vv^{(6)}_1 \vv^{(6)}_0 \vv^{(4)}_0 \vv^{(4)}_1 + 417792 (\vv^{(4)}_0)^6 + 417792 (\vv^{(4)}_1)^6 + 540 (\vv^{(6)}_0)^3 - 540 (\vv^{(6)}_1)^3 + 32 (\vv^{(8)}_0)^2 + 32 (\vv^{(8)}_1)^2 + 32 (\vv^{(8)}_2)^2 \,. 
\end{dmath*}  

\end{dgroup*} }